\documentclass[12pt]{article}
\pdfoutput=1
\usepackage[colorlinks,linkcolor=Blue,citecolor=Blue,bookmarks,bookmarksnumbered]{hyperref}
\usepackage[scaled=0.85]{helvet}
\usepackage{amsmath,amssymb,accents,mathrsfs,XoohmE}
\usepackage{graphicx,color}
\usepackage{subcaption}
\usepackage{booktabs}
\usepackage{multirow}
\usepackage{placeins}
\usepackage{amsmath}
\usepackage{comment}
\usepackage{XoohmE}

\definecolor{Green}  {rgb}{0.10,0.70,0.10} 
\definecolor{Orange} {rgb}{1.00,0.50,0.15} 
\definecolor{Red}    {rgb}{0.90,0.00,0.12} 
\definecolor{Purple} {rgb}{0.50,0.25,0.55} 
\definecolor{Turque} {rgb}{0.00,0.65,0.85} 
\definecolor{Blue}   {rgb}{0.00,0.00,1.00} 
\definecolor{Magenta}{rgb}{1.00,0.00,1.00} 
\definecolor{Gold}   {rgb}{1.00,0.75,0.25} 
\definecolor{Seaweed}{rgb}{0.01,0.24,0.09} 
\definecolor{Brown}  {rgb}{0.43,0.26,0.32} 
\definecolor{grey1}  {rgb}{0.20,0.20,0.20} 
\definecolor{grey2}  {rgb}{0.40,0.40,0.40} 
\definecolor{grey3}  {rgb}{0.60,0.60,0.60} 
\definecolor{grey4}  {rgb}{0.80,0.80,0.80} 
\definecolor{grey5}  {rgb}{0.90,0.90,0.90} 
\def\C#1#2{{\ifcase#1\or
             \color{Green}\or \color{Orange}\or \color{Red}\or
              \color{Purple}\or \color{Turque}\or \color{Blue}\or
               \color{Magenta}\or \color{Gold}\or \color{Seaweed}\or
                \color{Brown}\or\color{grey1}\or\color{grey2}\or
                 \color{grey3}\else\color{grey4}\fi#2}}

\definecolor{Slate} {rgb}{0.00,0.45,0.55}

\def\rI{{\rm I}}
\def\rJ{{\rm J}}
\def\rK{{\rm K}}
\def\rL{{\rm L}}

\def\hk{{\hat{k}}}

\def\fracm#1#2{\hbox{\large{${\frac{{#1}}{{#2}}}$}}}

\def\be{\begin{equation}}
\def\ee{\end{equation}}
\newcommand{\bea}{\begin{eqnarray}}
\newcommand{\eea}{\end{eqnarray}}
\newcommand{\ena}{\end{eqnarray}}

\def\pp{{\mathchoice
          {
              \kern 1pt%
              \raise 1pt
              \vbox{\hrule width5pt height0.4pt depth0pt
                    \kern -2pt
                    \hbox{\kern 2.3pt
                          \vrule width0.4pt height6pt depth0pt
                          }
                    \kern -2pt
                    \hrule width5pt height0.4pt depth0pt}%
                    \kern 1pt
           }
            {
              \kern 1pt%
              \raise 1pt
              \vbox{\hrule width4.3pt height0.4pt depth0pt
                    \kern -1.8pt
                    \hbox{\kern 1.95pt
                          \vrule width0.4pt height5.4pt depth0pt
                          }
                    \kern -1.8pt
                    \hrule width4.3pt height0.4pt depth0pt}%
                    \kern 1pt
            }
            {
              \kern 0.5pt%
              \raise 1pt
              \vbox{\hrule width4.0pt height0.3pt depth0pt
                    \kern -1.9pt  
                    \hbox{\kern 1.85pt
                          \vrule width0.3pt height5.7pt depth0pt
                          }
                    \kern -1.9pt
                    \hrule width4.0pt height0.3pt depth0pt}%
                    \kern 0.5pt
            }
            {
              \kern 0.5pt%
              \raise 1pt
              \vbox{\hrule width3.6pt height0.3pt depth0pt
                    \kern -1.5pt
                    \hbox{\kern 1.65pt
                          \vrule width0.3pt height4.5pt depth0pt
                          }
                    \kern -1.5pt
                    \hrule width3.6pt height0.3pt depth0pt}%
                    \kern 0.5pt
            }
        }}

\def\mm{{\mathchoice
                       {
                             \kern 1pt
               \raise 1pt    \vbox{\hrule width5pt height0.4pt depth0pt
                                  \kern 2pt
                                  \hrule width5pt height0.4pt depth0pt}
                             \kern 1pt}
                       {
                            \kern 1pt
               \raise 1pt \vbox{\hrule width4.3pt height0.4pt depth0pt
                                  \kern 1.8pt
                                  \hrule width4.3pt height0.4pt depth0pt}
                             \kern 1pt}
                       {
                            \kern 0.5pt
               \raise 1pt
                            \vbox{\hrule width4.0pt height0.3pt depth0pt
                                  \kern 1.9pt
                                  \hrule width4.0pt height0.3pt depth0pt}
                            \kern 1pt}
                       {
                           \kern 0.5pt
             \raise 1pt  \vbox{\hrule width3.6pt height0.3pt depth0pt
                                  \kern 1.5pt
                                  \hrule width3.6pt height0.3pt depth0pt}
                           \kern 0.5pt}
                       }}

\def\ad{{\kern0.5pt
                   \alpha \kern-5.05pt \raise5.8pt\hbox{$\textstyle.$}\kern
0.5pt}}

\def\bd{{\kern0.5pt
                   \beta \kern-5.05pt \raise5.8pt\hbox{$\textstyle.$}\kern
0.5pt}}

\def\qd{{\kern0.5pt
                   q \kern-5.05pt \raise5.8pt\hbox{$\textstyle.$}\kern
0.5pt}}
\def\Dot#1{{\kern0.5pt
     {#1} \kern-5.05pt \raise5.8pt\hbox{$\textstyle.$}\kern
0.5pt}}

\catcode`@=11
\def\un#1{\relax\ifmmode\@@underline#1\else
        $\@@underline{\hbox{#1}}$\relax\fi}
\catcode`@=12

\def\a{\alpha}
\def\b{\beta}
\def\c{\chi}
\def\d{\delta}
\def\e{\epsilon}

\def\g{\gamma}

\def\i{\iota}
\def\j{\psi}
\def\k{\kappa}
\def\l{\lambda}
\def\m{\mu}
\def\n{\nu}
\def\o{\omega}

\def\r{\rho}
\def\s{\sigma}
\def\t{\tau}

\def\L{\Lambda}
\def\O{\Omega}
\def\P{\Pi}

\def\S{\Sigma}

\def\ca{{\cal A}}

\def\cf{{\cal F}}

\def\dslash{\not{\hbox{\kern-2pt $\partial$}}}
\def\Dslash{\not{\hbox{\kern-4pt $D$}}}
\def\pslash{\not{\hbox{\kern-2.3pt $p$}}}
 \newtoks\slashfraction
 \slashfraction={.13}
 \def\slash#1{\setbox0\hbox{$ #1 $}
 \setbox0\hbox to \the\slashfraction\wd0{\hss \box0}/\box0 }
 
\def\kcr{{\hbox{\ro \char'170}}}                
\def\ktl{{\hbox{\ro \char'170}}}        
\def\ktr{{\hbox{\ro \char'170}}}        
\def\kbl{{\hbox{\ro \char'170}}}        
\def\kbr{{\hbox{\ro \char'170}}}        

\def\plpl{\raise-2pt\hbox{$\raise3pt\hbox{$_+$}\hskip-6.67pt\raise0.0pt
\hbox{$^+$}\hskip 0.01pt$}}
\def\mimi{\raise-2pt\hbox{$\raise3pt\hbox{$_-$}\hskip-6.67pt\raise0.0pt
\hbox{$^-$}\hskip 0.01pt$}} 

\def\bo{{\raise.15ex\hbox{\large$\Box$}}}               
\def\pa{\partial}                                       
\def\iff{\leftrightarrow}                               
\def\TH{{\raise.2ex\hbox{$\displaystyle \bigodot$}\mskip-4.7mu \llap H \;}}
\def\face{{\raise.2ex\hbox{$\displaystyle \bigodot$}\mskip-2.2mu \llap {$\ddot
        \smile$}}}                                      

\def\dt#1{\on{\hbox{\bf .}}{#1}}                
\def\Dot#1{\dt{#1}}

\def\Hat#1{\widehat{#1}}                        
\def\leftrightarrowfill{$\mathsurround=0pt \mathord\leftarrow \mkern-6mu
        \cleaders\hbox{$\mkern-2mu \mathord- \mkern-2mu$}\hfill
        \mkern-6mu \mathord\rightarrow$}
\def\dvec#1{\vbox{\ialign{##\crcr
        \leftrightarrowfill\crcr\noalign{\kern-1pt\nointerlineskip}
        $\hfil\displaystyle{#1}\hfil$\crcr}}}           
\def\dt#1{{\buildrel {\hbox{\LARGE .}} \over {#1}}}     

\def\fracm#1#2{\hbox{\large{${\frac{{#1}}{{#2}}}$}}}
\def\sfrac#1#2{{\vphantom1\smash{\lower.5ex\hbox{\small$#1$}}\over
        \vphantom1\smash{\raise.4ex\hbox{\small$#2$}}}} 
\def\bfrac#1#2{{\vphantom1\smash{\lower.5ex\hbox{$#1$}}\over
        \vphantom1\smash{\raise.3ex\hbox{$#2$}}}}       
\def\afrac#1#2{{\vphantom1\smash{\lower.5ex\hbox{$#1$}}\over#2}}    

\def\pa{\partial}      
\let\bm\relax
\newcommand{\bm}[1]{{\boldsymbol{#1}}}

\def\ad{{\dot{\alpha}}}
\def\bd{{\dot{\beta}}}

 \font\rOpe=cmsy10                        
 \def\ktl{{\hbox{\rOpe\char'170}}}        
 \def\kbl{{\hbox{\rOpe\char'170}}}        
 \def\kcr{{\reflectbox{\rOpe\char'170}}}        
 \def\ktr{{\reflectbox{\rOpe\char'170}}}        
 \def\kbr{{\reflectbox{\rOpe\char'170}}}        
 \def\Border{\vbox{\hsize0pt
        \setlength{\unitlength}{1mm}
        \newcount\xco
        \newcount\yco
        \xco=-21
        \yco=12
        \begin{picture}(0,0)(-7.5,0)
        \put(\xco,\yco){$\ktl$}
        \advance\yco by-1
        {\loop
        \put(\xco,\yco){$\kcr$}
        \advance\yco by-2
        \ifnum\yco>-240
        \repeat
        \put(\xco,\yco){$\kbl$}}
        \xco=170
        \yco=12
        \put(\xco,\yco){$\ktr$}
        \advance\yco by-1
        {\loop
        \put(\xco,\yco){$\kcr$}
        \advance\yco by-2
        \ifnum\yco>-240
        \repeat
        \put(\xco,\yco){$\kbr$}}
        \put(-19.5,13){\scalebox{.6065}{%
         University of Maryland Center for String and Particle  Theory \&\ Physics Department%
        |University of Maryland Center for String and Particle  Theory \&\ Physics Department}}
        \put(-19.5,-241.5){\scalebox{.5835}{%
         ****University of Maryland * Center for String and
         Particle  Theory* Physics Department****University of Maryland *Center
        for String and Particle  Theory* Physics Department}}
        \end{picture}
        \par\vskip-8mm}}
\definecolor{UMred}{rgb}{.9,.05,.2}
\definecolor{HUblue}{rgb}{.0,.3,.7}

\definecolor{Red}    {rgb}{0.90,0.00,0.12} 
\definecolor{Blue}   {rgb}{0.00,0.00,1.00} 
\definecolor{Green}  {rgb}{0.10,0.70,0.10} 
\definecolor{Turque} {rgb}{0.00,0.65,0.85} 
\definecolor{Orange} {rgb}{1.00,0.50,0.15} 
\definecolor{Magenta}{rgb}{1.00,0.00,1.00} 
\definecolor{Gold}   {rgb}{1.00,0.75,0.25} 
\definecolor{Seaweed}{rgb}{0.01,0.24,0.09} 
\definecolor{Purple} {rgb}{0.50,0.25,0.55} 
\definecolor{Brown}  {rgb}{0.43,0.26,0.32} 
\definecolor{grey1}  {rgb}{0.20,0.20,0.20} 
\definecolor{grey2}  {rgb}{0.40,0.40,0.40} 
\definecolor{grey3}  {rgb}{0.60,0.60,0.60} 
\definecolor{grey4}  {rgb}{0.80,0.80,0.80} 
\definecolor{grey5}  {rgb}{0.90,0.90,0.90} 
\def\C#1#2{{\ifcase#1\or
             \color{Red}\or \color{Green}\or \color{Blue}\or\
              \color{Turque}\or \color{Orange}\or \color{Magenta}\or 
               \color{Gold}\or \color{Seaweed}\or \color{Purple}\or
                \color{Brown}\or\color{grey1}\or\color{grey2}\or
                 \color{grey3}\else\color{grey4}\fi#2}}

\definecolor{Slate} {rgb}{0.00,0.45,0.55}

\newdimen\parshift\parshift=\parindent
\catcode`@=11
 \long\def\@footnotetext#1{\insert\footins{\reset@font\footnotesize
           \interlinepenalty\interfootnotelinepenalty\splittopskip%
            \footnotesep\splitmaxdepth\dp\strutbox\floatingpenalty\@MM%
             \hsize\columnwidth\addtolength{\hsize}{-2\parindent}
              \@parboxrestore\protected@edef\@currentlabel%
              {\csname p@footnote\endcsname\@thefnmark}%
                \color@begingroup%
                 \@makefntext{\rule\z@\footnotesep\ignorespaces#1%
                  \@finalstrut\strutbox}%
                \color@endgroup}}
 \long\def\@makefntext#1{\hglue\parshift%
           \vbox{\noindent\baselineskip=11pt plus.5pt minus.5pt\hb@xt@0em{\hss\@makefnmark\kern1pt}#1}}
\catcode`@=12

\newskip\humongous \humongous=0pt plus 1000pt minus 1000pt
\def\caja{\mathsurround=0pt}
\def\eqalign#1{\,\vcenter{\openup2\jot \caja
        \ialign{\strut \hfil$\displaystyle{##}$&$
        \displaystyle{{}##}$\hfil\crcr#1\crcr}}\,}
\newif\ifdtup

\makeatletter
\def\section{\@startsection{section}{1}{\z@}
        {3ex plus-1ex minus-.2ex}{1pt plus1pt}{\large\sf\bfseries\boldmath}}
\def\subsection{\@startsection{subsection}{2}{\z@}
         {1.5ex plus-1ex minus-.2ex}{0.01pt plus1pt}{\sf\slshape}}
\def\subsubsection{\@startsection{subsubsection}{3}{\z@}
          {1.5ex plus-1ex minus-.2ex}{0.01pt plus0.2pt}{\sf\boldmath}}
\def\paragraph{\@startsection{paragraph}{4}{\z@}
           {.75ex \@plus.5ex \@minus.2ex}{-2mm}{\sf\bfseries\boldmath}}
\makeatother

 \allowdisplaybreaks
 \seceq

\begin{document}

\thispagestyle{empty}
%
\noindent{\small
\hfill{HET-1770  \\ 
$~~~~~~~~~~~~~~~~~~~~~~~~~~~~~~~~~~~~~~~~~~~~~~~~~~~~~~~~~~~~~~~~~$
$~~~~~~~~~~~~~~~~~~~~~~~~~~~~~~~~~~~~~~~~~~~~~~~~~~~~~~~~~~~~~~~~~$
{}
}
\vspace*{8mm}
\begin{center}
{\large \bf
Examples of 4D, $ \mathcal{N}=2$ Holoraumy
}   \\   [12mm]
{\large {
S.\ James Gates, Jr.,\footnote{sylvester$_-$gates@brown.edu}$^{}$} 
and S.-N. Hazel Mak\footnote{sze$_-$ning$_-$mak@brown.edu}$^{}$
}
\\*[12mm]
\emph{
\centering
${}^{}$Department of Physics, Brown University,
\\[1pt]
Box 1843, 182 Hope Street, Barus \& Holley 545,
Providence, RI 02912, USA 
}
 \\*[112mm]
{ ABSTRACT}\\[4mm]
\parbox{142mm}{\parindent=2pc\indent\baselineskip=14pt plus1pt
We provide an introduction to the concepts of Holoraumy tensors, Lorentz
covariant four-dimensional ``Gadgets'', and Gadget angles within the 
context of minimal off-shell 4D, $\cal N$ = 2 supermultiplets.  This is followed 
by the calculation of the Holoraumy tensors, Gadgets, and Gadget angles for 
minimal off-shell supermultiplets. Four tetrahedrons in four 3D subspaces 
of the Holoraumy lattice space are found.
}
 \end{center}
\vfill
\vfill
\clearpage
%

\newpage
\section{Introduction}

Space-time supersymmetry theories in four dimensions can be dated to works of 
the early to middle seventies \cite{susyB1,susyB2,susyB3,susyB4,susyB5}.  Their introduction to 
the physics literature also marked the inauguration of investigations into new mathematical 
subjects - ``super Lie algebras'' and ``super Lie groups.''  As the names suggest, these are 
extensions of the more well established subjects of Lie algebras and Lie groups.  Currently 
is a time that is almost fifty years since the introduction of the space-time supersymmetry 
concept.  A comparison of the former concepts with the latter shows a marked distinction.

Group theory is a much older subject having emerged from algebraic equations, 
geometry, and number theory in the seventeen hundreds.  Thus the subject benefits from a 
long period of exploration and investigation.  In particular, the representation theory of
Lie algebras enjoys a highly developed  status.  This can partly be seen from the tight
nexus of structures involving matrix algebra and elements of graph theory (roots, weights, 
Dynkin diagrams and Young Tableaux) which can be  marshalled 
\cite{Grf0,Grf1,Grf2,Grf3,Grf4,Grf5,GCG,NS} to study issues that arise
surrounding Lie algebras and groups.   As Sophus Lie (1842-1899) was the pioneer who
researched the issue of discovering all group actions infinitesimally acting on manifolds,
the subject bares his name.  However, an impressive and long list of other mathematicians
drove the development of the subject to its current high level of sophistication.  In particular 
it was Cartan \cite{CAR} (1861-1959) who delivered a mathematically rigorous classification 
of {\em {all possible}} simple Lie algebras.  Toward the end of the nineteen seventies, Kac 
\cite{KAC} extended this result with a classification of simple super Lie algebras.  

The concepts of weights and roots play an important role in this line of mathematical
investigation.  However, these two concepts rely on the fact it is possible to partition
Lie algebras using the Jordan-Chevalley decomposition which splits all the generators of
the algebra into two sets.  In practice to accomplish this one must find the maximal commuting
set of generators among all of the generators and construct a set of simultaneous
eigenvectors (along with their corresponding eigenvalues) of this maximal commuting
set.  These eigenvalues then provide a basis for constructing vectors in the space of
weights and the roots (as differences in weight vectors) lead to Dynkin diagrams.

It is here that the spacetime supersymmetry algebra presents a challenge to this
tried and true method as there is no generally accepted definition about the concept
of eigenvectors with respect to supersymmetry generators.

It has been a goal of some of our research to develop tools that can be used to fill 
in the gap left by this absence of concepts related to eigenvectors and eigenvalues.  In
a sense our quest has been to find what structures in spacetime SUSY hold the data
that occurs in Lie algebra via use of eigenvectors and eigenvalues.   This has occurred 
in a continuing series of works that began with a concept of $\cal {GR}$(d, $\cal N$) 
algebras (or ``garden algebras'') \cite{GRana1,GRana2} as a foundation and subsequently 
led to the discover of ``adinkras'' \cite{adnk1}.  Adinkras are one dimensional graphical 
representations \cite{adnkM1,YZ,adnkM2,adnkGEO1,adnkGEO2} of the garden algebras 
and by their study has led to the concept of holoraumy \cite{adnk1dHoloR1,adnk1dHoloR2} 
as a tool for this purpose.  However, the concept of holoraumy possesses a natural 
extension from the context of 1D SUSY systems to higher dimensional ones and 
in particular to 4D SUSY representations \cite{adnk4dGdgt1,adnk4dGdgt2}.
So our proposition is the data normally accessed in Lie algebra (via use of eigenvectors 
and eigenvalues) within the context of spacetime SUSY theories is accessed via the
holoraumy.

The purpose of this work is to continue the extension of this exploration by 
presenting the first report of holoraumy in the context of 4D, $\cal N$ = 2
minimal supersymmetrical representations.

The subsequent presentation unfolds in the following manner.

Chapter two is devoted to a quick review of the tools of representation theory as applied
to the very familiar example of the su(3) algebra.  The usual Gell-Mann representation
of 3 $\times$ 3 matrices is utilized and leads to the usual structure constants and 
``d-coefficients.''  The traditional maximal set of commuting generators and the role
of their simultaneous eigenvectors and associated eigenvalues are noted as the foundation
that advances the understanding of the structure of the su(3) algebra. It is noted the 
partitioning of the generators into a semisimple portion (containing only the commuting 
generators) and a nilpotent portion achieves the Jordan-Chevally decomposition of su(3).   
These are the basis for the considerations of the roots and weights of the algebra.  By parallel 
transport of the roots, a lattice emerges and the vertices of the lattice are noted to be the 
weights of representations.  Finally, the existence of Casimir operators leading to a 
classification of the representations in terms of two integers $p$ and $q$ is observed.  
The integers are then related to the structure of Young Tableaux.

Chapter three contains a discussion of the concept of ``holoraumy'' in the context of 
representations of the 4D, $\cal N$-extended supersymmetry algebra.  The basic definition
of holoraumy is followed by describing a set of conventions and the structures that
emerge from the definition of the holoraumy operator are presented.  For a general
value of $\cal N$ it is noted that the irreducible (with respect to the covering algebra
of the Dirac matrices, generators of so($\cal N$), and the symmetric tensors in the defining
representation of so($\cal N$) appear in the holoraumy operator in precisely a manner
that leads to a set of so(4$\cal N$) generators in the reduction to one dimensional 
supersymmetrical systems.

The fourth chapter contains the new results of this work by presenting the explicit form
of holoraumy for the mininal off-shell supermultiplets that realize 4D, $\cal N$ = 2
supersymmetrical systems with a finite number of auxiliary fields.  The holoraumy operator 
is second order in the D-operators of SUSY and thus possesses an engineering dimension
of one.  It maps bosons to derivatives of bosons and separately fermions to derivatives of 
fermions. So evaluation of the holoraumy operator has two distinct parts.  This chapter only 
presents the evaluation on the fermionic fields.  The facts that holoraumy solely maps bosons 
to derivatives of bosons, separately fermions to derivatives of fermions, and possess an
engineering dimension of one ensures that a set of dimensionless numbers emerge in these
calculations.  These numbers are specific to each supersymmetical multiplet.  It is the
contention of this work that these dimensionless numbers are the analogs of  eigenvalues 
seen in the Chevalley-Jordan decomposition of ordinary Lie algebras.

The chapter reviews the known off-shell  4D, $\cal N$ = 2 supersymmetrical systems with a 
finite number of auxiliary fields.  The systems consist of the vector, tensor, relaxed 
hypermultiplet, supergravity, hyperplet, and higher spin supermultiplets.  The counting 
of component fields in each representation is given and the vector and tensor supermultiplets
(together with their parity duals) are identified as the minimal ones.  A brief discussion of 
some minimal on-shell representations is given to contrast with the off-shell constructions 
and notational conventions are set in place for the presentation of results.

The fifth chapter contains the information that is equivalent to that of the fourth chapter
but with the distinction that these results describe the realization of the holoraumy operator
solely on the bosonic fields.  Owing the disparate spins among the bosons that appear 
in the minimal supermultiplets, the results do not present any obvious interpretation as 
the equivalent one presented on the spinor fields.  For this reason, the spinorial ones 
have been and remain the prime focus of our study.

A sixth chapter sets into place an operator that takes pairs of the minimal supermultiplets
and maps such pairs into a real number.  In the past, this operator has been given the
name of the ``Gadget'' and the corresponding real number is referred to as ``the Gadget
value'' of the pair.  The Gadget essentially defines a ``dot product'' on the space of
supermultiplets.  We show there are a priori a number of definitions of the Gadget
that are consistent with Lorentz and so(2) covariance.  All such definitions are found
to lead to a possible matrix of dot product containing only three values which 
are denoted by ${\cal X}{}_1$,  ${\cal X}{}_2$,  and ${\cal X}{}_3$.

In the seventh chapter, all the results in the previous ones are reduced to the case
where all spatial dimensions are eliminated and thus links to adinkras can
be directly studied.  The corresponding ``L-matrices'' and ``R-matrices'' for the
adinkras with four colors, four open nodes, and four closed nodes are obtained as
substructures of adinkras with eight colors, eight open nodes, and eight closed nodes. 

The eighth chapter builds upon the work of the seventh chapter and constructs the
original 1D holoraumy matrices as conceptualized previously \cite{adnk1dHoloR1,adnk1dHoloR2} 
but now for the first time explicitly in the context of adinkras with eight colors, eight 
open nodes, and eight closed nodes.

The ninth chapter is used to construct the 1D, $N$ = 8 Gadget and Gadget values
associated with the adinkras constructed in the previous chapter.  it is shown that
for one special choice of the parameters ${\cal X}{}_1$,  ${\cal X}{}_2$,  and ${\cal 
X}{}_3$, the matrix of dot products for the 1D, $N$ = 8 Gadget and the 4D, $\cal N$ = 2 Gadget agree and thus realize the concept of ``SUSY holography''.

Chapter ten summarizes the results achieved in this work and examines the
challenges ahead in this line of study.

This work contains three appendices.  The first appendix simply lays out the explicit
basis of matrices and vectors used in this work.  The second chapter lays out the
D-algebra results which are equivalent to SUSY transformation laws.  These have
been presented and validated in previous work and provide the basis for the calculations
in chapter four.  The final appendix contains the explicit expression for the 1D, $N$
= 8 holoraumy matrices associated with the adinkras and used to calculate the
values of the Gadget dot products.

\newpage
\section{Review of Representation Tools for the su(3) Lie Algebra}

For discussion of the su(3) algebra, we use the standard Gell-Mann representation
matrices ${\bm \l}{}_{i}$ (see Appendix A) which satisfy the relations
\be
\left[ \, {\bm T}{}_{i}  ~,~ {\bm T}{}_{j} \, \right] ~=~ i \,f{}_{i \, j}{}^k \, {\bm T}{}_{k} 
~~~,
\label{su(3)alg}
\ee
where ${\bm T}{}_i$ = $\tfrac 12 {\bm \l}{}_{i}$ are the su(3) generators.  The only non-vanishing 
values of the totally anti-symmetric structure constant $ f{}_{i \, j}{}^k$ are completely specified 
by giving the values shown in (\ref{fss}). 
\be  \eqalign{
f_{1\,2\,3} ~&=~ 1, ~~,~~  
 f_{1\,4\,7} ~=~ f_{1\,6\,5} ~=~f_{2\,4\,6} ~=~f_{2\,5\,7} ~=~ f_{3\,4\,5} ~=~
f_{3\,7\,6} ~=~ \fracm 12 ~~,~~ \
f_{4\,5\,8} ~=~ f_{6\,7\,8} ~=~  \fracm {\sqrt 3}2 ~~,
} \label{fss}
\ee
and we use Kronecker deltas to raise and lower indices.  As ${\bm T}{}_3$ and ${\bm T}{}_8$
commute, we define three simultaneous eigenvectors of these generators in terms of their
eigenvalues $t_3$ , and $t_8$, and denoting the 
eigenvectors as $ | t_3 , \, t_8 \rangle$, and thus 
\be
{\bm T}{}_{3} | t_3 , \, t_8 \rangle  ~=~  t_3 \, | t_3 , \, t_8 \rangle ~~~,~~~ {\bm T}{}_{8}  | t_3 , \, t_8 \rangle ~=~  
t_8 \, | t_3 , \, t_8 \rangle  ~~~.
\label{TEs}
\ee
Given an arbitrary representation ($\cal R$) in this space of ``3-tuples'' we can write
\be
| ({\cal R}) \rangle ~=~ \sum_{t_3, \, t_8} \, c[({\cal R}):{t_3, \, t_8}] \, | t_3 , \, t_8 \rangle
~\to~
{\bm T}{}_{i}  \, | ({\cal R}) \rangle ~=~ \sum_{t_3, \, t_8} \, c[({\cal R}):{t_3, \, t_8}] \, {\bm T}{}_{i}  \,| t_3 , \, t_8 \rangle
~~~,
\label{Tmo}
\ee
where $c[({\cal R}):{t_3, t_8}]$  are simply a set of constants.

A well-known result about the matrices in (\ref{su3}) involves their anti-commutator
taking the form
\be
\left\{ \, {\bm T}{}_{i}  ~,~ {\bm T}{}_{j} \, \right\} ~=~ \fracm 13 \,  \d{}_{i \, j} \, {\bm {\rm I}}{}_{3\times3} ~+~
 d{}_{i \, j}{}^k \, {\bm T}{}_{k} 
~~~,
\label{ddee}
\ee
where the totally symmetric ``d-coefficients'' are completely specified 
by giving the values shown in (\ref{ds})
\be  \eqalign{
d{}_{1\,1\,8} ~&=~ d{}_{2\,2\,8} ~=~ d{}_{3\,3\,8} ~=~ - \,d{}_{8\,8\,8} ~=~ \fracm 1{\sqrt 3} ~~,~~  
d{}_{4\,4\,8} ~=~ d{}_{5\,5\,8} ~=~ d{}_{6\,6\,8} ~=~ \,d{}_{7\,7\,8} ~=~ -\, \fracm 1{2 \,{\sqrt 3}}  ~~,~~  \cr
d{}_{1\,4\,6} ~&=~ d{}_{1\,5\,7} ~=~ -\, d{}_{2\,4\,7} ~=~  d{}_{2\,5\,6}  
~=~ d{}_{3\,4\,4} ~=~  d{}_{3\,5\,5} ~=~  - \, d{}_{3\,6\,6} ~=~ - d{}_{3\,7\,7} ~=~\fracm 12 ~~.
}  \label{ds}
\ee
It is of great importance to note that the result in (\ref{ddee}) is not universal for all
representations unlike (\ref{su(3)alg}).

The results in (\ref{TEs}) also open the route to a deeper understanding of su(3) as these lead 
to the roots and weights of the representations via interpreting the results of any explicit evaluation
of (\ref{Tmo}) as describing ``motions'' in the space of the eigenvectors.  We illustrate these in Figure \ref{fig:root} to follow.
\begin{figure}[h!]
\centering
\includegraphics[width=3.3in]{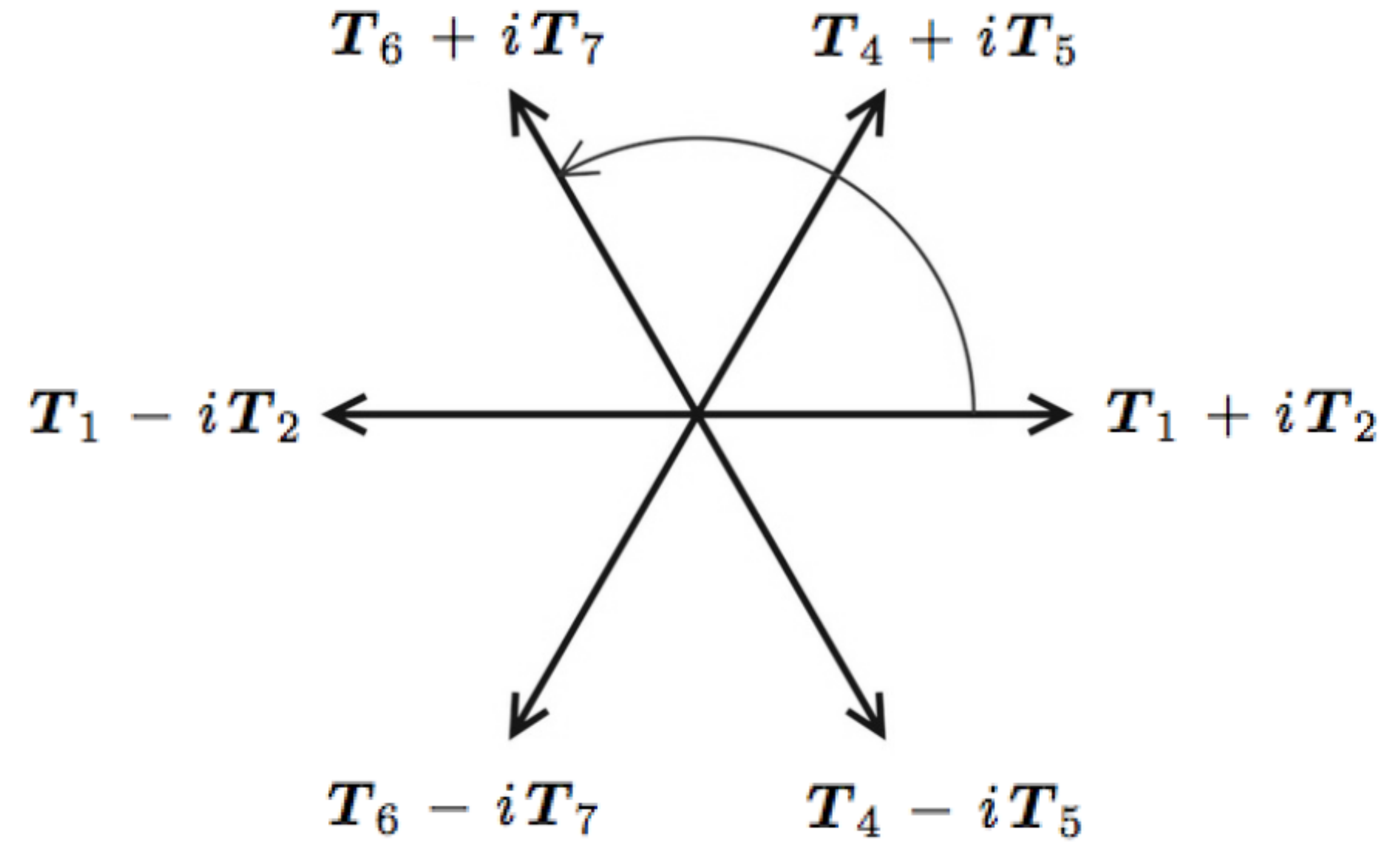}
\vspace*{-0.2cm}
\caption{Illustration of ``motions'' of su(3) generators.}
\label{fig:root}
\vspace*{-0.2cm}
\end{figure}

The angle shown above is equal to $2 \pi/3$ and the arrows show the motions of the 
generators which are the ``roots'' of the su(3) algebra.  By parallel transport of the roots 
along one another, one can generate \newpage \noindent
a lattice-like structure (the root space) shown in Figure \ref{fig:su3lattice}.

\begin{figure}[h!]
\centering
\begin{subfigure}{0.45\linewidth}
\centering
\includegraphics[width=0.63\linewidth]{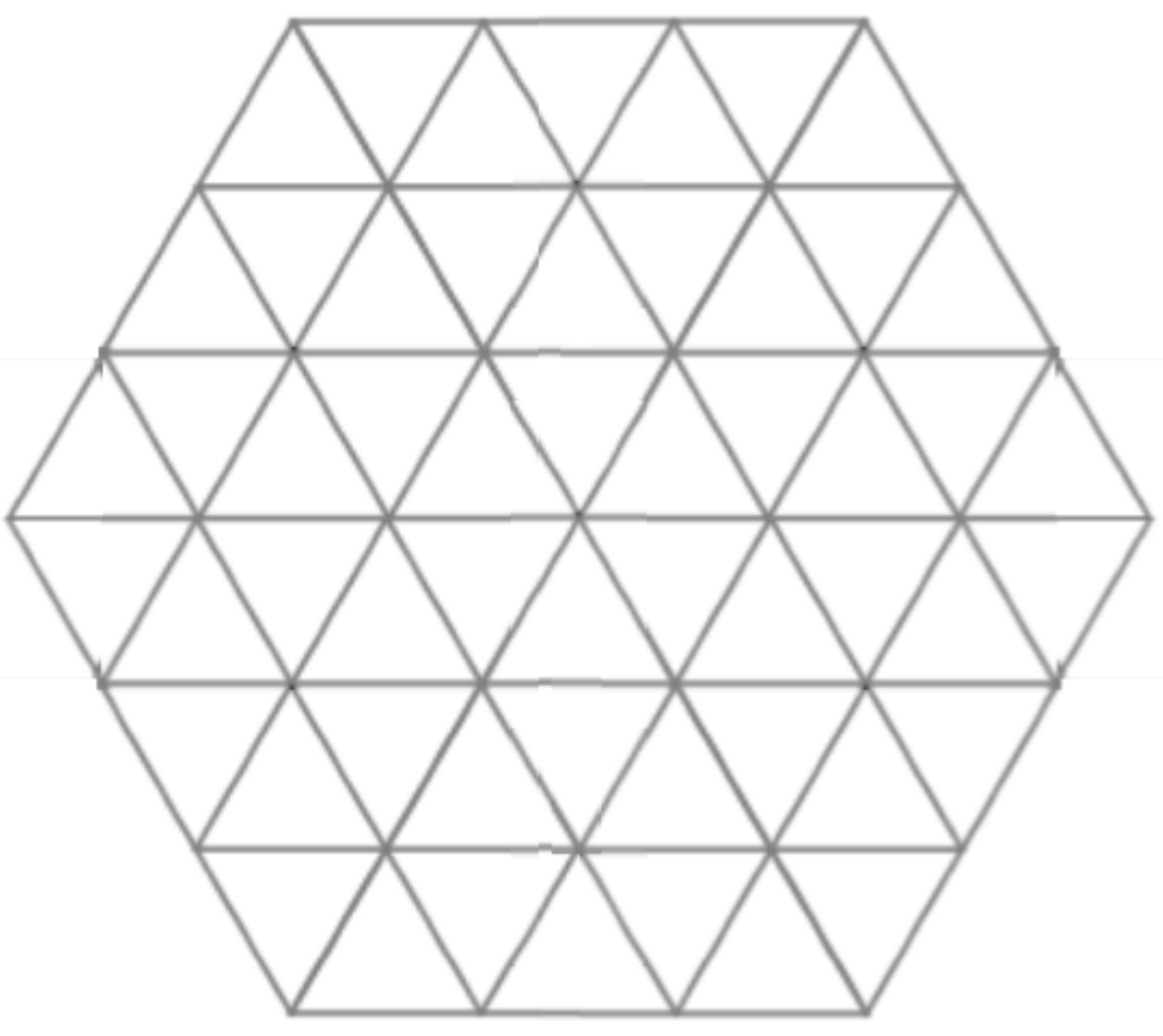}
\caption{}
\end{subfigure}
\begin{subfigure}{0.45\linewidth}
\centering
\includegraphics[width=0.67\linewidth]{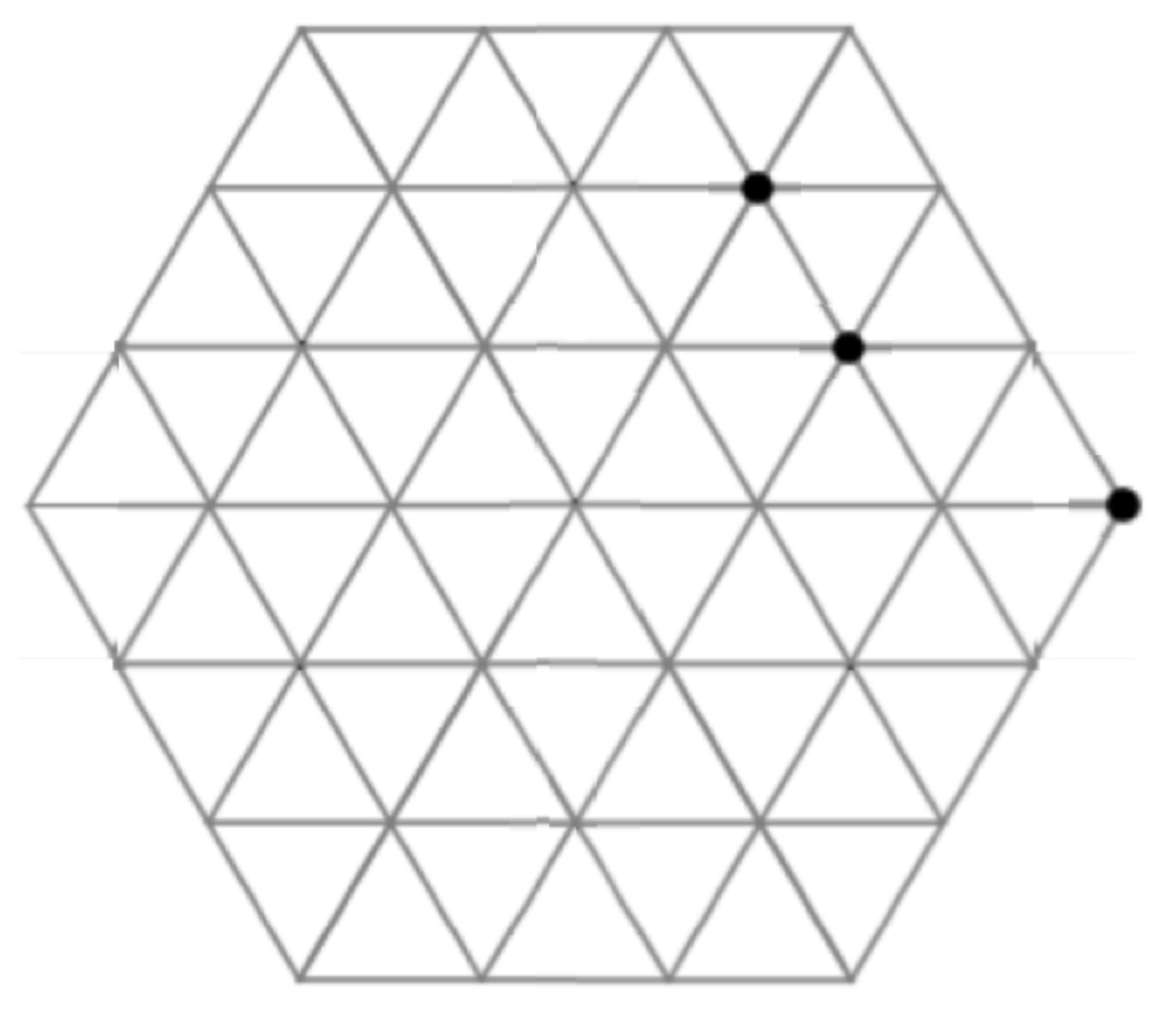}
\caption{}
\end{subfigure}
\caption{Illustrations of: (a) Lattice from su(3) motions and (b) three states on some vertices.}
\label{fig:su3lattice}
\end{figure}

The weights of states of any representation can be visualized the coordinates of points occupying the vertices of the lattice as shown to the right in Figure \ref{fig:su3lattice}.

It is well understood that the states of irreducible representations are not randomly distributed\footnote{It is this fact, that led to the very discovery of the role that Lie algebras play in fundamental physics.}, but instead fall into regular patterns as seen in Figure \ref{fig:su3plets}\footnote{In the diagrams shown in Figure \ref{fig:su3plets}, the quantity $s$ denotes strangeness while $q$ denotes electrical charge.}. 
\begin{figure}[h!]
\centering
\begin{subfigure}{0.48\linewidth}
\includegraphics[width=1.05\linewidth]{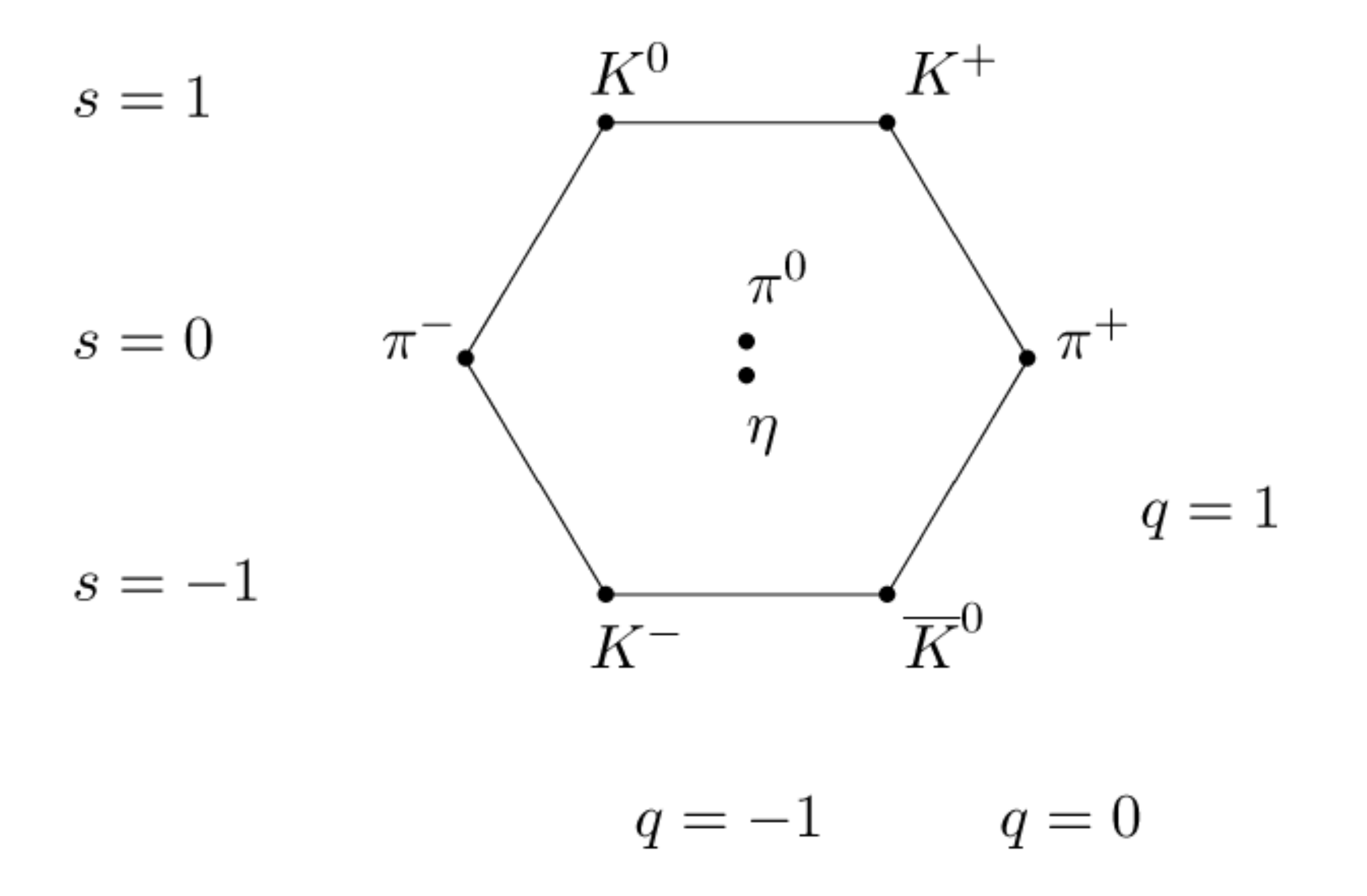}
\caption{}
\end{subfigure}
\begin{subfigure}{0.48\linewidth}
\vspace*{-0.4cm}
\centering
\includegraphics[width=1.05\linewidth]{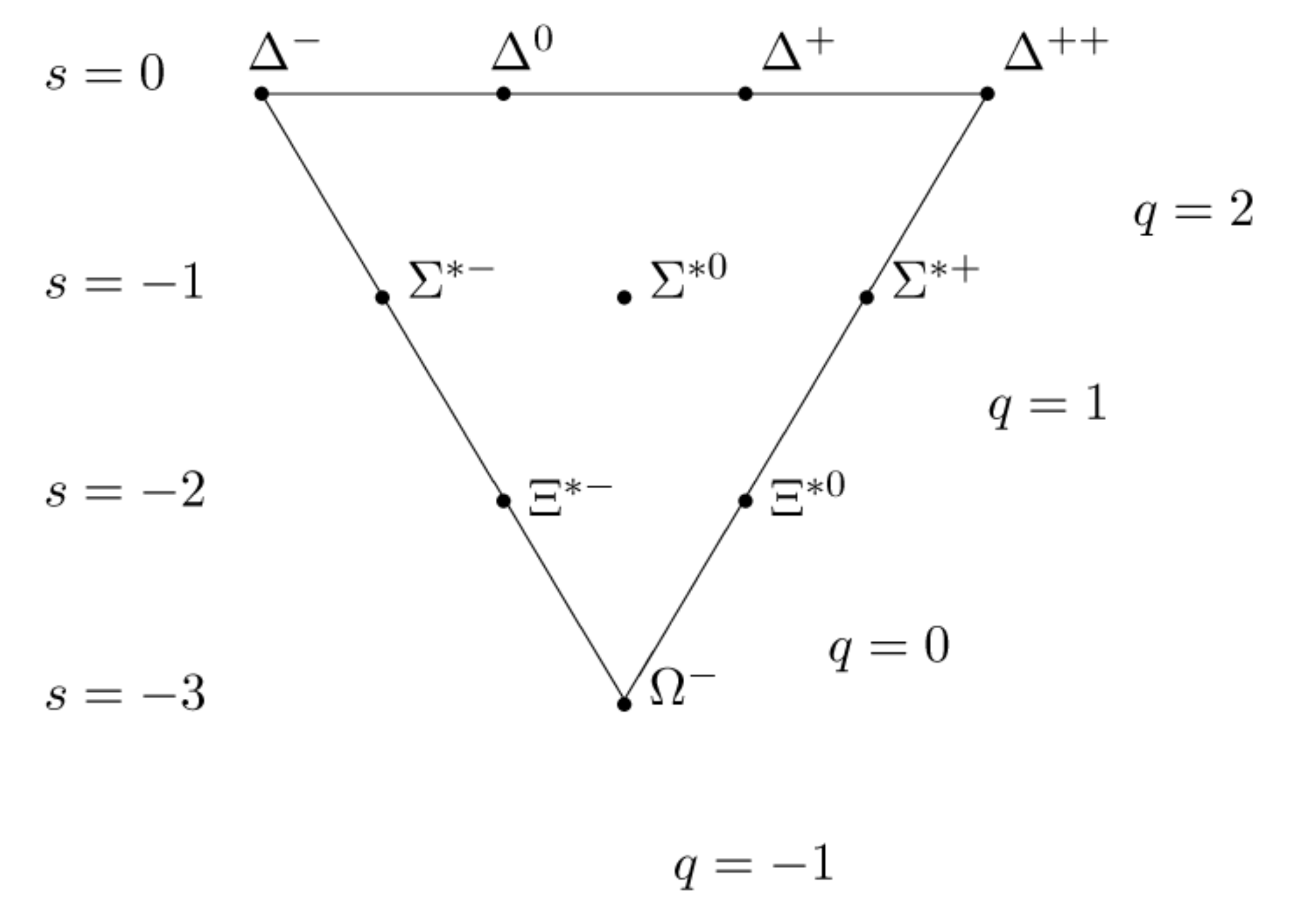}
\caption{}
\end{subfigure}
\caption{Illustrations of: (a) octet representation and (b) decuplet representation.}
\label{fig:su3plets}
\end{figure}

Two integers $p$ and $q$ can be found by examining the weight of
the state with the largest value of the eigenvalue $t{}_3$ and simultaneously looking
at the value of the eigenvalue of $t{}_8$ for this same state.  For this state, the relations
\be
t{}_3 ~=~ \fracm 12 \, \left(  \,  p ~+~ q \, \right)  ~~~,~~~ t{}_8 ~=~ \, \fracm 1{2{\sqrt 3 \,}} \, 
\left(  \,  p ~-~ q \, \right) ~~~,
\ee
are valid and thus through these eigenvalues of the maximal commuting group, the
integers $p$ and $q$ are determined quantities.

All irreducible representations are characterized by the two integers $p$ and $q$
so that the number of states in any irreducible representations described by
$p$ and $q$ is given by applying the Weyl dimension formula to su(3)\cite{WEY}
\be {
d(p, \, q) ~=~ \fracm 12 \, \left( \, p \,+\, 1  \, \right) \, \left( \,  q \,+\, 1\, \right)  \, \left( \, 
 p \,+\, q \,+\, 2 \, \right) ~~~.
 \label{dimnum}
} \ee


These same integers characterize the Quadratic Casimir via Freudenthal's formula 
\cite{FrL1,FrL2,FrL3}
\be
{C}{}_2(p, \, q) ~\propto~ \fracm 14 \,  \sum_{i = 1}^8 \,  {\rm {Tr}} \left( \, {\bm  \l}{}_i \, {\bm  \l}{}_i 
\, \right)
~=~ \fracm 13\, \left( \, p^2 \,+\, q^2 \,+\, p\, q \,+\, 3 \, p \,+\, 3\, q \,  \right)  ~~~,
\label{Frd4b}
\ee
as well as a Cubic Casimir \cite{Pais}
\be
{C}{}_3(p, \, q) ~\propto~ \fracm 18  \sum_{i = 1, j= 1, k = 1}^8 \, d{}_{i \, j \, k}\,  {\rm {Tr}} \left( \, 
 {\bm  \l}{}_i \, {\bm  \l}{}_j \, {\bm  \l}{}_k \, \right) ~=~ \fracm 1{18} \left( \,p\,-\, q \, \right) \,
 \left( \, 3\,+\, p \,+\, 2\, q \, \right) \, \left( \, 3 \,+\, q \, +\,2 \, p \, \right)  ~~~.
 \label{Frd4c}
\ee

The ubiquitous appearance of the two integers $p$ and $q$ point toward another aspect of the representation theory of su(3) as they are avatars for the existence of Young Tableaux.  As shown in Figure \ref{fig:pqyoung}, the integers $p$ and $q$ respectively describe the length of one-height rows and the length of two-height rows in any given irreducible representation.
\begin{figure}[h!]
\vspace*{-0.3cm}
\centering
\includegraphics[width=0.3\linewidth]{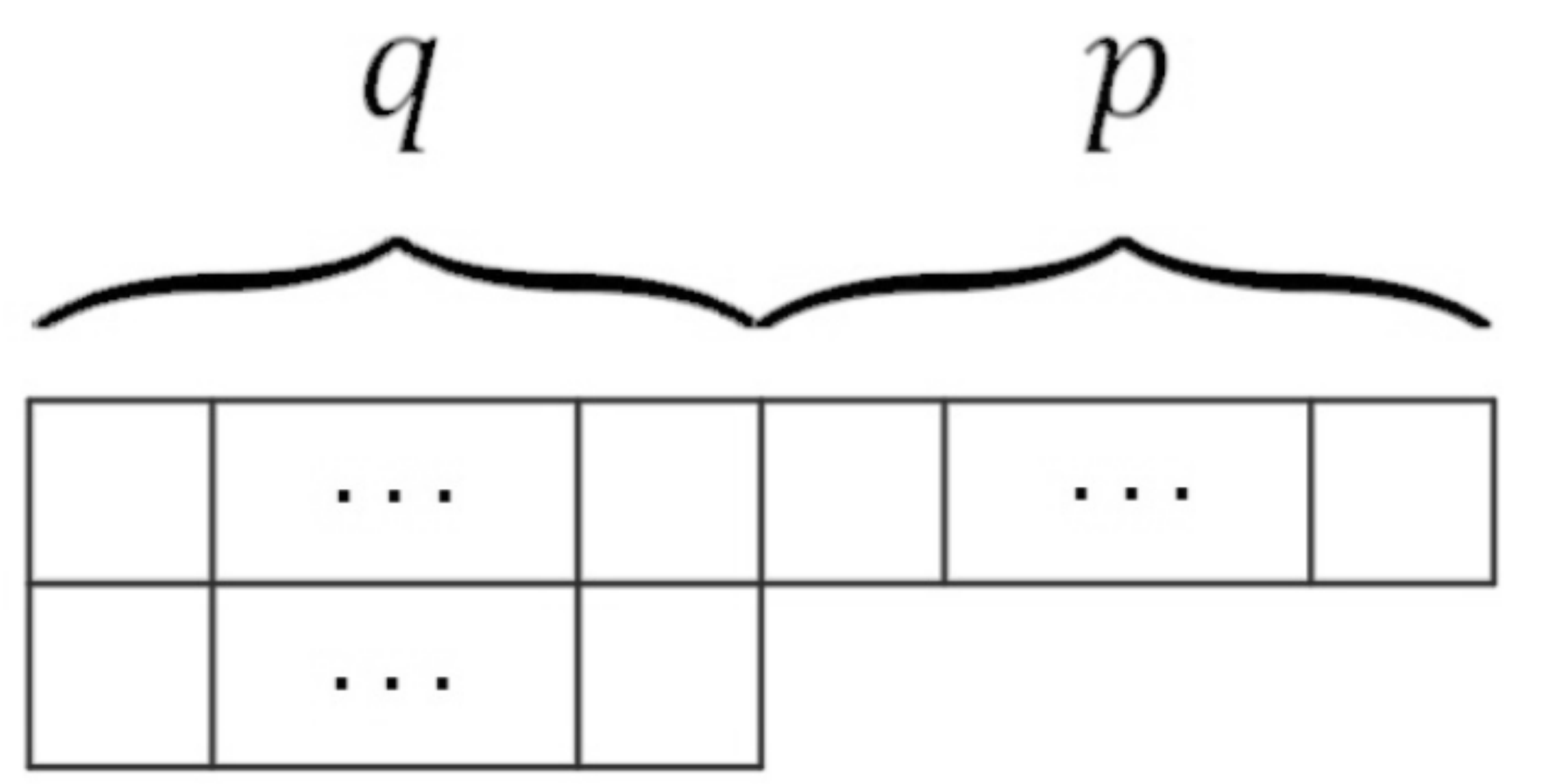}
\caption{Illustration of integers ${\bm p}$ and ${\bm q}$ within su(3) Young Tableau.}
\label{fig:pqyoung}
\end{figure}

As the discussion in this chapter has shown, the matrix representation of su(3)
leads, via eigenvectors and eigenvalues (\ref{TEs}) to roots and weights.  However,
via the Casimir values the matrix representation also leads to the Young Tableaux
representations.  All are tightly linked together.

We wish to close this discussion by concentrating on one more Casimir, the one
associated with a quartic Casimir value and given by
\be
{C}{}_4(p, \, q) ~=~ \fracm 1{16}   \sum_{i = 1, j= 1}^8 \,  {\rm {Tr}} \left( \, \left\{ {\bm \l}{}_i
~,~ {\bm  \l}{}_j \,  \right\} \, \left\{ {\bm  \l}{}^i ~,~ {\bm  \l}{}^j \,  \right\} \, \right)
~=~ d(p, \, q) \, C_2(p, \, q)  \,  \left[ \, 4 \,  C_2(p, \, q)  ~-~ 3 \, \right]~~~,
\label{su3GdGt}
\ee
and after reviewing the literature, we have not found a prior determination of the function 
to the far right in (\ref{su3GdGt}).  In a separate work \cite{GMX}, an argument is 
given for how this conclusion was reached.   

Whenever two representations (${\cal R}$) and  (${\cal R}^{\prime}$) characterized
respectively by $(p, \, q)$ and $(p^{\prime}, \, q^{\prime})$ satisfy the condition
\be
d(p, \, q) ~=~ d(p^{\prime}, \, q^{\prime})   ~~~,
\ee
where $d(p, \, q)$ is defined in Eq. (\ref{dimnum}), the matrices $ {\bm  \l}{}_i{}^{({\cal R})}$
associated with the first representation have the same size as matrices $ {\bm  \l}{}_i{}^{({
\cal R}^{\prime})}$ associated with the second representation.  In this case, we can deform
the quartic Casimir to define\footnote{In appendix A, this concept is considered in the
context of the C-K-M matrix.} 
\be
{C}{}_{4,\, \cal G} ({\cal R}, \, {\cal R}^{\prime}) ~=~ \fracm 1{16}   \sum_{i = 1, j= 1}^8 \,  
{\rm {Tr}} \left( \, \left\{ {\bm  \l}{}_i{}^{({\cal R})} ~,~ {\bm  \l}{}_j{}^{({\cal R})}  \,  \right\} \, 
\left\{ {\bm  \l}{}^i{}^{({\cal R}^{\prime})} ~,~ {\bm  \l}{}^j{}^{({\cal R}^{\prime})}  \,  \right\} \, 
\right)
~~~.
\label{quGdGt}
\ee
In the discussion of supersymmetry that follows a similar structure called ``the Gadget''
will play a role.

\newpage
\section{General ${\bm 4}{\bf D}$, $\cal N$-extended SUSY Holoraumy Structure }}

On the basis of four dimensional Lorentz- and as well so($\cal N$)-covariance, the 
work of \cite{adnk1dHoloR1} asserted that the most general equation for the holoraumy 
operator of a supersymmetrical multiplet must take the form\footnote{We have used a 
slightly different notation here in comparison to previous presentations.}
\be  \eqalign{ {\,}
[ \, {\rm D}{}_{a \, i}  ~,~ {\rm D}{}_{b \, j} \, ] =&~ 
i \, C_{a \, b} \,  \d {}_{i \, j} \, { {  \mathscr{H}}}{}^{(1)}  +  (\g^{5}){}_{a \, b} \, 
\d {}_{i \, j} \,  {  \mathscr{H}}{}^{(2)} +  (\g^{5} \g^{\m}){}_{a \, b} \, \d 
{}_{i \, j} \,  { \mathscr{H}}{}_{\m}{}^{(3)}   \cr
& + i \, C_{a \, b} \, (\,  \mathscr{S}{}_{i \, j}^{(S)} \cdot  { {\mathscr{H}}}{}^{(4:(S))}\,) + (\g^{5}){}_{a \, b} \, ( \,  \mathscr{S}{}_{i \, j}^{(S)} \cdot  {\mathscr{H}}{}^{(5: (S))} \, )  + (\g^{5} \g^{\m}){}_{a \, b} \, (\, \mathscr{S}{}_{i \, j}^{(S)} \cdot  {\mathscr{H}}{}_{\m}{}^{(6: (S))}  \, )  \cr
& + i\, (\g^{\m} ){}_{a \, b} \, ( \, \mathscr{A}{}_{i \, j}^{[A]} \cdot  \mathscr{H}{}_{\m}{}^{(7:[A])} \,) 
+ i\, \fracm 12 \,  ([\, \g^{\m} ~,~ \g^{\n}\,]){}_{a \, b} \,  ( \, \mathscr{A}{}_{i \, j}^{[A]} \cdot \mathscr{H}{}_{\m \, \n}{}^{(8:[A])}  \, )  ~~~,
} \label{H1}
\ee
where in this expression, the quantity $C{}_{a \, b}$ is the usual ``spinor metric'' and $(\g^{\m} ){}_{a \, b}$ denote the usual Dirac gamma matrices\footnote{See appendix A in the work of \cite{G-1} for our conventions.}. The quantities $\d {}_{i \, j}$, $ \mathscr{S}{}_{i \, j}^{(S)}$ and $\mathscr{A}
{}_{i \, j}^{[A]}$ are tensors in so($\cal N$). All of these tensors possess indices $i$ and $j$ taking on values 1,  $\dots$, $\cal N$. The first of these denotes the Kronecker delta tensor. 
The second collection of these objects denoted by $\mathscr{S}{}_{i \, j}^{(S)}$ satisfies $\mathscr{S}{}_{i \, j}^{(S)} = \mathscr{S}{}_{j \, i}^{(S)}$ and $\d{}_{i \, j} \mathscr{S}{}_{i \, j}^{(S)}$ = 0, where the index $(S)$ takes on values $1, \dots, ({\cal N} + 2) (\cal N - $1)/2.  The final collection of such objects denoted by $\mathscr{A}{}_{i \, j}^{[A]}$ satisfies $\mathscr{A}{}_{i \, j}^{[A]} = - \mathscr{A}{}_{j \, i}^{[A]}$, where the index $[A]$ takes on values 1, $\dots$, $\cal N(\cal N - $1)/2. 

We may regard the quantities $\mathscr{A}{}_{i \, j}^{[A]}$ as the generators of so($\cal N$). Under this interpretation, $\d {}_{i \, j}$ and $\mathscr{S}{}_{i \, j}^{(S)}$ are respectively the so($\cal N$) invariant and the second order traceless symmetric tensor representations, respectively, under the action of the so($\cal N$) generators $\mathscr{A}{}_{i \, j}^{[A]}$.  In the expression (\ref{H1}), the $4 {\cal N} (4 {\cal N} - 1)/2$ quantities
\be \eqalign{
&i C{}_{a \, b} \, \d {}_{i \, j}  ~~~~~~~\,~,~~~
(\g^{5} ){}_{a \, b} \, \d {}_{i \, j} ~~~~~\,~,~~~
(\g^{5} \g^{\m}){}_{a \, b} \,   \d {}_{i \, j}  ~~~\,~~, \cr
&i \, C_{a \, b} \,  \mathscr{S}{}_{i \, j}^{(S)} ~~~~~,~~~
(\g^{5}){}_{a \, b} \,  \mathscr{S}{}_{i \, j}^{(S)} ~\,~\,~,~~~
(\g^{5} \g^{\m}){}_{a \, b} \, \mathscr{S}{}_{i \, j}^{(S)} ~\,~,~~~ \cr
& i (\g^{\m} ){}_{a \, b} \,  \mathscr{A}{}_{i \, j}^{[A]}  ~~~,~~~
i \, \fracm 12 \,  ([\, \g^{\m} ~,~ \g^{\n}\,]){}_{a \, b} \, \mathscr{A}{}_{i \, j}^{[A]} 
 ~~~~~~~~~~~~\,~~~, 
} \label{songen}
\ee
are equal in number to the generators of so(4$\cal N$).  In fact, each of these quantities possesses
an antisymmetry property under the pairwise exchange of indices $a \, i$ $\iff$ $b \, j$.  So
the total collection (\ref{songen}) must provide a representation of so(4$\cal N$).  When such a
system is subjected to reduction on a torus to one dimension, this observation is the origin of the 
fact that an so(4$\cal N$) symmetry is present after the reduction.

Finally, in (\ref{H1}), the $4 {\cal N} (4 {\cal N} - 1)/2$ quantities 
\be \eqalign{
&\mathscr{H}{}_{\m}{}^{(1)}~~~~~,~~~ \mathscr{H}{}_{\m}{}^{(2)}~~~~~,~~~
{\mathscr{H}}{}_{\m}{}^{(3)}~~~~~,
\cr
&\mathscr{H}{}^{(4:S)}~~~~,~~~ \mathscr{H}^{(5:S)}~~~~~,~~~ \mathscr{H}{}_{\m}{}^{(6:S)}~~~,~~~
\cr
&\mathscr{H}{}_{\m}{}^{(7:A)}~~~,~~~
\mathscr{H}{}_{\m \, \n}{}^{(8:A)} ~~~~~~~~~~~~~~~~~~~,
} \label{Liegens}
\ee
correspond to Lie-algebra operators whose explicit forms depend on the supermultiplet on which the holoraumy operator is being evaluated.

A major purpose of this current work is to ``flesh out'' (i.\ e.\ present explicit results) the assertions made in writing (\ref{H1}).  As our previous discussions of 4D holoraumy were in the context of $\cal N$ = 1 supermultiplets \cite{adnk4dGdgt1,adnk4dGdgt2}, the quantities $\mathscr{ S}{}_{i \, j}^{(S)}$ and $\mathscr{A}{}_{i \, j}^{[A]}$ perforce vanish.  Thus, the work that follows provides the first opportunity to realize these more general 
structures by working out details within the context of a specific set of examples.   One 
such point of focus will be to examine the commutator algebra (appropriate for $\cal N
$ = 2) that follows from the generators shown in (\ref{songen}).

\newpage
\newpage
\section {Structure of ${\bm 4}{\bf D}$, $\cal N$ = 2 SUSY Fermionic Holoraumy For Minimal Multiplets}

To begin the discussion of the 4D, $\mathcal{N}$ = 2 fermionic holoraumy, we first need to
specify the definition and notation of the 4D, $\mathcal{N}$ = 2 supercovariant derivative 
operator via the equation
\be
  \text{D}^{i}_{a} ~=~ \begin{bmatrix}
\,  \text{D}^{1}_{a} \,  \\
\,  \text{D}^{2}_{a} \,
 \end{bmatrix}   ~~~~,~~~~    \text{D}_{a}{}_{i} ~=~ \d {}_{i \, j} \,  \text{D}^{j}_{a} 
   ~~~~,~~~~    \text{D}^{i}_{a}~=~ \d {}^{i \, j}  \,  \text{D}_{a}{}_{j}  ~~~,
\label{Ddef}  \ee
here $i$ and $j$ are isospin indices taking on two values.  Above $\text{D}^{1}_{a}$ and 
$  \text{D}^{2}_{a}$ are Majorana 4D operators.   We use the convention that such isospin 
indices are raised and/or lowered via the Kronecker delta symbol.  

The complete listing of known off-shell 4D, $\cal N$ = 2 supermultiplets, {\it {containing a finite number of auxiliary fields}} to our knowledge, is shown in Table \ref{tab:N2susy} where the initial literature presentation for each supermultiplet was given in: \newline \indent
(a.) vector supermultiplet \cite{Fy8,Wss}, \newline \indent
(b.) tensor supermultiplet \cite{Fy8,Wss}, \newline \indent
(c.) relaxed supermultiplet \cite{HST}, \newline \indent
(d.) supergravity supermultiplet \cite{FradVas,FradVas2,dWvH}, \newline \indent 
(e.) ``hyperplet'' supermultiplet \cite{N2matter,MF}, and \newline \indent
(f.) ``HS'', higher spin supermultiplet \cite{KuZSirb1,KuZSirb2}.
\newline \noindent
Clearly, only the vector and the tensor 4D, $\mathcal{N}$ = 2 supermultiplet are minimal\footnote{
In the presentation of this table, we have ignored the possible existence of ``variant representations'' 
of these supermultiplets.}.  We have also indicated the number of boson (alternately fermionic) degrees of 
freedom in Table \ref{tab:N2susy} for each of these supermultiplets\footnote{The astute reader will note that there
is no result reported in the table for a 4D, $\cal N$ = 2 supermultiplet with spins of (3/2, 1, 1, 1/2).
Although an on-shell description is available \cite{GatKos}, no {\it {off}}-{\it {shell}} description has been presented.}.
\begin{table}[h!]
\centering
\begin{tabular}{c|c}
4D, ${\cal N}=2$ Supermultiplet  &  Multiplicity of Bosons or Fermions \\ \hline 
vector & $\{8\}$  \\ 
tensor & $\{8\}$  \\ 
relaxed & $\{32\}$  \\ 
supergravity & $\{40\}$ \\
hyperplet & $\{96\}$ \\
HS (w. lowest spin of propagating field = integer $s > 2$) & $\{ 8s^{2} + 8s + 4 \}$ \\
\end{tabular}
\caption{Degrees of Freedom In 4D, ${\cal N}=2$ Supermultiplets.}
\label{tab:N2susy}
\end{table}

In the work of \cite{adnkKyeoh}, an exhaustive study was made of the most general construction of a 4D, $\mathcal{N}$ = 2 supermultiplet based on pairings of the 4D, $\mathcal{N}$ = 1 chiral, vector, or tensor supermultiplets.  As this involved the three distinct  4D,  $\mathcal{N
}$ = 1 supermultiplets, when taken in pairs, there might have existed as many as six combinations that form off-shell $\mathcal{N}$ = 2 supermultiplets.   For systems that only respect supersymmetry {\it {on-shell}}, any such pairing was found to work.  However and in
fact, it was found that only the pairings of 4D, $\mathcal{N}$ = 1 chiral$\oplus$vector or the chiral$\oplus$tensor occur as {\it {off-shell}} representations.  We will denote these representations by the symbols ($\widehat {2{\rm {VS}}}$) and ($\widehat {2{\rm {TS}}}$) for 
the respective pairings chiral$\oplus$vector and chiral$\oplus$tensor.  However, whenever the bosonic fields of a supermultiplet are not symmetrical with respect to parity, one can perform a parity transformation on all of the bosonic fields in the supermultiplet to obtain a new supermultiplet.  Thus performing this parity transformation on the bosons of ($\widehat {2{\rm {VS}}}$) and ($\widehat {2{\rm {TS}}}$), we obtain the representations ($\widehat {2{\rm {AVS}}}$) and ($\widehat {2{\rm {ATS}}}$).

As minimal off-shell 4D, $\mathcal{N}$ = 2 supermultiplets contain a doublet of Majorana spinors, we introduce a notation of the form
\be
\Psi^{({\widehat {2VS}})}_{c i} = \begin{bmatrix}
\psi_{c} \\
\lambda_{c}
\end{bmatrix}  ~~~,~~~
\Psi^{({\widehat {2TS}})}_{c i}
= \begin{bmatrix}
\psi_{c} \\
\chi_{c}
\end{bmatrix}  ~~~,~~~  
\Psi^{({\widehat {2AVS}})}_{c i}
= \begin{bmatrix}
\psi_{c} \\
{\tilde \lambda}{}_{c}
\end{bmatrix}  ~~~,~~~
\Psi^{({\widehat {2ATS}})}_{c i}
= \begin{bmatrix}
\psi_{c} \\
{\tilde \chi}{}_{c}
 \end{bmatrix}   ~~~,
\label{doubs} \ee
where $ \psi_{c}$ denotes the spinor field of the off-shell 4D, $\mathcal{N}$ = 1 chiral supermultiplet, $\lambda_{c}$ denotes the spinor field of the off-shell 4D, $\mathcal{N}$ = 1 vector supermultiplet, $\chi_{c}$ denotes the spinor field of the off-shell 4D, $\mathcal{N}$ = 1 tensor supermultiplet, ${\tilde \lambda}{}_{c}$ denotes the spinor field of the off-shell 4D, $\mathcal{N}$ = 1 axial-vector supermultiplet, and ${\tilde \chi}{}_{c}$ denotes the spinor field of the off-shell 4D, $\mathcal{N}$ = 1 axial-tensor supermultiplet.  The
``isospin'' index $i$ takes on values 1 and 2.  As in previous works \cite{adnk4dGdgt1,adnk4dGdgt2} we introduce a 4D supermultiplet ``representation index'' $({\Hat {\cal R}})$ that takes on 
the values of ($\widehat {2{\rm {VS}}}$), ($\widehat {2{\rm {TS}}}$), ($\widehat {2{\rm {AVS
}}}$), and ($\widehat {2{\rm {ATS}}}$).  This allows us to express the four spinors shown in 
(\ref{doubs}) collectively as $ \Psi_{c i}^{({\Hat {\cal R}})}$ and we are now in position to 
state the main results of this investigation to calculate its fermionic holoraumy.

\subsection{Results For Fermionic Holoraumy 4D, $\cal N$ = 2 On Minimal Representations}

Having dispensed with all the necessary preliminary set of review and statement of 
our problem, conventions, etc., we now move toward the presentation of final results 
for the 4D, $\cal N$ = 2 fermionic holoraumy on minimal representations.   In order to 
achieve a maximally concise notation, we introduce the notion of a ``lattice variable'',
denoted by $\Pi$, which is simply defined to be an ordered set of four integers $({\rm p}
,{\rm q},{\rm r},{\rm s} ) $.

After a set of calculation we find our answer can be expressed concisely by the equation,
\be
 [\, \text{D}^{i}_{a} ~,~ \text{D}^{j}_{b} \, ]\, \Psi^{({\Hat {\cal R}})}_{c k} ~=~ \big[ \hat{\bm 
 H}^{\m ({\Hat {\cal R}})} \big]^{ij}_{ab \, c k}{}^{d l} \pa_{\m} \Psi^{({\Hat {\cal R}})
 }_{d l} ~~~,
\label{4dn2holor} \ee
where
\be
\begin{split}
\big[ \hat{\bm H}^{\m ({\Hat {\cal R}})} \big]^{ij}_{ab \, c k}{}^{d l} =& - \delta^{ij}\delta_{k}{}^{l} \big[ \hat{\bm h}^{(0) \m} (\Pi{}_{0}^{({\Hat {\cal R}})})  \big]_{abc}{}^{d}  
~+~ (\sigma^{3})^{ij}(\sigma^{3})_{k}{}^{l} \big[ \hat{\bm h}^{(3) \m} (\Pi{}_{3}^{({\Hat {\cal 
R}})})  \big]_{abc}{}^{d}  
\\ 
& 
+ (\sigma^{1})^{ij}(\sigma^{1})_{k}{}^{l} \big[ \hat{\bm h}^{(1) \m} (\Pi{}_{1}^{({\Hat {\cal R}})})
\big]_{abc}{}^{d}  
~+~  (\sigma^{2})^{ij}(\sigma^{2})_{k}{}^{l} \big[ \hat{\bm h}^{(2) \m} (\Pi{}_{2}^{({\Hat {\cal R}})})
 \big]_{abc}{}^{d} ~~~,
\end{split}
\label{Hn2d4}
\ee
and 
$\big[ \hat{\bm h}^{(0) \m} (\Pi) \big]_{abc}{}^{d}$, $\big[ \hat{\bm h}^{(3) \m} (\Pi) \big]_{abc}{}^{d}$, 
and $\big[ \hat{\bm h}^{(1) \m} (\Pi) \big]_{abc}{}^{d}$ satisfy the equations
\be  \eqalign{  {~~~}
\big[ \hat{\bm h}^{(0) \m} (\Pi) \big]_{abc}{}^{d} ~&=~ \big[ \hat{\bm h}^{(3) \m} (\Pi) \big]_{abc}{}^{d} 
~=~ \big[ \hat{\bm h}^{(1) \m} (\Pi) \big]_{abc}{}^{d}  \cr
&=~ i \, {\big [} ~ {\rm p} \, C_{ab} \, (\g^{\m})_{c}{}^{d} ~+~ {\rm q}\,
(\g^{5})_{ab} \, (\g^{5}\g^{\m})_{c}{}^{d} \cr 
&{~~~~~~} ~~+~ {\rm r} \,  (\g^{5}\g^{\m})_{ab} \, (\g^{5})_{c}{}^{d} ~+~ \tfrac{1}{2} 
{\rm s} (\g^{5}\g_{\n})_{ab}\,  (\g^{5}[\g^{\m}, \g^{\n}])_{c}{}^{d} ~ {\big ]} ~
{\equiv~} \big[ \hat{\bm h}^{\m} (\Pi) \big]_{abc}{}^{d} 
~~~.
}   \label{Lilhs} \ee
The function $\big[ \hat{\bm h}^{ \m} (\Pi ) \big]_{abc}{}^{d}$ was first presented in the work on 4D, ${\cal N}=1$ minimal supermultiplets \cite{adnk4dGdgt2} previously and we see that the only distinction between $\big[ \hat{\bm h}^{(0) \m} 
\big]_{abc}{}^{d}$, $\big[ \hat{\bm h}^{(3) \m} \big]_{abc}{}^{d}$, and $\big[ \hat{\bm h}^{(1) \m} \big]_{abc}{}^{d}$, is that distinct lattice variables are utilized. 

The quantity $ \big[ \hat{\bm h}^{(2) \m}  (\Pi) \big]_{abc}{}^{d}$
is given by
\be  \eqalign{
 \big[ \hat{\bm h}^{(2) \m}  (\Pi ) \big]_{abc}{}^{d}  ~&=~
i\  {\big[} \tfrac{1}{2} {\rm p} \, (\g^{5}[\g^{\m},
\g^{\n}])_{ab} \, (\g^{5}\g_{\n})_{c}{}^{d} ~+~ \tfrac{1}{2} {\rm q} \, ([\g^{\m},\g^{\n}])_{ab} \,
(\g_{\n})_{c}{}^{d}  \cr
&{~~~~~~} ~~+~ \tfrac{1}{2} {\rm r} \,  (\g_{\n})_{ab} \, ([\g^{\m},
\g^{\n}])_{c}{}^{d}
~+~ {\rm s} \,  (\g^{\m})_{ab} \delta_{c}{}^{d}
{\big ]}  ~~~.
}   \label{Lilha} \ee

It can be seen there is a more substantial distinction between $\big[ \hat{\bm h}^{ \m} ( \Pi ) \big]_{abc}{}^{d}$ and $ \big[ \hat{\bm h}^{(2) \m}  ( \Pi) \big]_{abc}{}^{d}$ with regard to the 
symmetry on the exchange of the first two spinor indices on each.  Namely, these 
respectively satisfy the identities
\be  \eqalign{
\big[ \hat{\bm h}^{\m} (\Pi ) \big]_{abc}{}^{d} ~=~ -~ \big[ \hat{\bm h}^{\m} 
(\Pi) \big]_{bac}{}^{d}     ~~,~~
\big[ \hat{\bm h}^{(2) \m}  (\Pi ) \big]_{abc}{}^{d} ~=&~   +~ \big[ \hat{\bm h}^{(2) \m}  
(\Pi ) \big]_{bac}{}^{d}       ~~~.
} \ee
In addition, we also see the following  identities are valid.
\be  \eqalign{
\big[ \hat{\bm h}^{\m} (\Pi ) \big]{}_{a}{}^{b}{}_{c}{}^{d} ~&=~  -\,  (\g^5){}_{a}{}^{k}
 \, (\g^5){}_{c}{}^{m} \, \big[ \hat{\bm h}^{\m} (\Pi ) \big]{}_{k}{}^{l}{}_{m}{}^{n} \, 
 (\g^5){}_{l}{}^{b} \, (\g^5){}_{n}{}^{d} ~~~, \cr
\big[ \hat{\bm h}^{(2) \m} (\Pi ) \big]{}_{a}{}^{b}{}_{c}{}^{d} ~&=~ -\, (\g^5){}_{a}{}^{k} \, 
(\g^5){}_{c}{}^{m} \, \big[ \hat{\bm h}^{(2) \m} (\Pi ) \big]{}_{k}{}^{l}{}_{m}{}^{n} \,(\g^5){
}_{l}{}^{b} \, (\g^5){}_{n}{}^{d}
 ~~~.
}  \ee

In (\ref{Hn2d4}) the quantities denoted by ${\rm p}_{A}^{({\Hat {\cal R}})},{\rm q}_{A}^{({\Hat 
{\cal R}})},{\rm r}_{A}^{({\Hat {\cal R}})},{\rm s}_{A}^{({\Hat {\cal R}})}$, (i.\ e.\ $\Pi_{A}^{({\Hat 
{\cal R}})}$)  ${\rm {with}}~ A = 0, \ldots, 3$ are integers listed in Table \ref{tab:4DN2-pqrs}.  

\begin{table}[h!]
\centering
\begin{tabular}{c|cccc|cccc|cccc|cccc}
$({\Hat {\cal R}})$  & ${\rm p}_{0}$ & ${\rm q}_{0}$ & ${\rm r}_{0}$ & ${\rm s}_{0}$ & ${\rm 
p}_{3}$ & ${\rm q}_{3}$ & ${\rm r}_{3}$ & ${\rm s}_{3}$ & ${\rm p}_{1}$ & ${\rm q}_{1}$ & 
${\rm r}_{1}$ & ${\rm s}_{1}$ & ${\rm p}_{2}$ & ${\rm q}_{2}$ & ${\rm r}_{2}$ & ${\rm s}_{2}$ 
\\ \hline 
(${\widehat {{2VS}}}$) & 1 & 1 & 1 & -1  & 1 & 1 & 1 & 1  & 1 & 1 & 1 & 1  & 1 & 1 & 1 & 1  \\ 
(${\widehat {{2TS}}}$)& -1 & 1 & -1 & -1  & -1 & 1 & -1 & 1  & -1 & 1 & -1 & 1  & -1 & 1 & -1 & 1 \\ 
(${\widehat {{2AVS}}}$) & -1 & -1 & 1 & -1  & -1 & -1 & 1 & 1  & -1 & -1 & 1 & 1  & -1 & -1 & 1 & 1  \\ 
(${\widehat {{2ATS}}}$) & 1 & -1 & -1 & -1  & 1 & -1 & -1 & 1  & 1 & -1 & -1 & 1  & 1 & -1 & -1 & 1 \\
\end{tabular}
\caption{Holoraumy Integers For 4D, ${\cal N}=2$ Vector, Tensor, Axial-Vector and Axial-Tensor Supermultiplets.}
\label{tab:4DN2-pqrs}
\end{table}

To be more explicit, we can substitute all the above information to re-write the result for
each supermultiplet in the forms
\begin{align}
\begin{split}
\big[ \hat{\bm H}^{\m ({\widehat {2VS}})} \big]^{ij}_{ab \, c k}{}^{d l} =& - \delta^{ij}\delta_{k}{}^{l} \big[ \hat{\bm h}^{\m} (1,1,1,-1) \big]_{abc}{}^{d} 
+ (\sigma^{3})^{ij}(\sigma^{3})_{k}{}^{l} \big[ \hat{\bm h}^{\m} (1,1,1,1) \big]_{abc}{}^{d}      \\ 
&  + (\sigma^{1})^{ij}(\sigma^{1})_{k}{}^{l} \big[ \hat{\bm h}^{\m} (1,1,1,1) \big]_{abc}{}^{d}  
+ (\sigma^{2})^{ij}(\sigma^{2})_{k}{}^{l} \big[ \hat{\bm h}^{(2) \m} (1,1,1,1) \big]_{abc}{}^{d}
~~~,
\end{split}  \\ &  \nonumber \\
\begin{split}
\big[ \hat{\bm H}^{\m ({\widehat {2TS}})} \big]^{ij}_{ab \, c k}{}^{d l} =& - \delta^{ij}\delta_{k}{}^{l} \big[ \hat{\bm h}^{\m} (-1,1,-1,-1) \big]_{abc}{}^{d} 
+ (\sigma^{3})^{ij}(\sigma^{3})_{k}{}^{l} \big[ \hat{\bm h}^{\m} (-1,1,-1,1) \big]_{abc}{}^{d} 
\\ 
&  + (\sigma^{1})^{ij}(\sigma^{1})_{k}{}^{l} \big[ \hat{\bm h}^{\m} (-1,1,-1,1) \big]_{abc}{}^{d} 
+ (\sigma^{2})^{ij}(\sigma^{2})_{k}{}^{l} \big[ \hat{\bm h}^{(2) \m} (-1,1,-1,1) \big]_{abc}{}^{d}
~~~,
\end{split}  \\
\begin{split}
\big[ \hat{\bm H}^{\m ({\widehat {2AVS}})} \big]^{ij}_{ab \, c k}{}^{d l} =& - \delta^{ij}\delta_{k}{}^{l} \big[ \hat{\bm h}^{\m} (-1,-1,1,-1) \big]_{abc}{}^{d}  
+ (\sigma^{3})^{ij}(\sigma^{3})_{k}{}^{l} \big[ \hat{\bm h}^{\m} (-1,-1,1,1) \big]_{abc}{}^{d}  
\\ 
& + (\sigma^{1})^{ij}(\sigma^{1})_{k}{}^{l} \big[ \hat{\bm h}^{\m} (-1,-1,1,1) \big]_{abc}{}^{d} 
+ (\sigma^{2})^{ij}(\sigma^{2})_{k}{}^{l} \big[ \hat{\bm h}^{(2) \m} (-1,-1,1,1) \big]_{abc}{}^{d}
\end{split}  ~~~,  \\
\begin{split}
\big[ \hat{\bm H}^{\m ({\widehat {2ATS}})} \big]^{ij}_{ab \, c k}{}^{d l} =& - \delta^{ij}\delta_{k}{}^{l} \big[ \hat{\bm h}^{\m} (1,-1,-1,-1) \big]_{abc}{}^{d} 
+ (\sigma^{3})^{ij}(\sigma^{3})_{k}{}^{l} \big[ \hat{\bm h}^{\m} (1,-1,-1,1) \big]_{abc}{}^{d} 
\\ 
&  + (\sigma^{1})^{ij}(\sigma^{1})_{k}{}^{l} \big[ \hat{\bm h}^{\m} (1,-1,-1,1) \big]_{abc}{}^{d} 
+ (\sigma^{2})^{ij}(\sigma^{2})_{k}{}^{l} \big[ \hat{\bm h}^{(2) \m} (1,-1,-1,1) \big]_{abc}{}^{d}
~~~,
\end{split}
\end{align}
and we note that
\be
\begin{split}
\big[ \hat{\bm H}^{\m ({\widehat {2AVS}})} \big]^{ij}_{ab \, c k}{}^{d l} ~=~& (\g^{5})_{c}{}^{e} 
\big[ \hat{\bm H}^{\m ({\widehat {2VS}})} \big]^{ij}_{ab \, e k}{}^{f l} (\g^{5})_{f}{}^{d}  ~~~, \\
\big[ \hat{\bm H}^{\m ({\widehat {2ATS}})} \big]^{ij}_{ab \, c k}{}^{d l} ~=~& (\g^{5})_{c}{}^{e} 
\big[ \hat{\bm H}^{\m ({\widehat {2TS}})} \big]^{ij}_{ab \, e k}{}^{f l} (\g^{5})_{f}{}^{d}  ~~~,
\end{split}
\label{P1}
\ee
such that the parity flip involves
\be
\begin{split}
 {\rm p}_{A}^{({\Hat {\cal R}})} & ~\longrightarrow~ - {\rm p}_{A}^{({\Hat {\cal R}})} ~~,~~
 {\rm q}_{A}^{({\Hat {\cal R}})}  ~\longrightarrow~ - {\rm q}_{A}^{({\Hat {\cal R}})} ~~,~~
 {\rm r}_{A}^{({\Hat {\cal R}})}  ~\longrightarrow~ {\rm r}_{A}^{({\Hat {\cal R}})}  ~~,~~
 {\rm s}_{A}^{({\Hat {\cal R}})}  ~\longrightarrow~ {\rm s}_{A}^{({\Hat {\cal R}})} ~~.
\end{split}
\label{P2}
\ee
for all $A$.

It is useful at this point to review the results of the previous subsection within the context of our opening remarks.  In particular, the calculations carried out that support the results shown in the last subsection allow the identification of the general result proffered in
(\ref{songen}) and (\ref{Liegens}).  The constant tensors $\mathscr{S}{}_{i \, j}^{(S)}$ and $\mathscr{A}{}_{i \, j}^{[A]}$ are clearly determined.  Since $\cal N$ = 2, the $(S)$ index on $\mathscr{S}{}_{i \, j}^{(S)}$ takes on two values, which can be identified with
\be
\mathscr{S}{}_{i \, j}^{(3)} ~=~  ( \s{}^{3} ) {}_{i \, j}   ~~~,~~~ \mathscr{S}{}_{i \, j}^{(1)} ~=~    
( \s{}^{1} ) {}_{i \, j}  ~~~,
\label{so2ten}
\ee
while again since $\cal N$ = 2, the $[A]$ index 
on $\mathscr{A}{}_{i \, j}^{[A]}$ takes on a single value and we identify it as simply
\be
\mathscr{A}{}_{i \, j}^{[2]} ~=~ i\, ( \s{}^{2} ) {}_{i \, j} ~~~.
\ee
Next the Lie-algebra valued quantities define by  (\ref{H1}) can now be explicitly
demonstrated.  One simply evaluates the RHS of Eq. (\ref{H1}) on the representation
dependent fermion defined in (\ref{doubs}) This leads to four equations
\be \eqalign{  
& -\, \delta_{k}{}^{l} 
\big[ \hat{\bm h}^{ \m} ( \Pi {}_{0}^{({\Hat {\cal R}})}) \big]_{abc}{}^{d} \, \pa_{\m} \Psi^{({\Hat {\cal R}}) }_{d l} ~=~  
{\big [} \,  i \, C_{a \, b} \,   { {  \mathscr{H}}}{}^{(1)}  ~+~  (\g^{5}){}_{a \, b} \, 
 { \mathscr{H}}{}^{(2)} ~+~  (\g^{5} \g^{\m}){}_{a \, b} \, 
 { \mathscr{H}}{}_{\m}{}^{(3)}  \, {\big ]} \,  \Psi^{({\Hat {\cal R}}) }_{d l}  ~~~,
} \label{LiegensX1}
\ee
\be \eqalign{  
& ~(\sigma^{3})_{k}{}^{l} \big[ \hat{\bm h}^{\m} (  \Pi {}_{3}^{({\Hat {\cal R}})}
) \big]_{abc}{}^{d} \, \pa_{\m}  \Psi^{({\Hat {\cal 
R}}) }_{d l} 
~=~  ~~~~~~~~~~ \cr
&~~~~~~~~~~~~~~~~~~~{\big [} \,  i \, C_{a \, b} \, (\,   { {\mathscr{H}}}{}^{(4:3)}
\,) ~+~ (\g^{5}){}_{a \, b} \, ( \,  {\mathscr{H}}{}^{(5:3)} \, ) ~+~ (\g^{5} \g^{\m}){}_{a \, 
b} \, (\, {\mathscr{H}}{}_{\m}{}^{(6:3)}  \, )  \, {\big ]} \,   \Psi^{({\Hat {\cal R}}) }_{d l}  ~~~,
} \label{LiegensX2}
\ee
\be \eqalign{  
& ~(\sigma^{1})_{k}{}^{l} \big[ \hat{\bm h}^{\m} ( \Pi {}_{1}^{({\Hat {\cal R}})}
) \big]_{abc}{}^{d} \, \pa_{\m}  \Psi^{({\Hat {\cal 
R}}) }_{d l} 
~=~  ~~~~~~~~~~ \cr
&~~~~~~~~~~~~~~~~~~~{\big [} \,  i \, C_{a \, b} \, (\,   { {\mathscr{H}}}{}^{(4:1)}
\,) ~+~ (\g^{5}){}_{a \, b} \, ( \,  {\mathscr{H
}}{}^{(5:1)} \, ) ~+~ (\g^{5} \g^{\m}){}_{a \, b} \, (\, 
{\mathscr{H}}{}_{\m}{}^{(6:1)}  \, )  \, {\big ]} \,   \Psi^{({\Hat {\cal R}}) }_{d l}   ~~~,
} \label{LiegensX3}
\ee
\be \eqalign{  
& (\sigma^{2})_{k}{}^{l} \big[ \hat{\bm h}^{(2) \m} ( \Pi{}_{2}^{({\Hat {\cal R}})}
) \big]_{abc}{}^{d} \pa_{\m}  \Psi^{({\Hat {\cal R}}) }_{d l} ~ ~=~  
~ - \,{\big [} \,  (\g^{\m} ){}_{a \, b} \, ,  \mathscr{H}{
}_{\m}{}^{(7:2)}  ~+~  \fracm 12 \,  ([\, \g^{\m} ~,~ \g^{\n}\,]){}_{a \, 
b} \,  \mathscr{H}{}_{\m \, \n}{}^{(8:2)} \, {\big ]} \, \Psi^{({\Hat {\cal R}}) }_{d l}  
~~~.~~~~~~~~~~  \cr
} \label{LiegensX4}
\ee
that can be used to extract explicit expressions for the action of each of the Lie-algebra
generators.

\newpage
\newpage \noindent
\section{4D, $\cal N$ = 2 Bosonic Holoraumy Results For Minimal Multiplets}

Up to this point as well as in the bulk of our previous discussions on the topic of holoraumy,
the focus is on the results of the holoraumy calculations as evaluated on the fermions within 
supermultiplets.  However, one can also carry out calculation on the bosons within any 
supermultiplet.  The reason for focusing only on the fermions is one of convenience.  Let us 
demonstrate some examples.

In the following, there will be presented the results for the evaluation of the holoraumy on
the basis of the bosonic fields of each minimal off-shell 4D, $\cal N$ = 2 supermultiplet.
In particular, the evaluations allow for the comparison of the holoraumies of the pairs of 
supermultiplets (${\widehat {2VS}}$)-(${\widehat {2AVS}}$) and ($\widehat {2{\rm {TS}}}
$)-($\widehat {2{\rm {ATS}}}$).   By use of Eq. (\ref{P1}) and (\ref{P2}) it was demonstrated 
that  each member of these pairs are the ``parity flipped'' version of the other member.  This 
is encoded by simply performing a ``conjugation'' of the holoraumy tensors by use of the $
\g^5$-matrix.

In the remaining portions of this chapter, we simply present the results for the holoraumy 
operator in (\ref{H1}) explicitly evaluated on each bosonic component field in all the minimal 
4D, $\cal N$ = 2 off-shell supermultiplets.

\subsection{Bosonic 4D, $\mathcal{N}$ = 2 Vector Supermultiplet Holoraumy}

\begin{align}
\begin{split}
{\big [}\, \text{D}^{i}_{a} ~,~ \text{D}^{j}_{b} \, {\big ]} A ~=~& 
- 2 \delta^{ij}  \left[ (\g^{5}\g^{\m})_{ab} \pa_{\m} B + i C_{ab} F \right]  
+ 2 (\sigma^{3})^{ij} (\g^{5})_{ab} G  \\ 
&  + 2 (\sigma^{1})^{ij} (\g^{5})_{ab} d  
+ (\sigma^{2})^{ij} ([\g^{\m},\g^{\n}])_{ab} \pa_{\m} A_{\n}  ~~~,
\end{split}  \\
\begin{split}
{\big [}\, \text{D}^{i}_{a} ~,~ \text{D}^{j}_{b} \, {\big ]} B  ~=~&  
2 \delta^{ij}  \left[  (\g^{5}\g^{\m})_{ab} \pa_{\m} A + (\g^{5})_{ab} F  \right] 
+ i 2 (\sigma^{3})^{ij} C_{ab} G \\ 
& + i 2 (\sigma^{1})^{ij} C_{ab} d 
+ i (\sigma^{2})^{ij}  (\g^{5}[\g^{\m},\g^{\n}])_{ab} \pa_{\m} A_{\n}  ~~~,
\end{split}  \\
\begin{split}
{\big [}\, \text{D}^{i}_{a} ~,~ \text{D}^{j}_{b} \, {\big ]} F  ~=~&  
2 \delta^{ij}  \left[ - i C_{ab} \Box A + (\g^{5})_{ab} \Box B \right] 
- 2 (\sigma^{3})^{ij}  (\g^{5}\g^{\m})_{ab} \pa_{\m} G  \\
& - 2 (\sigma^{1})^{ij} (\g^{5}\g^{\m})_{ab} \pa_{\m} d
- 2 (\sigma^{2})^{ij} (\g^{\m})_{ab} \pa^{\n} F_{\m\n}    ~~~,
\end{split}  \\
\begin{split}
{\big [}\, \text{D}^{i}_{a} ~,~ \text{D}^{j}_{b} \, {\big ]} G ~=~&  2 (\sigma^{3})^{ij} \left[  (\g^{5})_{ab} \Box A + i  C_{ab} \Box B + (\g^{5 }\g^{\m})_{ab} \pa_{\m} F  \right]   \\ 
&  - 2 (\sigma^{1})^{ij} (\g^{5}\g^{\m})_{ab} \pa^{\n} F_{\m\n}  
+ 2 (\sigma^{2})^{ij} (\g^{\m})_{ab} 
\pa_{\m} d    ~~~,
\end{split}  \\
\begin{split}
{\big [}\, \text{D}^{i}_{a} ~,~ \text{D}^{j}_{b} \, {\big ]} A_{\m} ~=~&  - 2 \delta^{ij} \epsilon_{\m}{}^{\n\alpha\beta} (\g^{5}\g_{\n})_{ab} \pa_{\alpha} A_{\beta}  
- 2 (\sigma^{3})^{ij} (\g^{5}\g_{\m})_{ab} d  \\ 
&  + 2  (\sigma^{1})^{ij}(\g^{5}\g_{\m})_{ab} G      \\ 
&  +  (\sigma^{2})^{ij} \left[ ([\g_{\m},\g^{\n}])_{ab} \pa_{\n} A + i  ( \g^{5}[\g_{\m},\g^{\n}])_{ab} \pa_{\n} B + 2 (\g_{\m} )_{ab} F \right]   ~~~,
\end{split}  \\
\begin{split}
{\big [}\, \text{D}^{i}_{a} ~,~ \text{D}^{j}_{b} \, {\big ]} d ~=~&  
2 (\sigma^{3})^{ij} (\g^{5}\g^{\m})_{ab} \pa^{\n} F_{\m\n}    \\ 
&  +  2 (\sigma^{1})^{ij} \left[ (\g^{5})_{ab} \Box A + i  C_{ab} \Box B +  (\g^{5}\g^{\m})_{ab} \pa_{\m} F \right]    \\
&  -  2 (\sigma^{2})^{ij} (\g^{\m})_{ab} \pa_{\m} G ~~~.
\end{split}
\end{align}

\subsection{Bosonic 4D, $\mathcal{N}$ = 2 Tensor Supermultiplet Holoraumy}
 
\begin{align}
\begin{split}
{\big [}\, \text{D}^{i}_{a} ~,~ \text{D}^{j}_{b} \, {\big ]} A ~=~&  
2 (\sigma^{3})^{ij} \left[  - (\g^{5}\g^{\m})_{ab} \pa_{\m} B - i C_{ab} F + (\g^{5})_{ab} G  \right]    \\ 
&  + 2 (\sigma^{1})^{ij} \epsilon_{\m}{}^{\n\alpha\beta} (\g^{5}\g^{\m})_{ab} \pa_{\n} B_{\alpha\beta} 
+ 2 (\sigma^{2})^{ij} (\g^{\m})_{ab} \pa_{\m} \varphi  ~~~,
\end{split}  \\
\begin{split}
{\big [}\, \text{D}^{i}_{a} ~,~ \text{D}^{j}_{b} \, {\big ]} B  ~=~&  
2  \delta^{ij}  \left[  (\g^{5})_{ab} F + i C_{ab} G  \right] 
+ 2 (\sigma^{3})^{ij} (\g^{5}\g^{\m})_{ab} \pa_{\m} A   \\ 
&  + 2 (\sigma^{1})^{ij} (\g^{5}\g^{\m})_{ab} \pa_{\m} \varphi - 2 (\sigma^{2})^{ij}  \epsilon_{\m}{}^{\n\alpha\beta} (\g^{\m})_{ab} \pa_{\n} B_{\alpha\beta}  ~~~,
\end{split}  \\
\begin{split}
{\big [}\, \text{D}^{i}_{a} ~,~ \text{D}^{j}_{b} \, {\big ]} F  ~=~&  
2 \delta^{ij}  \left[  (\g^{5})_{ab} \Box B -  (\g^{5}\g^{\m})_{ab} \pa_{\m} G  \right] 
- i 2 (\sigma^{3})^{ij} C_{ab} \Box A    \\ 
&  -  i 2 (\sigma^{1})^{ij} C_{ab} \Box \varphi + i (\sigma^{2})^{ij}  \epsilon_{\m}{}^{\n\alpha\beta} (\g^{5}[\g^{\m},\g^{\rho}])_{ab} \pa_{\n} \pa_{\rho} B_{\alpha\beta}   ~~~,
\end{split}  \\
\begin{split}
{\big [}\, \text{D}^{i}_{a} ~,~ \text{D}^{j}_{b} \, {\big ]} G ~=~&  
2 \delta^{ij} \left[  i C_{ab} \Box B + (\g^{5}\g^{\m})_{ab} \pa_{\m} F  \right]   
+  2 (\sigma^{3})^{ij} (\g^{5})_{ab} \Box A  
\\ 
&  +  2 (\sigma^{1})^{ij} (\g^{5})_{ab} \Box \varphi  - (\sigma^{2})^{ij} \epsilon_{\m}{}^{\n\alpha\beta} ([\g^{\m},\g^{\rho}])_{ab} \pa_{\n} \pa_{\rho} B_{\alpha\beta}  ~~~,
\end{split}  \\
\begin{split}
{\big [}\, \text{D}^{i}_{a} ~,~ \text{D}^{j}_{b} \, {\big ]} \varphi ~=~& - 2 (\sigma^{3})^{ij} \epsilon_{
\m}{}^{\n\alpha\beta} (\g^{5}\g^{\m})_{ab} \pa_{\n} B_{\alpha\beta}  
\\ 
&  + 2  (\sigma^{1})^{ij} \left[  - (\g^{5}\g^{\m})_{ab} \pa_{\m} B - i C_{ab} F + (\g^{5})_{ab} G  \right] - 2 (\sigma^{2})^{ij} (\g^{\m})_{ab} \pa_{\m} A  ~~~,
\end{split}  \\
\begin{split}
{\big [}\, \text{D}^{i}_{a} ~,~ \text{D}^{j}_{b} \, {\big ]} B_{\m\n} ~=~&  - \delta^{ij} \epsilon_{[\m}{}^{\rho\alpha\beta} (\g^{5}\g_{\n]})_{ab} \pa_{\rho} B_{\alpha\beta} 
+ (\sigma^{3})^{ij} \epsilon_{\m\n}{}^{\alpha\beta} (\g^{5}\g_{\alpha})_{ab}  \pa_{\beta} \varphi
 \\ 
&  - (\sigma^{1})^{ij} \epsilon_{\m\n}{}^{\alpha\beta} (\g^{5}\g_{\alpha})_{ab} \pa_{\beta} A     \\ 
&  + (\sigma^{2})^{ij} \left[ \epsilon_{\m\n}{}^{\alpha\beta} (\g^{5}\g_{\alpha})_{
ab} \pa_{\beta} B - \tfrac{1}{2} ([\g_{\m},\g_{\n}])_{ab} F - i \tfrac{1}{2} (\g^{5}[\g_{\m},\g_{\n}])_{ab} G  \right]   ~~~.
\end{split}
\end{align}

\subsection{Bosonic 4D, $\mathcal{N}$ = 2 Axial-Vector Supermultiplet Holoraumy}

\begin{align}
\begin{split}
{\big [}\, \text{D}^{i}_{a} ~,~ \text{D}^{j}_{b} \, {\big ]} A ~=~&   
2   \delta^{ij}  \left[ -(\g^{5}\g^{\m})_{ab} \pa_{\m} B + (\g^{5})_{ab} G \right] - i 2 (\sigma^{3})^{ij} C_{ab} F     \\ 
&  +  i 2 (\sigma^{1})^{ij} C_{ab} \tilde{d}  + i (\sigma^{2})^{ij} (\g^{5}[\g^{\m},\g^{\n}])_{ab} \pa_{\m} U_{\n}  ~~~,
\end{split}  \\
\begin{split}
{\big [}\, \text{D}^{i}_{a} ~,~ \text{D}^{j}_{b} \, {\big ]} B  ~=~&  2  \delta^{ij}  \left[  (\g^{5}\g^{\m})_{ab} \pa_{\m} A + i C_{ab} G \right]  + 2 (\sigma^{3})^{ij} (\g^{5})_{ab} F   \\ 
&  - 2 (\sigma^{1})^{ij} (\g^{5})_{ab} \tilde{d} - (\sigma^{2})^{ij}  ([\g^{\m},\g^{\n}])_{ab} \pa_{\m} U_{\n}   ~~~,
\end{split}  \\
\begin{split}
{\big [}\, \text{D}^{i}_{a} ~,~ \text{D}^{j}_{b} \, {\big ]} F ~=~&   2 (\sigma^{3})^{ij} \left[ - i C_{ab} \Box A + (\g^{5})_{ab} \Box B - (\g^{5}\g^{\m})_{ab} \pa_{\m} G  \right]   \\ 
&  + 2 (\sigma^{1})^{ij} (\g^{5}\g^{\m})_{ab} \pa^{\n} F_{\m\n}  - 2 (\sigma^{2})^{ij} (\g^{\m})_{ab} \pa_{\m} \tilde{d}    ~~~,
\end{split}  \\
\begin{split}
{\big [}\, \text{D}^{i}_{a} ~,~ \text{D}^{j}_{b} \, {\big ]} G ~=~&  2 \delta^{ij}  \left[ (\g^{5})_{ab} \Box A + i C_{ab} \Box B \right] + 2 (\sigma^{3})^{ij} (\g^{5}\g^{\m})_{ab} \pa_{\m} F     \\ 
&  - 2 (\sigma^{1})^{ij} (\g^{5}\g^{\m})_{ab} \pa_{\m} \tilde{d} - 2 (\sigma^{2})^{ij} (\g^{\m})_{ab} \pa^{\n} F_{\m\n}   ~~~,
\end{split}  \\
\begin{split}
{\big [}\, \text{D}^{i}_{a} ~,~ \text{D}^{j}_{b} \, {\big ]} U_{\m} ~=~&  - 2 \delta^{ij} \epsilon_{\m}{}^{\n\alpha\beta} (\g^{5}\g_{\n})_{ab} \pa_{\alpha} U_{\beta}  - 2 (\sigma^{3})^{ij} (\g^{5}\g_{\m})_{ab} \tilde{d}    \\ 
&  - 2 (\sigma^{1})^{ij} (\g^{5}\g_{\m})_{ab} F    \\ 
&  +  i (\sigma^{2})^{ij} \left[ (\g^{5}[\g_{\m},\g^{\n}])_{ab} \pa_{\n} A + i  ([\g_{\m},\g^{\n}])_{ab} \pa_{\n} B - i 2 (\g_{\m})_{ab} G \right]    ~~~,
\end{split}  \\
\begin{split}
{\big [}\, \text{D}^{i}_{a} ~,~ \text{D}^{j}_{b} \, {\big ]} \tilde{d} ~=~& 2 (\sigma^{3})^{ij} (\g^{5}\g^{\m})_{ab} \pa^{\n} F_{\m\n}    \\ 
&  + 2 (\sigma^{1})^{ij} \left[  i C_{ab} \Box A - (\g^{5})_{ab} \Box B + (\g^{5}\g^{\m})_{ab} \pa_{\m} G \right]     \\ 
&  +  2 (\sigma^{2})^{ij} (\g^{\m})_{ab} \pa_{\m} F   ~~~.
\end{split}
\end{align}

\subsection{Bosonic 4D, $\mathcal{N}$ = 2 Axial-Tensor Supermultiplet Holoraumy}

\begin{align}
\begin{split}
{\big [}\, \text{D}^{i}_{a} ~,~ \text{D}^{j}_{b} \, {\big ]} A ~=~&   
2  \delta^{ij}  \left[  - i C_{ab} F + (\g^{5})_{ab} G \right] 
- 2 (\sigma^{3})^{ij} (\g^{5}\g^{\m})_{ab} \pa_{\m} B 
 \\ 
&  - 2 (\sigma^{1})^{ij} (\g^{5}\g^{\m})_{ab} \pa_{\m} \tilde{\varphi} 
+ 2 (\sigma^{2})^{ij} \epsilon_{\m}{}^{\n\alpha\beta} (\g^{\m})_{ab} \pa_{\n} C_{\alpha\beta}      ~~~, \end{split}  \\
\begin{split}
{\big [}\, \text{D}^{i}_{a} ~,~ \text{D}^{j}_{b} \, {\big ]} B  ~=~&   
2 (\sigma^{3})^{ij} \left[ (\g^{5}\g^{\m})_{ab} \pa_{\m} A + (\g^{5})_{ab} F + i C_{ab} G  \right]    \\
&  + 2 (\sigma^{1})^{ij}  \epsilon_{\m}{}^{\n\alpha\beta} (\g^{5}\g^{\m})_{ab} \pa_{\n} C_{\alpha\beta} 
+ 2 (\sigma^{2})^{ij} (\g^{\m})_{ab} \pa_{\m} \tilde{\varphi}   ~~~, 
\end{split}   \\ 
\begin{split}
{\big [}\, \text{D}^{i}_{a} ~,~ \text{D}^{j}_{b} \, {\big ]} F  ~=~&  
- 2 \delta^{ij}  \left[ i C_{ab} \Box A  +  (\g^{5}\g^{\m})_{ab} \pa_{\m} G  \right]  
+  2 (\sigma^{3})^{ij} (\g^{5})_{ab} \Box B     \\ 
&  +  2 (\sigma^{1})^{ij} (\g^{5})_{ab} \Box \tilde{\varphi} - (\sigma^{2})^{ij} \epsilon_{\m}{}^{\n\alpha\beta} ([\g^{\m},\g^{\rho}])_{ab} \pa_{\n} \pa_{\rho} C_{\alpha\beta}    ~~~,
\end{split}  \\
\begin{split}
{\big [}\, \text{D}^{i}_{a} ~,~ \text{D}^{j}_{b} \, {\big ]} G ~=~&  
2 \delta^{ij} \left[  (\g^{5})_{ab} \Box A + (\g^{5}\g^{\m})_{ab} \pa_{\m} F  \right]  
+  i 2 (\sigma^{3})^{ij} C_{ab} \Box B    \\ 
&  +  i 2 (\sigma^{1})^{ij} C_{ab} \Box \tilde{\varphi}  
- i (\sigma^{2})^{ij}  \epsilon_{\m}{}^{\n\alpha\beta} (\g^{5}[\g^{\m},
\g^{\rho}])_{ab} \pa_{\n} \pa_{\rho} C_{\alpha\beta}    ~~~,
\end{split}  \\
\begin{split}
{\big [}\, \text{D}^{i}_{a} ~,~ \text{D}^{j}_{b} \, {\big ]} \tilde{\varphi} ~=~&  - 2 (\sigma^{3})^{ij} \epsilon_{\m}{}^{\n\alpha\beta} (\g^{5}\g^{\m})_{ab} \pa_{\n} C_{\alpha\beta}   \\
&  + 2  (\sigma^{1})^{ij} \left[  (\g^{5}\g^{\m})_{ab} \pa_{\m} A + (\g^{5})_{ab} F + i C_{ab} G  \right]  
- 2 (\sigma^{2})^{ij} (\g^{\m})_{ab} \pa_{\m} B  
 ~~~,
\end{split}  \\
\begin{split}
{\big [}\, \text{D}^{i}_{a} ~,~ \text{D}^{j}_{b} \, {\big ]} C_{\m\n} ~=~&  - \delta^{ij} \epsilon_{[\m}{}^{\rho\alpha\beta} (\g^{5}\g_{\n]})_{ab} \pa_{\rho} C_{\alpha\beta} 
+ (\sigma^{3})^{ij} \epsilon_{\m\n}{}^{\alpha\beta} (\g^{5}\g_{\alpha})_{ab} \pa_{\beta} \tilde{\varphi}      \\ 
&   - (\sigma^{1})^{ij} \epsilon_{\m\n}{}^{\alpha\beta} (\g^{5}\g_{\alpha})_{ab} 
\pa_{\beta} B    \\ 
&   + i (\sigma^{2})^{ij} \left[ i \epsilon_{\m\n}{}^{\alpha\beta} (\g_{\alpha})_{ab} 
\pa_{\beta} A - \tfrac{1}{2} (\g^{5}[\g_{\m},\g_{\n}])_{ab} F - i 
\tfrac{1}{2} ([\g_{\m},\g_{\n}])_{ab} G  \right]  ~~~.
\end{split}
\end{align}

The actual derivation of all of these results follow from the forms of the 4D, $\cal N$ = 2 supercovariant derivative operator when evaluated on each field.  The explicit expressions for this evaluation were presented in \cite{adnkKyeoh} and in order to streamline our presentation,
we have given these in an appendix.

\newpage
\newpage
\section{A Lorentz-Covariant \& so(2)-Covariant 4D, $\cal N$ = 2 Supermultiplet Gadget}

In the works of \cite{adnk4dGdgt1,adnk4dGdgt2}, the concept of a spacetime Gadget function for 
supermultiplets in 4D was introduced.  By definition, a spacetime Gadget is a bilinear function whose 
domain is pairs of 4D supermultiplet representations $({\Hat {\cal R}}{}_1)$ and $({\Hat {\cal 
R}}{}_2)$ such that its range is the real numbers.  In a sense a spacetime Gadget introduces a metric on 
the space of supermultiplets.  We typically use the notation $ \widehat{\mathcal{G}}[({\Hat 
{\cal R}}{}_1), \,({\Hat {\cal R}}{}_2)]$ for the spacetime Gadget and in all previous known examples, 
this metric has two properties:  \vskip 2pt  \indent
(a.) when $({\Hat {\cal R}}{}_1)$ = $({\Hat {\cal R}}{}_2)$, the spacetime Gadget value is non-negative,
and
\newline \indent
(b.)
it only maps to zero if either $({\Hat {\cal R}}{}_1)$ or $({\Hat {\cal R}}{}_2)$ is the
zero supermultiplet.  \vskip 2pt  \noindent
Previously this result has been achieved via the spacetime Gadget being constructed from a quadratic that uses
the holoraumy operator evaluated on the fermionic fields of a supermultiplet.

Following the spirit of the 4D, $\mathcal{N} = 1$ case, we define the 4D, $\mathcal{N} = 2$ 
Gadget as
\be
\begin{split}
\widehat{\mathcal{G}}[({\Hat {\cal R}}),({\Hat {\cal R}}')]  =& \quad\,  m_{1}  \big[ \hat{\bm H}^{
\m ({\Hat {\cal R}})} \big]_{a i \, b j \, c k}{}^{d l}  \big[ \hat{\bm H}_{\m}{}^{({\Hat {\cal R}}')} 
\big]^{a i \, b j}{}_{d l}{}^{c k}   \\
&  +  m_{2}  (\g^{5})_{c}{}^{e}  \big[ \hat{\bm H}^{\m ({\Hat {\cal R}})} \big]_{a i \, b j \, 
e k}{}^{f l} (\g^{5})_{f}{}^{d}   \big[ \hat{\bm H}_{\m}{}^{({\Hat {\cal R}}')} \big]^{a i \, b j
}{}_{d l}{}^{c k}   \\
&  +  m_{3}  (\g^{5}\g^{\alpha})_{c}{}^{e}  \big[ \hat{\bm H}^{\m ({\Hat {\cal R}})} 
\big]_{a i \, b j \, e k}{}^{f l} (\g^{5}\g_{\alpha})_{f}{}^{d}   \big[ \hat{\bm H}_{\m}{
}^{({\Hat {\cal R}}')} \big]^{a i \, b j}{}_{d l}{}^{c k}    \\
&  +  m_{4}  (\g^{\alpha})_{c}{}^{e}  \big[ \hat{\bm H}^{\m ({\Hat {\cal R}})} \big]_{a i \, 
b j \, e k}{}^{f l} (\g_{\alpha})_{f}{}^{d}   \big[ \hat{\bm H}_{\m}{}^{({\Hat {\cal R}}')} \big
]^{a i \, b j}{}_{d l}{}^{c k}    \\
&  +  m_{5}  ([\g^{\alpha},\g^{\beta}])_{c}{}^{e}  \big[ \hat{\bm H}^{\m ({\Hat {\cal 
R}})} \big]_{a i \, b j \, e k}{}^{f l} ([\g_{\alpha},\g_{\beta}])_{f}{}^{d}   \big[ \hat{\bm 
H}_{\m}{}^{({\Hat {\cal R}}')} \big]^{a i \, b j}{}_{d l}{}^{c k}   ~~~, \\  
\end{split}
\label{gdt1}  \ee
with the introduction (as of now) undetermined parameters $m_1$, $\dots$, $m_5$.   Let us 
note that it is possible to introduce more such parameters.  The parameters in (\ref{gdt1})
are associated with performing all possible conjugations of the $a$ and $b$ spinor indices 
of the holoraumy with respect to the universal covering algebra of the $\g$-matrices.  However,
it should be recognized that such a set of conjugations may be taken with respect to the
$i$ and $j$ isospinor indices of the holoraumy with respect to the su(2) algebra.

To continue we note it is convenient to decompose the calculation into several parts that
depend respectively on $\big[ \hat{\bm h}^{(0) \m} \big]_{abc}{}^{d}$, $\big[ \hat{\bm h}^{(3) \m} \big]_{abc}{}^{d}$, $\big[ \hat{\bm h}^{(1) \m} \big]_{abc}{}^{d}$, and $\big[ \hat{\bm h}^{(2) \m} \big]_{abc}{}^{d}$.  So as an intermediate expression we write
\be
\begin{split}
\widehat{\mathcal{G}}[({\Hat {\cal R}}),({\Hat {\cal R}}')]  
\equiv &  \,   4 \big\{ \widehat{\mathcal{G}}_{0}[({\Hat {\cal R}}),({\Hat {\cal R}}')] + \widehat{
\mathcal{G}}_{3}[({\Hat {\cal R}}),({\Hat {\cal R}}')] + \widehat{\mathcal{G}}_{1}[({\Hat {\cal 
R}}),({\Hat {\cal R}}')] + \widehat{\mathcal{G}}_{2}[({\Hat {\cal R}}),({\Hat {\cal R}}')]  \big\} 
~~~,
\end{split}
\ee
where
\be
\begin{split}
 \widehat{\mathcal{G}}_{A}[({\Hat {\cal R}}),({\Hat {\cal R}}')]  =&  \quad\,  m_{1}  \big[ \hat{\bm h}^{(A)\m}{}^{({\Hat {\cal R}})} \big]_{abc}{}^{d}  \big[ \hat{\bm h}^{(A)}{}_{\m}{}^{({\Hat {\cal 
 R}}')} \big]^{ab}{}_{d}{}^{c}   \\
 &  +  m_{2}  (\g^{5})_{c}{}^{e}  \big[ \hat{\bm h}^{(A)\m}{}^{({\Hat {\cal R}})} \big]_{abe}{}^{
 f} (\g^{5})_{f}{}^{d}   \big[ \hat{\bm h}^{(A)}{}_{\m}{}^{({\Hat {\cal R}}')} \big]^{ab}{}_{d}{}^{c}   \\
 &  +  m_{3}  (\g^{5}\g^{\alpha})_{c}{}^{e}  \big[ \hat{\bm h}^{(A)\m}{}^{({\Hat {\cal R}})} 
 \big]_{abe}{}^{f} (\g^{5}\g_{\alpha})_{f}{}^{d}   \big[ \hat{\bm h}^{(A)}{}_{\m}{}^{({\Hat {\cal R}}')
 } \big]^{ab}{}_{d}{}^{c}    \\
 &  +  m_{4}  (\g^{\alpha})_{c}{}^{e}  \big[ \hat{\bm h}^{(A)\m}{}^{({\Hat {\cal R}})} \big]_{abe }{}^{f} (\g_{\alpha})_{f}{}^{d}   \big[ \hat{\bm h}^{(A)}{}_{\m}{}^{({\Hat {\cal R}}')} \big]^{ab}{}_{d}{}^{c}  \\
&  +  m_{5}  ([\g^{\alpha},\g^{\beta}])_{c}{}^{e}  \big[ \hat{\bm h}^{(A)\m}{}^{({\Hat {\cal R}})} \big]_{abe}{}^{f} ([\g_{\alpha},\g_{\beta}])_{f}{}^{d}   \big[ \hat{\bm h}^{(A)}{}_{\m}{}^{({\Hat 
{\cal R}}')} \big]^{ab}{}_{d}{}^{c}  ~~~,
\end{split}
\ee

Recognizing the relations shown in (\ref{Lilhs}), it becomes clear that in order to evaluate the
first three terms is equivalent to only having to evaluate the first term explicitly and one can
later make the appropriate substitutions for the lattice variables.

So we have
\be
\begin{split}
\widehat{\mathcal{G}}_{0}[({\Hat {\cal R}}),({\Hat {\cal R}}')]  =&  4^{3} \big\{  {\rm p}_{0}^{({\Hat 
{\cal R}})} {\rm p}_{0}^{({\Hat {\cal R}}')} ( - m_{1} + m_{2} + 2 m_{3} + 2 m_{4} )  \\
&  +  {\rm q}_{0}^{({\Hat {\cal R}})} {\rm q}_{0}^{({\Hat {\cal R}}')} ( m_{1} - m_{2} + 2 m_{3} + 2 
m_{4} )  \\
&  +  {\rm r}_{0}^{({\Hat {\cal R}})} {\rm r}_{0}^{({\Hat {\cal R}}')} ( m_{1} + m_{2} + 4 m_{3} - 4 
m_{4} - 48 m_{5} )  \\
&  +  3 {\rm s}_{0}^{({\Hat {\cal R}})} {\rm s}_{0}^{({\Hat {\cal R}}')} ( - m_{1} - m_{2} - 16 m_{5} )  
\big\}  ~~~,
\end{split}
\ee
and which also follow for $\widehat{\mathcal{G}}_{3}$ and $\widehat{\mathcal{G}}_{1}$.
We continue by performing the evaluate of $\widehat{\mathcal{G}}_{2}$ and find
\be
\begin{split}
\widehat{\mathcal{G}}_{2}[({\Hat {\cal R}}),({\Hat {\cal R}}')]  =&  4^{3} \big\{  3 {\rm p}_{2}^{({\Hat 
{\cal R}})} {\rm p}_{2}^{({\Hat {\cal R}}')} ( m_{1} - m_{2} + 2 m_{3} + 2 m_{4} )   \\
&  +  3 {\rm q}_{2}^{({\Hat {\cal R}})} {\rm q}_{2}^{({\Hat {\cal R}}')} ( - m_{1} + m_{2} + 2 m_{3} + 2 
m_{4} )  \\
&  +  3 {\rm r}_{2}^{({\Hat {\cal R}})} {\rm r}_{2}^{({\Hat {\cal R}}')} ( - m_{1} - m_{2} - 16 m_{5} ) \\
&  +  {\rm s}_{2}^{({\Hat {\cal R}})} {\rm s}_{2}^{({\Hat {\cal R}}')} ( m_{1} + m_{2} - 4 m_{3} + 4 m_{4} - 48 m_{5} )   
\big\}  ~~~.
\end{split}
\ee
Therefore, adding all the contributions, we obtain
\be
\begin{split}
\widehat{\mathcal{G}}[({\Hat {\cal R}}),({\Hat {\cal R}}')]  
=&~  4^{4} \big\{  ( {\rm p}_{0}^{({\Hat {\cal R}})} {\rm p}_{0}^{({\Hat {\cal R}}')} + {\rm p}_{3}^{({\Hat {\cal R}})} {\rm p}_{3}^{({\Hat {\cal R}}')} + {\rm p}_{1}^{({\Hat {\cal R}})} {\rm p}_{1}^{({\Hat {\cal R}}')} 
 ) ( - m_{1} + m_{2} + 2 m_{3} + 2 m_{4} )
\\
&  ~~~~~~~+ ( 3 {\rm p}_{2}^{({\Hat {\cal R}})} {\rm p}_{2}^{({\Hat {\cal R}}')} ) ( m_{1} - m_{2} + 2 m_{3} + 2 m_{4} )   \\
&  ~~~~~+  ( {\rm q}_{0}^{({\Hat {\cal R}})} {\rm q}_{0}^{({\Hat {\cal R}}')} + {\rm q}_{3}^{({\Hat {\cal 
R}})} {\rm q}_{3}^{({\Hat {\cal R}}')} + {\rm q}_{1}^{({\Hat {\cal R}})} {\rm q}_{1}^{({\Hat {\cal R}}')} ) ( m_{1} - m_{2} + 2 m_{3} + 2 m_{4} )   \\
& ~~~~~~~+ (3 {\rm q}_{2}^{({\Hat {\cal R}})} {\rm q}_{2}^{({\Hat {\cal R}}')} ) ( - m_{1} + m_{2} + 2 m_{3} + 2 m_{4} )    \\
& ~~~~~+ ( {\rm r}_{0}^{({\Hat {\cal R}})} {\rm r}_{0}^{({\Hat {\cal R}}')} + {\rm r}_{3}^{({\Hat {\cal R}})} 
{\rm r}_{3}^{({\Hat {\cal R}}')} + {\rm r}_{1}^{({\Hat {\cal R}})} {\rm r}_{1}^{({\Hat {\cal R}}')} ) ( m_{1} + 
m_{2} + 4 m_{3} - 4 m_{4} - 48 m_{5} )    \\
& ~~~~~~~+ ( 3  {\rm r}_{2}^{({\Hat {\cal R}})} {\rm r}_{2}^{({\Hat {\cal R}}')} ) ( - m_{1} - m_{2} - 16 m_{5} ) \\
& ~~~~~+ ( 3 {\rm s}_{0}^{({\Hat {\cal R}})} {\rm s}_{0}^{({\Hat {\cal R}}')} + 3 {\rm s}_{3}^{({\Hat {\cal 
 R}})} {\rm s}_{3}^{({\Hat {\cal R}}')} + 3 {\rm s}_{1}^{({\Hat {\cal R}})} {\rm s}_{1}^{({\Hat {\cal R}}')} 
) ( - m_{1} - m_{2} - 16 m_{5} ) \\
& ~~~~~~~+  ( {\rm s}_{2}^{({\Hat {\cal R}})} {\rm s}_{2}^{({\Hat {\cal R}}')} ) ( m_{1} + m_{2} - 4 m_{3} + 4 m_{4} - 48 m_{5} )   
  \big\} ~~~.
\end{split}
\label{mtrcM1}
\ee
Let us also note for all of the supermultiplet representations $({\Hat {\cal R}})$ in Table 2 we find
\be
\begin{split}
 \widehat{\mathcal{G}}[({\Hat {\cal R}}),({\Hat {\cal R}})]   
 =&~  4^{4} \big\{ 
 6 \,( - m_{1} + m_{2} + 2 m_{3} + 2 m_{4} ) +  6 \,( m_{1} - m_{2} + 2 m_{3} + 2 m_{4} )
 \\
 & ~~~~~+  3 \, ( m_{1} + m_{2} + 4 m_{3} - 4 m_{4} - 48 m_{5} ) 
 \\
 & {} ~~~~~+ ( m_{1} + m_{2}  - 4 m_{3} + 4 m_{4} - 48 m_{5} ) 
 \\
 & ~~~~~+  12 \, ( - m_{1} - m_{2} - 16 m_{5} )  \big\} \\
 =&~ -\,  4^{4}\, \times 8 \, \big\{  m_1 + m_2 - 4 \, m_3 - \, 2 \,  m_{4} + 48 m_{5}   \big\}  ~~~.
\end{split}
\label{mtrcM2}
\ee
Thus, this analysis shows the requirement that $ \widehat{\mathcal{G}}[({\Hat {\cal R}}),({\Hat {\cal 
R}})]$ should be independent of which minimal representation (which in principle could lead to 
four conditions) only leads to one condition.  Continuing tha analysis when ($\widehat {\cal R}$) 
$\ne$ (${\widehat {\cal R}}{}^{\prime}$) we find
\be
\begin{split}
\widehat{\mathcal{G}}[(\widehat {2VS}),(\widehat {2TS})]  &=
\widehat{\mathcal{G}}[(\widehat {2VS}),(\widehat {2ATS})]  =
\widehat{\mathcal{G}}[(\widehat {2AVS}),(\widehat {2TS})]  =
\widehat{\mathcal{G}}[(\widehat {2AVS}),(\widehat {2ATS})]  = \\ 
&= -\, 4^{4}\, \times 8 \, \big\{  m_1 + m_2 + 2 \, m_3 - \, 2 \,  m_{4}  \big\}  ~~~,
\end{split}
\label{mtrcM3}
\ee
and
\be
\begin{split}
\widehat{\mathcal{G}}[(\widehat {2VS}),(\widehat {2AVS})]  =
\widehat{\mathcal{G}}[(\widehat {2TS}),(\widehat {2ATS})]  = 
-\, 4^{4}\, \times 8 \, \big\{  m_1 + m_2 + 2 \, m_3 + 4\,  m_{4} + 48 m_{5}   \big\}   ~~~,
\end{split}
\label{mtrcM4}
\ee 

So at this stage of the analysis we find
\be
\begin{split}
\widehat{\mathcal{G}}[({\Hat {\cal R}}),({\Hat {\cal R}}')]  =~&
\begin{bmatrix}
{\cal X}_1 &  {\cal X}_2 &  {\cal X}_3 &  {\cal X}_2 \\
{\cal X}_2 &  {\cal X}_1 &  {\cal X}_2 &  {\cal X}_3 \\
{\cal X}_3 &  {\cal X}_2 &  {\cal X}_1 &  {\cal X}_2 \\
{\cal X}_2 &  {\cal X}_3 &  {\cal X}_2 &  {\cal X}_1
 \end{bmatrix}  ~~~,
\end{split}
\label{mtrcM5}
\ee 
where
\be
\begin{split}
{\cal X}_1 =~& -\,  4^{4}\, \times 8 \, \big\{  m_1 + m_2 - 4 \, m_3 - \, 2 \,  m_{4} + 48 m_{5}   \big\}   ~~~, \\
{\cal X}_2 =~&   -\, 4^{4}\, \times 8 \, \big\{  m_1 + m_2 + 2 \, m_3 - \, 2 \,  m_{4}  \big\}  ~~~, \\
{\cal X}_3 =~& -\, 4^{4}\, \times 8 \, \big\{  m_1 + m_2 + 2 \, m_3 + 4\,  m_{4} + 48 m_{5}   \big\}  ~~~. \\
\end{split}
\label{Xs}
\ee 
Let us comment on role of the lattice variables in reaching this result.  Looking back at
(\ref{mtrcM1}), one would have said imposing a set of values for the diagonal and
upper triangular entries would have led to ten equations on the five m-parameters, i.e. 
an over-constrained system.  Instead and precisely due to how the lattice variables 
enter the Holoraumy, we are led to only three equations on the five $m$-parameters, 
i.e. an under-constrained system.

Taking advantage of this and as an analog to 4D, $\mathcal{N}=1$ case, we demand\footnote{
We will discuss the reason for this in our conclusion section.}
\be
\begin{split}
\widehat{\mathcal{G}}[({\Hat {\cal R}}),({\Hat {\cal R}}')]  =~&  \frac{1}{28} \big\{  {\rm p}_{0
}^{({\Hat {\cal R}})} {\rm p}_{0}^{({\Hat {\cal R}}')} + {\rm q}_{0}^{({\Hat {\cal R}})} {\rm q}_{0
}^{({\Hat {\cal R}}')} + {\rm r}_{0}^{({\Hat {\cal R}})} {\rm r}_{0}^{({\Hat {\cal R}}')} + 3 {\rm s}_{
0}^{({\Hat {\cal R}})} {\rm s}_{0}^{({\Hat {\cal R}}')}   \\
&  {~~~~~}+ {\rm p}_{3}^{({\Hat {\cal R}})} {\rm p}_{3}^{({\Hat {\cal R}}')} + {\rm q}_{3}^{({\Hat 
{\cal R}})} {\rm q}_{3}^{({\Hat {\cal R}}')} + {\rm r}_{3}^{({\Hat {\cal R}})} {\rm r}_{3}^{({\Hat {\cal 
R}}')} + 3 {\rm s}_{3}^{({\Hat {\cal R}})} {\rm s}_{3}^{({\Hat {\cal R}}')}   \\
&  {~~~~~}+ {\rm p}_{1}^{({\Hat {\cal R}})} {\rm p}_{1}^{({\Hat {\cal R}}')} + {\rm q}_{1}^{({\Hat 
{\cal R}})} {\rm q}_{1}^{({\Hat {\cal R}}')} + {\rm r}_{1}^{({\Hat {\cal R}})} {\rm r}_{1}^{({\Hat {\cal 
R}}')} + 3 {\rm s}_{1}^{({\Hat {\cal R}})} {\rm s}_{1}^{({\Hat {\cal R}}')}   \\
&  {~~~~~}+ 3 {\rm p}_{2}^{({\Hat {\cal R}})} {\rm p}_{2}^{({\Hat {\cal R}}')} + 3 {\rm q}_{2}^{
({\Hat {\cal R}})} {\rm q}_{2}^{({\Hat {\cal R}}')} + 3 {\rm r}_{2}^{({\Hat {\cal R}})} {\rm r}_{2
}^{({\Hat {\cal R}}')} + {\rm s}_{2}^{({\Hat {\cal R}})} {\rm s}_{2}^{({
\Hat {\cal R}}')} 
 \big\} ~~~,
\end{split}
\label{mtrc}
\ee
by choosing $- m_{1} = - m_{2} = m_{3} = m_{4} = - 8m_{5} = \tfrac{1}{7\cdot 4^{6}}$.
Then we can put all the 4D Gadget values in a matrix as follows:
\be
 \widehat{\mathcal{G}}[({\Hat {\cal R}}),({\Hat {\cal R}}')] = \begin{bmatrix}
  1 & \tfrac{1}{7} & \tfrac{1}{7} & \tfrac{1}{7} \\
  \tfrac{1}{7} & 1 & \tfrac{1}{7} & \tfrac{1}{7} \\
  \tfrac{1}{7} & \tfrac{1}{7} & 1 & \tfrac{1}{7} \\
  \tfrac{1}{7} & \tfrac{1}{7} & \tfrac{1}{7} & 1
 \end{bmatrix}  ~~~,
 \label{mtrc2}
\ee
where the row and column indices run from ($\widehat {2{\rm {VS}}}$), ($\widehat {2{\rm {TS}}}$), 
($\widehat {2{\rm {AVS}}}$), and ($\widehat {2{\rm {ATS}}}$).

Given the metric defined by $ \widehat{\mathcal{G}}[({\Hat {\cal R}}),({\Hat {\cal R}}')] $ in (\ref{mtrc}),
we can explore the geometry of the hexadecimal dimensional space where the numbers ($ \Pi{}_{A}^{( {\Hat {\cal R}})}$) define a set of coordinates.  The results shown in (Table \ref{tab:4DN2-pqrs}) imply 
that each of the supermultiplets can be associated with a point in this space.  As can be seen 
our choice of normalization implies that $ \widehat{\mathcal{G}}[({\Hat {\cal R}}),({\Hat {\cal R}})] 
$ = 1 for each of the representations.  This means we may regard each supermultiplet as being 
a distance of one away from the origin.  The fact that all the off diagonal values of $ \widehat{
\mathcal{G}}[({\Hat {\cal R}}),({\Hat {\cal R}}')] $ in (\ref{mtrc2}) are equal informs us that the angles 
between lines drawn from the origin of this space to each of the points are all the same.  By taking 
the differences between the hexadecimal coordinates and using the metric in (\ref{mtrc}), we find 
that the length of any side joining the points describing the location of any supermultiplet is given 
by $\sqrt{12/7}$.

The most important result of this chapter is that in (\ref{mtrc}) together with the specification
of the $m$-parameters stated just below the equation and the expression shown in (\ref{gdt1}).
Taken all together these show there exist a Lorentz covariant  and so(2) covariant derivation
from (\ref{mtrc}) to (\ref{gdt1}) or vice versa.

\newpage
\section{Reduction to 0-Brane}

Thus far in this work, all our calculations have strictly been in the realm of 4D.  However, 
these calculations are informed by structures related to adinkras \cite{adnk1}.  The entire concept of
Holoraumy arose from this source \cite{adnk1dHoloR1,adnk1dHoloR2} and our proposal that 
there is a type of holography that connects the two seemingly separate domains.  Under 
this circumstance, we think it is prudent to view the results shown in earlier sections of 
this work for the perspective of 1D, $N$ = 8 adinkras.

In preparation for the discussion, we need to set in place some conventions.

There are two sets of the ``general real algebra of dimension d and extension $N$", or ${\cal GR}({\rm d}, N)$ or alternately ``Garden Algebras'' relevant in the following.
The first of these corresponds to the values of d = $N$ = 8 and when expressed as $8 \times 8$ matrices ${\bm {\rm L}}^{(\cal R)}_\rI$ and ${\bm {\rm R}}^{(\cal R)}_\rJ$ we 
require
\be { \eqalign{
{\bm {\rm L}}^{(\cal R)}_\rI \,  {\bm {\rm R}}^{(\cal R)}_\rJ ~+~ 
{\bm {\rm L}}^{(\cal R)}_\rJ \,  {\bm {\rm R}}^{(\cal R)}_\rI 
&=  2\,\d_{\rI \, \rJ}\,   \bm{{\rm I}} {}_{8 \times 8}
 ~~,\cr
{\bm {\rm R}}^{(\cal R)}_\rI \,  {\bm {\rm L}}^{(\cal R)}_\rJ ~+~ 
{\bm {\rm R}}^{(\cal R)}_\rJ \,  {\bm {\rm L}}^{(\cal R)}_\rI 
& =   2\,\d_{\rI \, \rJ}\,   \bm{{\rm I} } {}_{8 \times 8} ~~,\cr
 {\bm {\rm L}}^{(\cal R)}_\rI  ~=~ [ {\bm {\rm R}}^{(\cal R)}_\rI  ]{}^t   &~~,
}}\label{GarDalg8}  
\ee
where the subscript indices I and J take on values of 1, $\dots$, 8 since $N$ = 8 and
d = 8, these are eight  $8 \times 8$ matrices.  Finally, there are different representations of 
these matrices.   In order to indicate this, we use an ``adinkra representation label'' denoted
by ($\cal R$)\footnote{It should be noted that in order to distinguish the  ``supermultiplet  
representation label'' from the  ``adinkra representation label'', we use the symbol (${\widehat 
{\cal R}}$) for the former and (${ {\cal R}}$) for the latter.}.

In order to take advantage of our previous extensive studies of ${\cal GR}(4, 4)$ algebras,
we also introduce a set of $4 \times 4$ ``Garden Algebras'' matrices ${\bm {\rm {\un L}}}^{(\cal R)}_{\underline {\rI}}$ and ${\bm {\rm {\un R}}}^{(\cal R)}_{\underline {\rJ}}$ for the case where d = $N$ = 4.
\be { \eqalign{
{ {\bm {\rm {\un L}}}}^{(\cal R)}_{\underline {\rI}} \,  {\bm {\rm {\un R}}}^{(\cal R)}_{\underline {\rJ}} ~+~ 
{ {\bm {\rm {\un L}}}}^{(\cal R)}_{\underline {\rJ}} \,  {\bm {\rm {\un R}}}^{(\cal R)}_{\underline {\rI}} 
&=  2\,\d_{{\underline {\rI}} \, {\underline {\rJ}}}\,   \bm{{\rm I}} {}_{4 \times 4}
 ~~,\cr
{\bm {\rm {\un R}}}^{(\cal R)}_{\underline {\rI}} \,  { {\bm {\rm {\un L}}}}^{(\cal R)}_{\underline {\rJ}} ~+~ 
{\bm {\rm {\un R}}}^{(\cal R)}_{\underline {\rJ}} \,  { {\bm {\rm {\un L}}}}^{(\cal R)}_{\underline {\rI}} 
& =   2\,\d_{{\underline {\rI}} \, {\underline {\rJ}}}\,   \bm{{\rm I} } {}_{4 \times 4} ~~,\cr
 { {\bm {\rm {\un L}}}}^{(\cal R)}_{\underline {\rI}}  ~=~ [ {\bm {\rm {\un R}}}^{(\cal R)}_{\underline 
 {\rI}}  ]{}^t   &~~.
}}\label{GarDalg4}  
\ee
where the underlined subscript indices $\underline {\rm I}$ and $\underline {\rm J}$ take on 
values of 1, $\dots$, 4 since $N$ = 4 and d = 4.  The advantage afforded from this is it 
allows the explicit construction of the $8 \times 8$ ${\bm {\rm L}}^{(\cal R)}_\rI$ and ${\bm {\rm 
R}}^{(\cal R)}_\rJ$ matrices expressed in terms of $4 \times 4$ ${\bm {\rm {\un L}}}^{(\cal R)}_{
\underline {\rI}}$ and ${\bm {\rm {\un R}}}^{(\cal R)}_{\underline {\rJ}}$matrices .  Careful attention 
should be noted that there is the additional underline as in ${\bm {\rm {\un L}}}$ and ${\bm {\rm 
{\un R}}}$ to distinguish the strictly $4 \times 4$ matrices from the $8 \times 8$ matrices.

The 1D, $N$ = 8 supercovariant derivative operator  $ \text{D}^{}_{\rI}$ can be described as a pair of 1D, $N$ = 4  supercovariant derivatives $ \text{D}^{}_{\rI}$ = ($\text{D}^{1}_{\underline {\rI}}$, $\text{D}^{2}_{\underline {\rI}}$) where
\be
 \text{D}^{}_{\rI} = \begin{cases}
  \text{D}^{1}_{\underline {\rI}} ~~~ , & \text{if } \rI = {\un \rI}   \\
  \text{D}^{2}_{\underline {\rI}} ~~~ , & \text{if } \rI = {\un \rI} + 4 
 \end{cases}   ~~~.
\ee
Using this definition, when acting on a set of 1D valise adinkra/superfields we can write,
\be
\begin{split}
 \text{D}^{1}_{\un {\rI}} \Phi_{i}  ~=&~  i ({\bm {\rm L}}_{\un {\rI}})_{i\hat{k}} \Psi_{\hat{k}}  ~~~\,~~~,~~~~
  \text{D}^{1}_{\un {\rI}} \Psi_{\hat{k}}  ~=~  ({\bm {\rm R}}_{\un {\rI}})_{\hat{k}i} \pa_{0} \Phi_{i} ~~~~~~,  \\
 \text{D}^{2}_{\un {\rI}} \Phi_{i}  ~=&~  i ({\bm {\rm L}}_{{\un {\rI}}+4})_{i\hat{k}} \Psi_{\hat{k}}  ~~~~,~~~~
 \text{D}^{2}_{\un {\rI}} \Psi_{\hat{k}}  ~=~  ({\bm {\rm R}}_{{\un {\rI}}+4})_{\hat{k}i} \pa_{0} \Phi_{i} ~~~,
\end{split}
\ee
where ${\un \rI}=1,\ldots,4$ on the doublet of supercovariant derivative operators and the subscript of ${\bm{\rm L}}_{\rI}$ runs from $\rI=1, \ldots, 8$. 

From how we construct the $\mathcal{N}=2$ vector / tensor / axial-vector / axial-tensor supermultiplets 
from $\mathcal{N}=1$ chiral supermultiplet plus $\mathcal{N}=1$ vector / tensor / axial-vector / 
axial-tensor supermultiplets, we observe that
\be 
\begin{aligned}
{\bm  {\rm L}}^{(2\mathcal{R})}_{{\un \rI}} =& 
 \begin{bmatrix}
  {\bm  {\rm {\un L}}}_{\underline {\rI}}^{(CS)} & 0 \\
   0 &{\bm  {\rm {\un L}}}_{\underline {\rI}}^{(\mathcal{R})} 
 \end{bmatrix} ~~,~~ &
{\bm  {\rm L}}^{(2\mathcal{R})}_{{\un \rI}+4} =& 
 \begin{bmatrix}
   0 & {\bm {\cal {\un S}}}_{1}^{(2\mathcal{R})}{\bm  {\rm {\un L}}}_{\underline {\rI}}^{(CS)} \\
   {\bm {\cal {\un S}}}_{2}^{(2\mathcal{R})}{\bm  {\rm {\un L}}}_{\underline {\rI}}^{(\mathcal{R})} & 0 
 \end{bmatrix}  ~\,~, \\
{\bm  {\rm R}}^{(2\mathcal{R})}_{\un {\rI}} =& 
 \begin{bmatrix}
  {\bm  {\rm {\un R}}}_{\underline {\rI}}^{(CS)} & 0 \\
   0 &{\bm  {\rm {\un R}}}_{\underline {\rI}}^{(\mathcal{R})} 
 \end{bmatrix} ~~,~~  &
{\bm  {\rm {R}}}^{(2\mathcal{R})}_{{\un {\rI}} +4} =& 
 \begin{bmatrix}
   0 &{\bm  {\rm {\un R}}}_{\underline {\rI}}^{(\mathcal{R})} {\bm {\cal {\un S}}}_{2}^{(2\mathcal{R})} \\
  {\bm  {\rm {\un R}}}_{\underline {\rI}}^{(CS)} {\bm {\cal {\un S}}}_{1}^{(2\mathcal{R})} & 0 
 \end{bmatrix} ~~. \\
\end{aligned}
\label{8-4mtrx}
\ee
where the $N$ = 4 adinkra representation label, $(\mathcal{R}),$ takes its values as (VS), (TS), 
(AVS), and (ATS);  while $N$ = 8 adinkra representation label, $(2\mathcal{R})$ takes its values 
as (2VS), (2TS), (2AVS), (2ATS).  

Also in writing (\ref{8-4mtrx}), we have utilized another notational device (``Boolean Factors'') 
introduced previously in the work of \cite{permutadnk}.  We define Boolean Factors as real diagonal 
matrices whose square is the identity.  Thus, any Boolean Factor has the form of the matrix shown 
in (\ref{BF})
\be
 {\bm {\cal S}}{}_b = \begin{bmatrix}
 (-1)^{b_{1}} & 0 & 0 & \cdots & 0 \\
 0 & (-1)^{b_{2}} & 0 & \cdots & 0 \\
 0 & 0 & (-1)^{b_{3}} & \cdots & 0 \\
 \vdots & \vdots & \vdots & \ddots & 0 \\
 0 & 0 & 0 & 0 & (-1)^{b_{d}}
 \end{bmatrix}
 \qquad \Leftrightarrow \qquad
 [b_{1}b_{2}b_{3}\cdots b_{d}]_{2} = \left( \sum_{i=1}^{d} b_{i} 2^{i-1} \right)_{b}
 ~~.
 \label{BF}
\ee
where the diagonal entries may be expressed as exponentials of ($-1$) to a set of bits, i.\ e.\ variables
that only take on values of one or zero, indicated by $b_1$, $\dots$ $b_d$.  On the RHS of (\ref{BF}),
we have shown how the bits that occur in a Boolean Factor can be written as an integer using
base two.  Thus any specific Boolean Factor can be denoted by such an integer as we show in
(\ref{BF1}) 
\be
\begin{aligned}
 {\bm {\cal {\un S}}}_{1}^{(2VS)} =& [0001]_{2} = (8)_{b} ~~,~~ &
 {\bm {\cal {\un S}}}_{2}^{(2VS)} =& [1110]_{2} = (7)_{b} ~~~\,,~~ \\
 {\bm {\cal {\un S}}}_{1}^{(2TS)} =& [1000]_{2} = (1)_{b} ~~,~~ &
 {\bm {\cal {\un S}}}_{2}^{(2TS)} =& [0111]_{2} = (14)_{b}~~,~~ \\
 {\bm {\cal {\un S}}}_{1}^{(2AVS)} =& [0010]_{2} = (4)_{b} ~~,~~&
 {\bm {\cal {\un S}}}_{2}^{(2AVS)} =& [1110]_{2} = (7)_{b}~~\,~,~~ \\
 {\bm {\cal {\un S}}}_{1}^{(2ATS)} =& [0100]_{2} = (2)_{b} ~~,~~&
 {\bm {\cal {\un S}}}_{2}^{(2ATS)} =& [0111]_{2} = (14)_{b} ~~.
\end{aligned}
 \label{BF1}
\ee

The L-matrices, themselves are simply signed permutation matrices, so we will express the as
Boolean Factors times permutation factors of order four.  We will make use of a notation 
$\langle n_1 n_2 n_3 n_4 \rangle$ as in \cite{permutadnk} to specify the permutation factors.  Thus at the 
end of this discussion, we will be able to express the $8 \times 8$ ${\bm {\rm L}}^{(\cal 
R)}_\rI $ matrices in terms of Boolean Factors times permutation elements denoted by $\langle n_1 n_2 n_3 n_4 \rangle$ for some integers $n_1$, $n_2$, $n_3$, and $n_4$.  These integers themselves are 
simply a reordering of the integers 1, 2, 3, and 4.

\subsection{L-matrices from the 4D, $\mathcal{N}=2$ Vector Supermultiplet}

For the 4D, $\cal N$ = 2 vector supermultiplet, the bosons are $A$, $B$, $F$, and $G$ from the
chiral 4D, $\cal N$ = 1 chiral supermultiplet and the spatial components of $A_{\m}$ and $d$ from
the 4D, $\cal N$ = 1 vector supermultiplet.  To define the bosons in the 1D, $N$ = 8 $(2VS)$ adinkra
representation we define the bosons $\Phi{}_i$ via
\be
\begin{aligned}
 \Phi_{1} ~=&~ A ~~~, & ~~~  \Phi_{2} ~=&~ B ~~~, & ~~~  \pa_{0}\Phi_{3} ~=&~ F ~~~, 
 & ~~~  \pa_{0}\Phi_{4} ~=&~ G  ~~~,  \\
 \Phi_{5} ~=&~ A_{1} ~~, & ~~~  \Phi_{6} ~=&~ A_{2} ~~, & ~~~  \Phi_{7} ~=&~ A_{3} ~~, 
 & ~~~  \pa_{0}\Phi_{8} ~=&~ d ~~~~,
\end{aligned}
\ee
and the fermions $\Psi_{\hk}$ in the 1D, $N$ = 8 $(2VS)$ adinkra representation via
\be
\begin{aligned}
 i \Psi_{1} ~=&~ \psi_{1}  ~~, & ~~~  i \Psi_{2} ~=&~ \psi_{2} ~~, & ~~~  i \Psi_{3} ~=&~ \psi_{3} ~~, 
 & ~~~  i \Psi_{4} ~=&~ \psi_{4}  ~~~, \\
 i \Psi_{5} ~=&~ \lambda_{1} ~~, & ~~~  i \Psi_{6} ~=&~ \lambda_{2} ~~, & ~~~  i \Psi_{7} ~=&~ 
 \lambda_{3} ~~, & ~~~  i \Psi_{8} ~=&~ \lambda_{4} ~~~.
\end{aligned}
\ee

The L-matrices of the $\mathcal{N}=2$ vector supermultiplet are
\be
\begin{aligned}
{\bm  {\rm L}}^{(2VS)}_{1} =& 
 \begin{bmatrix}
   (10)_{b} \langle 1423 \rangle & 0 \\
   0 & (10)_{b} \langle 2413 \rangle
 \end{bmatrix} ~~~, &
{\bm  {\rm L}}^{(2VS)}_{2} =& 
 \begin{bmatrix}
   (12)_{b} \langle 2314 \rangle & 0 \\
   0 & (12)_{b} \langle 1324 \rangle
 \end{bmatrix} ~~~, \\
{\bm  {\rm L}}^{(2VS)}_{3} =& 
 \begin{bmatrix}
   (6)_{b} \langle 3241 \rangle & 0 \\
   0 & (0)_{b} \langle 4231 \rangle 
\end{bmatrix} ~~~~~~,  &
{\bm  {\rm L}}^{(2VS)}_{4} =& 
 \begin{bmatrix}
   (0)_{b} \langle 4132 \rangle & 0 \\
   0 & (6)_{b} \langle 3142 \rangle 
 \end{bmatrix} ~~~~~~,  \\
{\bm  {\rm L}}^{(2VS)}_{5} =& 
 \begin{bmatrix}
   0 & (2)_{b} \langle 1423 \rangle \\
   (13)_{b} \langle 2413 \rangle & 0
 \end{bmatrix} ~~~\,~,  &
{\bm  {\rm L}}^{(2VS)}_{6} =& 
 \begin{bmatrix}
   0 & (4)_{b} \langle 2314 \rangle \\
   (11)_{b} \langle 1324 \rangle & 0 
 \end{bmatrix}~~~\,~,  \\
{\bm  {\rm L}}^{(2VS)}_{7} =& 
 \begin{bmatrix}
   0 & (14)_{b} \langle 3241 \rangle \\
   (7)_{b} \langle 4231 \rangle & 0 
\end{bmatrix} ~\,~~~, &
{\bm  {\rm L}}^{(2VS)}_{8} =& 
 \begin{bmatrix}
   0 & (8)_{b} \langle 4132 \rangle \\
   (1)_{b} \langle 3142 \rangle & 0 
 \end{bmatrix} ~~~~~~.
\end{aligned}
\ee

\subsection{L-matrices from 4D, $\mathcal{N}=2$ Tensor Supermultiplet}

For the 4D, $\cal N$ = 2 tensor supermultiplet, the bosons are $A$, $B$, $F$, and $G$ from the
chiral 4D, $\cal N$ = 1 chiral supermultiplet and the spatial components of $B_{\m \, \n}$ and $\varphi$ 
from the 4D, $\cal N$ = 1 tensor supermultiplet.  To define the bosons in the 1D, $N$ = 8 $(2TS)$ adinkra
representation we define the bosons $\Phi{}_i$ via
\be
\begin{aligned}
 \Phi_{1} ~=&~ A ~~~, & ~~~  \Phi_{2} ~=&~ B ~~~~\,~, & ~~~  \pa_{0}\Phi_{3} ~=&~ F ~~~\,~~, & ~~~  
 \pa_{0}\Phi_{4} ~=&~ G  ~~~\,~~~~,  \\
 \Phi_{5} ~=&~ \varphi ~~~, & ~~~  \Phi_{6} ~=&~ 2B_{12} ~~, & ~~~  \Phi_{7} ~=&~ 2B_{23} ~~, & ~~~  
\Phi_{8} ~=&~ 2B_{31} ~~~~,
\end{aligned}
\ee
and the fermions $\Psi_{\hk}$ in the 1D, $N$ = 8 $(2TS)$ adinkra representation via
\be
\begin{aligned}
 i \Psi_{1} ~=&~ \psi_{1}  ~~, & ~~~  i \Psi_{2} ~=&~ \psi_{2} ~~, & ~~~  i \Psi_{3} ~=&~ \psi_{3} ~~, 
 & ~~~  i \Psi_{4} ~=&~ \psi_{4} ~~~,  \\
 i \Psi_{5} ~=&~ \chi_{1} ~~, & ~~~  i \Psi_{6} ~=&~ \chi_{2} ~~, & ~~~  i \Psi_{7} ~=&~ \chi_{3} ~~, 
 & ~~~  i \Psi_{8} ~=&~ \chi_{4} ~~~.
\end{aligned}
\ee

The L-matrices of the $\mathcal{N}=2$ tensor supermultiplet are
\be
\begin{aligned}
{\bm  {\rm L}}^{(2TS)}_{1} =& 
 \begin{bmatrix}
   (10)_{b} \langle 1423 \rangle & 0 \\
   0 & (14)_{b} \langle 1342 \rangle
 \end{bmatrix} ~~~,  &
{\bm  {\rm L}}^{(2TS)}_{2} =& 
 \begin{bmatrix}
   (12)_{b} \langle 2314 \rangle & 0 \\
   0 & (4)_{b} \langle 2431 \rangle
 \end{bmatrix}   ~~~~~, \\
{\bm  {\rm L}}^{(2TS)}_{3} =& 
 \begin{bmatrix}
   (6)_{b} \langle 3241 \rangle & 0 \\
   0 & (8)_{b} \langle 3124 \rangle 
\end{bmatrix}  ~~~~~~, &
{\bm  {\rm L}}^{(2TS)}_{4} =& 
 \begin{bmatrix}
   (0)_{b} \langle 4132 \rangle & 0 \\
   0 & (2)_{b} \langle 4213 \rangle 
 \end{bmatrix}  ~~~~~\,~, \\
{\bm  {\rm L}}^{(2TS)}_{5} =& 
 \begin{bmatrix}
   0 & (11)_{b} \langle 1423 \rangle \\
   (0)_{b} \langle 1342 \rangle & 0
 \end{bmatrix}  ~~\,~~, &
{\bm  {\rm L}}^{(2TS)}_{6} =& 
 \begin{bmatrix}
   0 & (13)_{b} \langle 2314 \rangle \\
   (10)_{b} \langle 2431 \rangle & 0 
 \end{bmatrix} ~~\,~, \\
{\bm  {\rm L}}^{(2TS)}_{7} =& 
 \begin{bmatrix}
   0 & (7)_{b} \langle 3241 \rangle \\
   (6)_{b} \langle 3124 \rangle & 0 
\end{bmatrix} ~~~~~~, &
{\bm  {\rm L}}^{(2TS)}_{8} =& 
 \begin{bmatrix}
   0 & (1)_{b} \langle 4132 \rangle \\
   (12)_{b} \langle 4213 \rangle & 0 
 \end{bmatrix} ~~~~~.
\end{aligned}
\ee

\subsection{L-matrices from the 4D, $\mathcal{N}=2$ Axial-Vector Supermultiplet}

For the 4D, $\cal N$ = 2 axial-vector supermultiplet, the bosons are $A$, $B$, $F$, and $G$ from the
chiral 4D, $\cal N$ = 1 chiral supermultiplet and the spatial components of $U_{\m}$ and $\tilde d$ from
the 4D, $\cal N$ = 1 axial-vector supermultiplet.  To define the bosons in the 1D, $N$ = 8 $(2AVS)$ adinkra
representation we define the bosons $\Phi{}_i$ via
\be
\begin{aligned}
 \Phi_{1} ~=&~ A ~~~, & ~~~  \Phi_{2} ~=&~ B ~~~, & ~~~  \pa_{0}\Phi_{3} ~=&~ F ~~~, 
 & ~~~  \pa_{0}\Phi_{4} ~=&~ G  ~~~,  \\
 \Phi_{5} ~=&~ U_{1} ~~, & ~~~  \Phi_{6} ~=&~ U_{2} ~~, & ~~~  \Phi_{7} ~=&~ U_{3} ~~, 
 & ~~~  \pa_{0}\Phi_{8} ~=&~ {\tilde d} ~~~~,
\end{aligned}
\ee
and the fermions $\Psi_{\hk}$ in the 1D, $N$ = 8 $(2AVS)$ adinkra representation via
\be
\begin{aligned}
 i \Psi_{1} ~=&~ \psi_{1}  ~~, & ~~~  i \Psi_{2} ~=&~ \psi_{2} ~~, & ~~~  i \Psi_{3} ~=&~ \psi_{3} ~~, 
 & ~~~  i \Psi_{4} ~=&~ \psi_{4}  ~~~, \\
 i \Psi_{5} ~=&~ {\tilde \lambda}_{1} ~~, & ~~~  i \Psi_{6} ~=&~  {\tilde \lambda}_{2} ~~, & ~~~  
 i \Psi_{7} ~=&~  {\tilde \lambda}_{3} ~~, & ~~~  i \Psi_{8} ~=&~  {\tilde \lambda}_{4} ~~~.
\end{aligned}
\ee

The L-matrices of the $\mathcal{N}=2$ axial-vector supermultiplet are
\be
\begin{aligned}
{\bm  {\rm L}}^{(2AVS)}_{1} =& 
 \begin{bmatrix}
   (10)_{b} \langle 1423 \rangle & 0 \\
   0 & (9)_{b} \langle 3142 \rangle
 \end{bmatrix}  ~~~, &
{\bm  {\rm L}}^{(2AVS)}_{2} =& 
 \begin{bmatrix}
   (12)_{b} \langle 2314 \rangle & 0 \\
   0 & (0)_{b} \langle 4231 \rangle
 \end{bmatrix}   ~~~, \\
{\bm  {\rm L}}^{(2AVS)}_{3} =& 
 \begin{bmatrix}
   (6)_{b} \langle 3241 \rangle & 0 \\
   0 & (3)_{b} \langle 1324 \rangle 
\end{bmatrix} ~~~\,~, &
{\bm  {\rm L}}^{(2AVS)}_{4} =& 
 \begin{bmatrix}
   (0)_{b} \langle 4132 \rangle & 0 \\
   0 & (10)_{b} \langle 2413 \rangle 
 \end{bmatrix} ~~~, \\
{\bm  {\rm L}}^{(2AVS)}_{5} =& 
 \begin{bmatrix}
   0 & (14)_{b} \langle 1423 \rangle \\
   (14)_{b} \langle 3142 \rangle & 0
 \end{bmatrix} \,~,  &
{\bm  {\rm L}}^{(2AVS)}_{6} =& 
 \begin{bmatrix}
   0 & (8)_{b} \langle 2314 \rangle \\
   (7)_{b} \langle 4231 \rangle & 0 
 \end{bmatrix} ~~~\,~,  \\
{\bm  {\rm L}}^{(2AVS)}_{7} =& 
 \begin{bmatrix}
   0 & (2)_{b} \langle 3241 \rangle \\
   (4)_{b} \langle 1324 \rangle & 0 
\end{bmatrix} ~~~~, &
{\bm  {\rm L}}^{(2AVS)}_{8} =& 
 \begin{bmatrix}
   0 & (4)_{b} \langle 4132 \rangle \\
   (13)_{b} \langle 2413 \rangle & 0 
 \end{bmatrix} ~~~.
\end{aligned}
\ee

\subsection{L-matrices from the 4D, $\mathcal{N}=2$ Axial-Tensor Supermultiplet}

For the 4D, $\cal N$ = 2 axial-tensor supermultiplet, the bosons are $A$, $B$, $F$, and $G$ from the
chiral 4D, $\cal N$ = 1 chiral supermultiplet and the spatial components of $C_{\m \, \n}$ and ${\tilde 
\varphi}$ from the 4D, $\cal N$ = 1 axial-tensor supermultiplet.  To define the bosons in the 1D, $N$ 
= 8 $(2ATS)$ adinkra representation we define the bosons $\Phi{}_i$ via
\be
\begin{aligned}
 \Phi_{1} ~=&~ A ~~~, & ~~~  \Phi_{2} ~=&~ B ~~~~\,~, & ~~~  \pa_{0}\Phi_{3} ~=&~ F ~~~\,~~, & ~~~  
 \pa_{0}\Phi_{4} ~=&~ G  ~~~\,~~~~,  \\
 \Phi_{5} ~=&~ \tilde{\varphi} ~~~, & ~~~  \Phi_{6} ~=&~ 2C_{12} ~~, & ~~~  \Phi_{7} ~=&~ 2C_{23} ~~, & ~~~  
\Phi_{8} ~=&~ 2C_{31} ~~~~,
\end{aligned}
\ee
and the fermions $\Psi_{\hk}$ in the 1D, $N$ = 8 $(2ATS)$ adinkra representation via
\be
\begin{aligned}
 i \Psi_{1} ~=&~ \psi_{1}  ~~, & ~~~  i \Psi_{2} ~=&~ \psi_{2} ~~, & ~~~  i \Psi_{3} ~=&~ \psi_{3} ~~, & 
 ~~~  i \Psi_{4} ~=&~ \psi_{4} ~~~,  \\
 i \Psi_{5} ~=&~ {\tilde \chi}_{1} ~~, & ~~~  i \Psi_{6} ~=&~ {\tilde \chi}_{2} ~~, & ~~~  i \Psi_{7} ~=&~ 
 {\tilde \chi}_{3} ~~, & ~~~  i \Psi_{8} ~=&~ {\tilde \chi}_{4} ~~~.
\end{aligned}
\ee

The L-matrices of the $\mathcal{N}=2$ axial-tensor supermultiplet are
\be
\begin{aligned}
{\bm  {\rm L}}^{(2ATS)}_{1} =& 
 \begin{bmatrix}
   (10)_{b} \langle 1423 \rangle & 0 \\
   0 & (13)_{b} \langle 4213 \rangle
 \end{bmatrix}  ~~~, &
{\bm  {\rm L}}^{(2ATS)}_{2} =& 
 \begin{bmatrix}
   (12)_{b} \langle 2314 \rangle & 0 \\
   0 & (8)_{b} \langle 3124 \rangle
 \end{bmatrix}  ~~~,  \\
{\bm  {\rm L}}^{(2ATS)}_{3} =& 
 \begin{bmatrix}
   (6)_{b} \langle 3241 \rangle & 0 \\
   0 & (11)_{b} \langle 2431 \rangle 
\end{bmatrix}  ~~~~, &
{\bm  {\rm L}}^{(2ATS)}_{4} =& 
 \begin{bmatrix}
   (0)_{b} \langle 4132 \rangle & 0 \\
   0 & (14)_{b} \langle 1342 \rangle 
 \end{bmatrix} ~~~,  \\
{\bm  {\rm L}}^{(2ATS)}_{5} =& 
 \begin{bmatrix}
   0 & (8)_{b} \langle 1423 \rangle \\
   (3)_{b} \langle 4213 \rangle & 0
 \end{bmatrix}   ~~~~~\,,  &
{\bm  {\rm L}}^{(2ATS)}_{6} =& 
 \begin{bmatrix}
   0 & (14)_{b} \langle 2314 \rangle \\
   (6)_{b} \langle 3124 \rangle & 0 
 \end{bmatrix}  ~~~,  \\
{\bm  {\rm L}}^{(2ATS)}_{7} =& 
 \begin{bmatrix}
   0 & (4)_{b} \langle 3241 \rangle \\
   (5)_{b} \langle 2431 \rangle & 0 
\end{bmatrix}   ~~~~\,~, &
{\bm  {\rm L}}^{(2ATS)}_{8} =& 
 \begin{bmatrix}
   0 & (2)_{b} \langle 4132 \rangle \\
   (0)_{b} \langle 1342 \rangle & 0 
 \end{bmatrix} ~~\,~~.
\end{aligned}
\ee

\newpage

\section{Adinkra 1D, $N$ = 8 Holoraumy Matrices}

We define the bosonic holoraumy matrix ${\bm V}^{(\cal R)}_{\rI\rJ}$ and
the fermionic holoraumy matrix $ {\bm {\tilde{V}}}^{(\cal R)}_{\rI\rJ} $
via the respective equations shown in
(\ref{vees})
\be 
\begin{split}
{\bm {\rm L}}^{(\cal R)}_\rI \,  {\bm {\rm R}}^{(\cal R)}_\rJ ~-~ 
{\bm {\rm L}}^{(\cal R)}_\rJ \,  {\bm {\rm R}}^{(\cal R)}_\rI   
~=&~ i 2 \, {\bm V}^{(\cal R)}_{\rI\rJ}  ~~~, \\
{\bm {\rm R}}^{(\cal R)}_\rI \,  {\bm {\rm L}}^{(\cal R)}_\rJ ~-~ 
{\bm {\rm R}}^{(\cal R)}_\rJ \,  {\bm {\rm L}}^{(\cal R)}_\rI 
 ~=&~ i 2 \, {\bm {\tilde{V}}}^{(\cal R)}_{\rI\rJ} ~~~,
\end{split}
\label{vees}
\ee
in 1D, $N$ = 8 systems.  Thus, if we assemble the bosonic and fermionic components 
into $ \Phi {}^{(\cal R)}$ and $\Psi {}^{(\cal R)}$ according to the definitions
\be \eqalign{
\Phi {}^{(\cal R)} ~=&~  
\left[\begin{array}{c}
~\Phi {}^{(\cal R)}_{1} \\
\vdots~\\
~\Phi {}^{(\cal R)}_{8} \\
\end{array}\right]
~~,~~
\Psi {}^{(\cal R)} ~=~
\left[\begin{array}{c}
~\Psi {}^{(\cal R)}_{\hat 1} \\
\vdots~\\
~\Psi {}^{(\cal R)}_{\hat 8} \\
\end{array}\right] ~~~,
} \ee
then we have
\be
\begin{split}
[ \,  \text{D}_{\rI} ~,~ \text{D}_{\rJ} \, ]  \, \Phi {}^{(\cal R)} ~=~ 2 \, {\bm V}^{(\cal R)}_{
\rI\rJ} \, \pa_{0} \Phi {}^{(\cal R)}  
~~~,~~~
[ \,  \text{D}_{\rI} ~,~ \text{D}_{\rJ} \, ]  \, \Psi {}^{(\cal R)}  ~=~ 2 \, {\bm {\tilde{V}}}^{(\cal 
R)}_{\rI\rJ}\,  \pa_{0} \Psi {}^{(\cal R)}  ~~~.
\end{split}
\label{heses}
\ee
The key point about the equations in (\ref{heses}) is that same quantities (i.\ e.\ $\Phi {}^{(\cal R)}$
and $\Psi {}^{(\cal R)}$) appear on both sides of each respective equation.  So given differentiable
functions $\Phi {}^{(\cal R)} $ and $\Psi {}^{(\cal R)}$, these equations allow the determination of
the holoraumy quantities  ${\bm V}^{(\cal R)}_{\rI\rJ} $ and ${\bm {\tilde{V}}}^{(\cal R)}_{\rI\rJ}$.
Furthermore due to the engineering dimensions of the D-operators, $\pa_0$, and the functions
$\Phi {}^{(\cal R)}$ and $\Psi {}^{(\cal R)}$, the holoraumy quantities are dimensionless.   Aside
from the fact that temporal derivatives appear on the RHS of each of these equations, they have
exactly the form of equations that define eigenvalues within the context of simple Lie algebras.  
This is the observation that suggests a representation theory description of 1D supermultiplets can 
be constructed on the basis of the constants that appear in the holoraumy quantities  ${\bm V}^{
(\cal R)}_{\rI\rJ} $ and ${\bm {\tilde {V}}}^{(\cal R)}_{\rI\rJ}$.

Other implications of the equations in (\ref{heses}) is they can be used to derive the results
\be
\begin{split}
\pa_{0} \Phi {}^{(\cal R)} ~=~ \fracm 1{28} \, {\bm V}^{\rI\rJ (\cal R)} \,  
\left[ \,  \text{D}_{\rI} \, \text{D}_{\rJ} \,  \Phi {}^{(\cal R)}  \, \right] ~~~,~~~
 \pa_{0}  \Psi {}^{(\cal R)} ~=~ \fracm 1{28} \, {\bm {\tilde{V}}}^{\rI\rJ (\cal R)}\,  
\left[ \,  \text{D}_{\rI} \, \text{D}_{\rJ} \,  \Psi {}^{(\cal R)}  \, \right]    ~~~.
\end{split}
\label{Heses}
\ee
Written in this form, the holoraumy quantities  ${\bm V}^{(\cal R)}_{\rI\rJ} $ and ${\bm {\tilde{V
}}}^{(\cal R)}_{\rI\rJ}$ together with the D-operators generate time evolution of the  bosonic
and fermionic variables $\Phi {}^{(\cal R)}$ and $\Psi {}^{(\cal R)}$.

The bosonic holoraumy matrix ${\bm V}^{(\cal R)}_{\rI\rJ}$ and the fermionic holoraumy matrix $ {\bm {\tilde{V}}}^{(\cal R)}_{\rI\rJ} $ are both defined in such a way as to be hermitian,
\be
\begin{split}
[  \,  {\bm V}_{\rI\rJ}^{(\cal R)} \,]{}^{\dagger} ~=&~ {\bm V}_{\rI\rJ}^{(\cal R)}  ~~~,~~~
[  \,  {\bm {\tilde{V}}}_{\rI\rJ}^{(\cal R)}  \,]{}^{\dagger}  ~=~  {\bm {\tilde{V}}}_{\rI\rJ}^{(\cal R)}
~~~.
\end{split}
\ee
Now we focus on the fermionic holoraumy. From the definition (\ref{vees}), we have
\be
{\bm {\tilde V}}{}_{\rI\rJ} ~=~ - i \tfrac{1}{2} \, ( \, 
{\bm {\rm R}}{}_\rI \,  {\bm {\rm L}}{}_\rJ \,-\, 
{\bm {\rm R}}{}_\rJ \,  {\bm {\rm L}}{}_\rI  \,)  ~~~,
\ee
and we have ``dropped'' the adinkra representation label (${\cal R}$) as the following
will be valid for all such representations.

We can write
\be
\begin{split}
{\bm {\tilde{V}}}{}_{\rI\rJ}  \, {\bm {\tilde{V}}}{}_{\rK\rL} 
~=&~ -\, \tfrac{1}{4} \, ( \, {\bm {\rm R}}{}_\rI \,  {\bm {\rm L}}{}_\rJ \,-\, 
{\bm {\rm R}}{}_\rJ \,  {\bm {\rm L}}{}_\rI  \,) \, ( \, {\bm {\rm R}}{}_\rK \,  {\bm {\rm L}}{}_\rL \,-\, 
{\bm {\rm R}}{}_\rL \,  {\bm {\rm L}}{}_\rK  \,)        \\
=&~  - \tfrac{1}{4} ( {\bm {\rm R}}{}_{\rI}{\bm {\rm L}}{}_{\rJ}{\bm {\rm R}}{}_{\rK}{\bm {\rm L}}
{}_{\rL} - {\bm {\rm R}}{}_{\rI}{\bm {\rm L}}{}_{\rJ}{\bm {\rm R}}{}_{\rL}{\bm {\rm L}}{}_{\rK} ) - 
(\rI \leftrightarrow \rJ)       \\ 
=&~  - \tfrac{1}{4} ( {\bm {\rm R}}{}_{\rK}{\bm {\rm L}}{}_{\rL}{\bm {\rm R}}{}_{\rI}{\bm {\rm L}}
{}_{\rJ} + 2 \delta_{\rJ\rK}{\bm {\rm R}}{}_{\rI}{\bm {\rm L}}{}_{\rL} - 2 \delta_{\rI\rK}{\bm {\rm 
R}}{}_{\rJ}{\bm {\rm L}}{}_{\rL} + 2 \delta_{\rJ\rL}{\bm {\rm R}}{}_{\rK}{\bm {\rm L}}{}_{\rI} - 2 
\delta_{\rI\rL}{\bm {\rm R}}{}_{\rK}{\bm {\rm L}}{}_{\rJ}  \\
&~  \quad  - {\bm {\rm R}}{}_{\rL}{\bm {\rm L}}{}_{\rK}{\bm {\rm R}}{}_{\rI}{\bm {\rm L}}{}_{\rJ} 
- 2 \delta_{\rJ\rL}{\bm {\rm R}}{}_{\rI}{\bm {\rm L}}{}_{\rK} + 2 \delta_{\rI\rL}{\bm {\rm R}}{}_{\rJ
}{\bm {\rm L}}{}_{\rK} - 2 \delta_{\rJ\rK}{\bm {\rm R}}{}_{\rL}{\bm {\rm L}}{}_{\rI} + 2 \delta_{\rI\rK
}{\bm {\rm R}}{}_{\rL}{\bm {\rm L}}{}_{\rJ} ) - (\rI \leftrightarrow \rJ)        \\ 
=&~  - \tfrac{1}{4} ( {\bm {\rm R}}{}_{\rK}{\bm {\rm L}}{}_{\rL} - {\bm {\rm R}}{}_{\rL}{\bm {\rm 
L}}{}_{\rK} ) ( {\bm {\rm R}}{}_{\rI}{\bm {\rm L}}{}_{\rJ} - {\bm {\rm R}}{}_{\rJ}{\bm {\rm L}}{}_{\rI} )   \\ 
&  \quad  + \delta_{\rI\rK} ( {\bm {\rm R}}{}_{\rJ}{\bm {\rm L}}{}_{\rL} - {\bm {\rm R}}{}_{\rL}{\bm {\rm 
L}}{}_{\rJ} ) - \delta_{\rJ\rK} ( {\bm {\rm R}}{}_{\rI}{\bm {\rm L}}{}_{\rL} - {\bm {\rm R}}{}_{\rL}{\bm {\rm 
L}}{}_{\rI} )        \\ 
&  \quad  + \delta_{\rI\rL} ( {\bm {\rm R}}{}_{\rK}{\bm {\rm L}}{}_{\rJ} - {\bm {\rm R}}{}_{\rJ}{\bm {\rm 
L}}{}_{\rK} ) - \delta_{\rJ\rL} ( {\bm {\rm R}}{}_{\rK}{\bm {\rm L}}{}_{\rI} - {\bm {\rm R}}{}_{\rI}{\bm {\rm 
L}}{}_{\rK} )        \\ 
=&~ {\bm {\tilde{V}}}{}_{\rK\rL}  {\bm {\tilde{V}}}{}_{\rI\rJ} + i 2 \delta_{\rI\rK}  {\bm {\tilde{V}}}{}_{\rJ\rL
} - i 2 \delta_{\rJ\rK}  {\bm {\tilde{V}}}{}_{\rI\rL} + i 2 \delta_{\rI\rL}  {\bm {\tilde{V}}}{}_{\rK\rJ} - i 2 
\delta_{\rJ\rL}  {\bm {\tilde{V}}}{}_{\rK\rI}  ~~~.
\end{split}
\ee
Therefore,
\be
[ {\bm {\tilde{V}}}{}_{\rI\rJ},  {\bm {\tilde{V}}}{}_{\rK\rL}] = i 2 \delta_{\rI\rK}  {\bm {\tilde{V}}}{}_{\rJ\rL} 
- i 2 \delta_{\rJ\rK}  {\bm {\tilde{V}}}{}_{\rI\rL} + i 2 \delta_{\rI\rL}  {\bm {\tilde{V}}}{}_{\rK\rJ} - i 2 \delta_{
\rJ\rL}  {\bm {\tilde{V}}}{}_{\rK\rI}  ~~~,
\ee
which shows that ${\bm {\tilde{V}}}_{\rI\rJ}$ belongs to the spinor representation of so($N$) algebra. 
For the systems that we are considering in this paper, it is a so(8) algebra.

By using equation (\ref{8-4mtrx}), the results seen in (\ref{eqn:blockV}) and (\ref{eqn:blockV2}) below are derived.
\be \label{eqn:blockV}
\begin{aligned}
& {\bm {\tilde{V}}}{}_{{\un {\rI}}{\un {\rJ}}}^{(2\mathcal{R})} = 
\begin{bmatrix}
{\bm {\tilde{{\un V}}}}{}_{{\un {\rI}}{\un {\rJ}}}^{(CS)} & 0 \\
0 & {\bm {\tilde{{\un V}}}}{}_{{\un {\rI}}{\un {\rJ}}}^{(\mathcal{R})} 
\end{bmatrix}  
\qquad \qquad , \qquad
{\bm {\tilde{{V}}}}{}_{{\un {\rI}}+4,{\un {\rJ}}+4}^{(2\mathcal{R})} =
\begin{bmatrix}
{\bm {\tilde{{\un V}}}}{}_{{\un {\rI}}{\un {\rJ}}}^{(\mathcal{R})} & 0 \\
0 & {\bm {\tilde{{\un V}}}}{}_{{\un {\rI}}{\un {\rJ}}}^{(CS)} 
\end{bmatrix} & ~~~, \\
\end{aligned}\ee
\be  \label{eqn:blockV2}
\begin{split}
{\bm {\tilde{{V}}}}{}_{{\un {\rI}},{\un {\rJ}}+4}^{(\mathcal{R})} & ~=~ -
{\bm {\tilde{{V}}}}{}_{{{\un {\rJ}}+4},{\un {\rI}}}^{(\mathcal{R})}  \\
~=~ & \begin{bmatrix}
 0 & - i \tfrac{1}{2} \left( {\bm {\rm {\un R}}}{}_{{\un {\rI}}}^{(CS)} {\bm  {\cal {\un S}}}_{1}^{(2\mathcal{R})} {\bm {\rm 
{\un L}}}{}_{{\un {\rJ}}}^{(CS)} - {\bm {\rm {\un R}}}{}_{{\un {\rJ}}}^{(\mathcal{R})} {\bm {\cal {\un S}}}_{2}^{(2\mathcal{R
})} {\bm {\rm {\un L}}}{}_{{\un {\rI}}}^{(\mathcal{R})} \right)  \\
- i \tfrac{1}{2} \left( {\bm {\rm {\un R}}}{}_{{\un {\rI}}}^{(\mathcal{R})} {\bm  {\cal {\un S}}}_{1}^{(2\mathcal{R})} {\bm {\rm 
{\un L}}}{}_{{\un {\rJ}}}^{(\mathcal{R})} - {\bm {\rm {\un R}}}{}_{{\un {\rJ}}}^{(CS)} {\bm {\cal {\un S}}}_{2}^{(2\mathcal{R
})} {\bm {\rm {\un L}}}{}_{{\un {\rI}}}^{(CS)} \right) & 0
\end{bmatrix}
~~.
\end{split}
\ee
The ``off diagonal'' terms of the final expression (\ref{eqn:blockV2}) is very revealing in comparison to the two expressions for the ``diagonal'' terms on the first line of (\ref{eqn:blockV}).  The terms involving
${\bm {\tilde{{\un V}}}}{}_{{\un {\rI}}{\un {\rJ}}}^{(CS)}$ and ${\bm {\tilde{{\un V}}}}{}_{{\un {\rI}}{\un {\rJ}}}^{(\mathcal{R})}$ perforce all are elements of so(4).  On the other hand, the ``off diagonal'' terms
in (\ref{eqn:blockV2}) must lie in the coset so(8)/so(4), otherwise the quartet $ {\bm {\tilde{V}}}{}_{{\un 
{\rI}}{\un {\rJ}}}^{(2\mathcal{R})} $, ${\bm {\tilde{V}}}{}_{{\un {\rI}}+4,{\un {\rJ}}+4}^{(2\mathcal{R})}$,
${\bm {\tilde{V}}}{}_{{\un {\rI}},{\un {\rJ}}+4}^{(2\mathcal{R})}$, and ${\bm {\tilde{V}}}{}_{{{\un {\rJ}}+4},{\un 
{\rI}}}^{(2\mathcal{R})}$ cannot form a representation of so(8).

By this means, we can reinterpret the condition for when two 1D, $N$ = 4 adinkras can be combined
to form a single 1D, $N$ = 8 adinkra.  In the work of \cite{adnkKyeoh}, it was shown that this condition was
determined by the calculation of the quantity $\chi{}_{\rm o}$ on the two 1D, $N$ = 4 adinkras and
demand that the some of this quantity for the two 1D, $N$ = 4 adinkras must vanish.  The discussion
surrounding  (\ref{eqn:blockV}) and (\ref{eqn:blockV2}) show an equivalent condition that the
chiral 1D, $N$ = 4 adinkra can be combined with another 1D, $N$ = 4 adinkra ($\cal R$) if the
quantity at the end of the equations shown in  (\ref{eqn:blockV2}) is an element of the so(8)/so(4)
coset.

Complete sets of $\tilde{V}$-matrices for the 4D, $\mathcal{N}=2$ vector, tensor, axial-vector and 
axial-tensor supermultiplets are explicitly listed in Appendix \ref{appen:V}.

\newpage
\section{Adinkra 1D, $N$ = 8 Gadget}

In the works \cite{adnkBill1,adnkBill2}, the 1D, $N$ = 4 Gadget was defined.  We need to extend this in the current discussion.  So for our present purposes we define the 1D Gadget value as the following:
\be
\mathcal{G}[(\mathcal{R}),(\mathcal{R}')] = \frac{2}{N(N-1)d_{\min}(N)} \sum_{\rI,\rJ} \text{Tr} 
\Big[ {\bm {\tilde{V}}}_{\rI\rJ}^{(\mathcal{R})} {\bm {\tilde{V}}}_{\rI\rJ}^{(\mathcal{R}')} \Big] ~~~,
\label{adnkGRR}
\ee
where the normalization factor has the term
\be
 d_{\min}(N) = \begin{cases}
  2^{\tfrac{N-1}{2}} , & \quad N \equiv 1, 7 \mod 8  \\
  2^{\tfrac{N}{2}} , & \quad N \equiv 2, 4, 6 \mod 8  \\
  2^{\tfrac{N+1}{2}} , & \quad N \equiv 3, 5 \mod 8  \\
  2^{\tfrac{N-2}{2}} , & \quad N \equiv 8 \mod 8
 \end{cases}   ~~~,
\ee
and $N$ is the number of color, which is 8 in our case. Therefore, our normalization factor is $\tfrac{1}{224}$.

We can then build the matrix
\be
 \mathcal{G}[(\mathcal{R}),(\mathcal{R}')] = \begin{bmatrix}
  1 & \tfrac{1}{7} & \tfrac{1}{7} & \tfrac{1}{7} \\
  \tfrac{1}{7} & 1 & \tfrac{1}{7} & \tfrac{1}{7} \\
  \tfrac{1}{7} & \tfrac{1}{7} & 1 & \tfrac{1}{7} \\
  \tfrac{1}{7} & \tfrac{1}{7} & \tfrac{1}{7} & 1
 \end{bmatrix}   \label{adnkMT}   ~~~,
\ee
where the row and column indices run from ($2VS$), ($2TS$), ($2AVS$), ($2ATS$).  Note that 
it agrees completely with the 4D Gadget, i.e. 
\be
\widehat{\mathcal{G}}[({\widehat {\cal R}}),({\widehat {\cal R}}')] = \mathcal{G}[(\mathcal{R}),
(\mathcal{R}')]  ~~~.
\label{holography}
\ee

Define the angle between different representations as
\be
\cos \{ \theta[(\mathcal{R}),(\mathcal{R}')] \} = \frac{\mathcal{G}[(\mathcal{R}),(
\mathcal{R}')]}{\sqrt{\mathcal{G}[(\mathcal{R}),(\mathcal{R})]} \sqrt{\mathcal{G}
[(\mathcal{R}'),(\mathcal{R}')]}}  ~~~, 
\ee
and $\cos^{-1} \left( \tfrac{1}{7} \right) \approx 81.8^{\circ}$.  Thus in the space of 1D, $N$ = 8 minimal 
adinkras we are lead to propose that the four distinct adinkras that arise from the dimensional
reduction to the corresponding 1D representations $(2VS)$, $(2TS)$, $(2AVS)$, $(2ATS)$ should be 
regarded as four points in a hexadecimal dimensional space that has the topology of a tetrad, 
but with a metric defined by the 1D Gadget.

Let us close this chapter with the observation and emphasis that the derivation of the result 
in (\ref{adnkMT}) is very different from that of (\ref{mtrc2}).  Due to the parameters $m_1$, 
$\dots$, $m_5$ that appear in the definition of (\ref{gdt1}), there is a large parameter space 
that can be used to ``engineer'' the choice of a metric given a set of Holoraumy tensors for 
a set of 4D supermultiplets.  There are no such parameters in the 1D construction.  The metric 
in (\ref{adnkMT}) follows from (\ref{adnkGRR}), which follows entirely from the form of the L-matrices and R-matrices and the definition
of of the 1D Gadget.

\newpage
\section{Conclusion}

\begin{figure}[h]
\centering
\begin{subfigure}{0.35\linewidth}
\centering
 \begin{tabular}{c|cccc}
  $(\widehat{\mathcal{R}})$  & ${\rm p}$ & ${\rm q}$ & ${\rm r}$ & ${\rm s}$ \\ \hline
  $(\widehat{\text{CS}})$ & 0 & 0 & 0 & -1  \\ 
  $(\widehat{\text{VS}})$ & 1 & 1 & 1 & 0  \\
  $(\widehat{\text{TS}})$ & -1 & 1 & -1 & 0   \\ 
  $(\widehat{\text{AVS}})$ & -1 & -1 & 1 & 0  \\
  $(\widehat{\text{ATS}})$ &  1 & -1 & -1 & 0  \\
 \end{tabular}
\end{subfigure}
\begin{subfigure}{0.3\textwidth}
 \centering
 \includegraphics[width=1.0\linewidth]{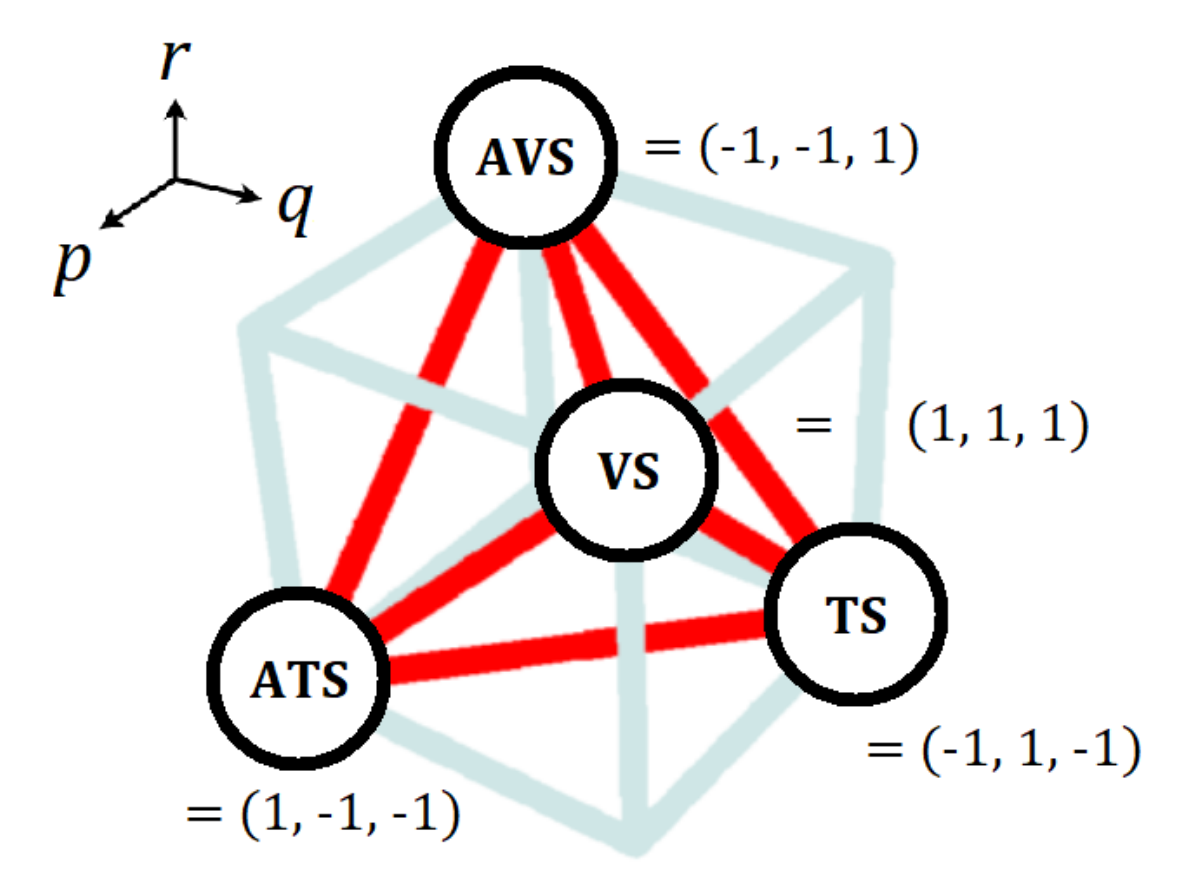}
\end{subfigure}
\caption{Illustrations of tetrahedral geometry in $\cal N$ = 1 minimal supermultiplets.}
\label{fig:N1tetra}
\end{figure}

In previous work \cite{adnk4dGdgt2}, an observation was introduced into the literature about 
4D, $\cal N$ = 1 minimal off-shell supermultiplets.  Namely, when one determines the lattice
parameters $\Pi$ = (${\rm p}$, ${\rm q}$, ${\rm r}$, ${\rm s}$) associated with each supermultiplet via a holoraumy 
calculation one is led to the function
\be  \eqalign{  {~}
\big[ \hat{\bm h}^{\m} (\Pi) \big]_{abc}{}^{d} 
~&=~ i \, {\big [} ~ {\rm p} \, C_{ab} \, (\g^{\m})_{c}{}^{d} ~+~ {\rm q}\,
(\g^{5})_{ab} \, (\g^{5}\g^{\m})_{c}{}^{d} 
~+~ {\rm r} \,  (\g^{5}\g^{\m})_{ab} \, (\g^{5})_{c}{}^{d}
\cr 
&{~~~~~~} ~ ~+~ \tfrac{1}{2} 
{\rm s} (\g^{5}\g_{\n})_{ab}\,  (\g^{5}[\g^{\m}, \g^{\n}])_{c}{}^{d} ~ {\big ]} ~
~~~,
}   \ee
and the values of the lattice parameters are shown in Figure \ref{fig:N1tetra}.

Taking a three dimensional projection of the four dimensional parameter space defined by ``dropping'' the last column, one is led to the tetrahedron shown above.

\begin{figure}[h]
\centering
\begin{subfigure}{0.9\linewidth}
\resizebox{\columnwidth}{!}{
\centering
\begin{tabular}{c|cccc|cccc|cccc|cccc}
$({\Hat {\cal R}})$  & ${\rm p}_{0}$ & ${\rm q}_{0}$ & ${\rm r}_{0}$ & ${\rm s}_{0}$ & ${\rm 
p}_{3}$ & ${\rm q}_{3}$ & ${\rm r}_{3}$ & ${\rm s}_{3}$ & ${\rm p}_{1}$ & ${\rm q}_{1}$ & 
${\rm r}_{1}$ & ${\rm s}_{1}$ & ${\rm p}_{2}$ & ${\rm q}_{2}$ & ${\rm r}_{2}$ & ${\rm s}_{2}$ 
\\ \hline 
(${\widehat {{2VS}}}$) & 1 & 1 & 1 & -1  & 1 & 1 & 1 & 1  & 1 & 1 & 1 & 1  & 1 & 1 & 1 & 1  \\ 
(${\widehat {{2TS}}}$)& -1 & 1 & -1 & -1  & -1 & 1 & -1 & 1  & -1 & 1 & -1 & 1  & -1 & 1 & -1 & 1 \\ 
(${\widehat {{2AVS}}}$) & -1 & -1 & 1 & -1  & -1 & -1 & 1 & 1  & -1 & -1 & 1 & 1  & -1 & -1 & 1 & 1  \\ 
(${\widehat {{2ATS}}}$) & 1 & -1 & -1 & -1  & 1 & -1 & -1 & 1  & 1 & -1 & -1 & 1  & 1 & -1 & -1 & 1 \\
\end{tabular}}
\end{subfigure}
\begin{subfigure}{1.0\textwidth}
 \centering
 \includegraphics[width=1.0\linewidth]{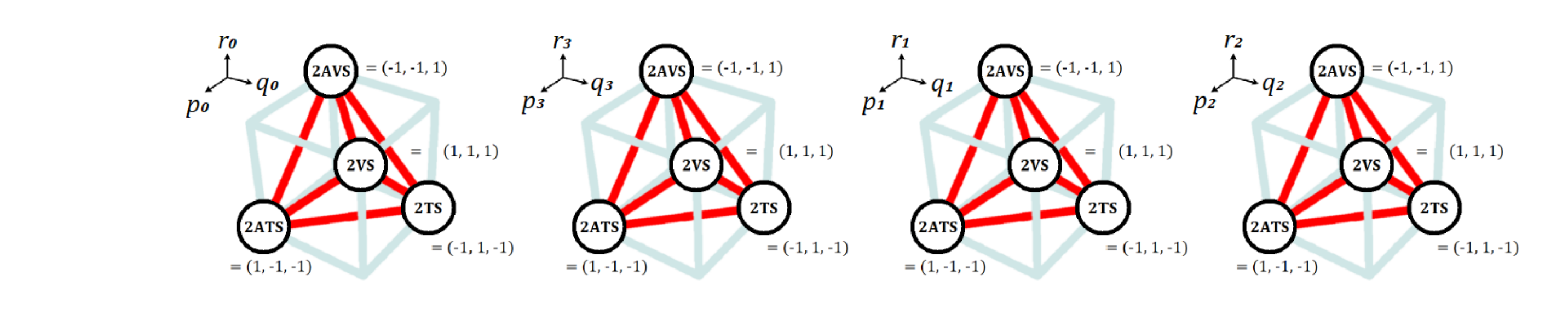}
\end{subfigure}
\caption{Illustrations of tetrahedral geometry in $\cal N$ = 2 minimal supermultiplets.}
\label{fig:N2tetra}
\end{figure}

Having completed the holoraumy calculations in the case of the minimal 4D, $\cal N$ = 2 supermultiplets, we can examine these results from the perspective of the tetrahedron geometry. The holoraumy tensors we have found for each of the minimal supermultiplets look as
\begin{align}
\begin{split}
\big[ \hat{\bm H}^{\m ({\widehat {2VS}})} \big]^{ij}_{ab \, c k}{}^{d l} =& - \delta^{ij}\delta_{k}{}^{l} \big[ \hat{\bm h}^{\m} (1,1,1,-1) \big]_{abc}{}^{d} 
+ (\sigma^{3})^{ij}(\sigma^{3})_{k}{}^{l} \big[ \hat{\bm h}^{\m} (1,1,1,1) \big]_{abc}{}^{d}      \\ 
&  + (\sigma^{1})^{ij}(\sigma^{1})_{k}{}^{l} \big[ \hat{\bm h}^{\m} (1,1,1,1) \big]_{abc}{}^{d}  
+ (\sigma^{2})^{ij}(\sigma^{2})_{k}{}^{l} \big[ \hat{\bm h}^{(2) \m} (1,1,1,1) \big]_{abc}{}^{d}
~~~,
\end{split}  \\ &  \nonumber \\
\begin{split}
\big[ \hat{\bm H}^{\m ({\widehat {2TS}})} \big]^{ij}_{ab \, c k}{}^{d l} =& - \delta^{ij}\delta_{k}{}^{l} \big[ \hat{\bm h}^{\m} (-1,1,-1,-1) \big]_{abc}{}^{d} 
+ (\sigma^{3})^{ij}(\sigma^{3})_{k}{}^{l} \big[ \hat{\bm h}^{\m} (-1,1,-1,1) \big]_{abc}{}^{d} 
\\ 
&  + (\sigma^{1})^{ij}(\sigma^{1})_{k}{}^{l} \big[ \hat{\bm h}^{\m} (-1,1,-1,1) \big]_{abc}{}^{d} 
+ (\sigma^{2})^{ij}(\sigma^{2})_{k}{}^{l} \big[ \hat{\bm h}^{(2) \m} (-1,1,-1,1) \big]_{abc}{}^{d}
~~~,
\end{split}  \\
\begin{split}
\big[ \hat{\bm H}^{\m ({\widehat {2AVS}})} \big]^{ij}_{ab \, c k}{}^{d l} =& - \delta^{ij}\delta_{k}{}^{l} \big[ \hat{\bm h}^{\m} (-1,-1,1,-1) \big]_{abc}{}^{d}  
+ (\sigma^{3})^{ij}(\sigma^{3})_{k}{}^{l} \big[ \hat{\bm h}^{\m} (-1,-1,1,1) \big]_{abc}{}^{d}  
\\ 
& + (\sigma^{1})^{ij}(\sigma^{1})_{k}{}^{l} \big[ \hat{\bm h}^{\m} (-1,-1,1,1) \big]_{abc}{}^{d} 
+ (\sigma^{2})^{ij}(\sigma^{2})_{k}{}^{l} \big[ \hat{\bm h}^{(2) \m} (-1,-1,1,1) \big]_{abc}{}^{d}
\end{split}  ~~~,  \\
\begin{split}
\big[ \hat{\bm H}^{\m ({\widehat {2ATS}})} \big]^{ij}_{ab \, c k}{}^{d l} =& - \delta^{ij}\delta_{k}{}^{l} \big[ \hat{\bm h}^{\m} (1,-1,-1,-1) \big]_{abc}{}^{d} 
+ (\sigma^{3})^{ij}(\sigma^{3})_{k}{}^{l} \big[ \hat{\bm h}^{\m} (1,-1,-1,1) \big]_{abc}{}^{d} 
\\ 
&  + (\sigma^{1})^{ij}(\sigma^{1})_{k}{}^{l} \big[ \hat{\bm h}^{\m} (1,-1,-1,1) \big]_{abc}{}^{d} 
+ (\sigma^{2})^{ij}(\sigma^{2})_{k}{}^{l} \big[ \hat{\bm h}^{(2) \m} (1,-1,-1,1) \big]_{abc}{}^{d}
~~~,
\end{split}
\end{align}
It is immediately apparent that the numbers of lattice parameters for the ${\cal N}=2$ case are four 
times more than for the ${\cal N}=1$ case. The explicit values of the lattice parameters
are shown in Figure \ref{fig:N2tetra}. The reason for this is clear.  In the ${\cal N}=1$ case as there is no ``isospin'' space for the supercovariant derivative, effectively one
only has to consider a factor of $\d {}_{1 \, 1}$.  Whereas in the  $\cal N$ = 2 case there are
factors of $\d {}_{i \, j}$, $ \mathscr{S}{}_{i \, j}^{(S)}$, and $\mathscr{A} {}_{i \, j}^{[A]}$,
which are four independent structures and each has an associated set of lattice parameters.

From the explicit values of the lattice parameters associated with each of the independent
isospin structures, it can be seen that the tetrahedral structure persists in the case of the
minimal $\cal N$ = 2 supermultiplets.  This is once more seen by neglecting respectively
the ${\rm s}_{0}$, ${\rm s}_{3}$, ${\rm s}_{1}$ and ${\rm s}_{2}$ coordinate to obtain three 
dimensional subspaces in each of the corresponding isospin subsectors. In all subsectors, 
the tetrahedrons affiliated with each ${\cal N} = 2$ supermultiplets are identical to that of 
the ${\cal N}=1$ case.

Using the facts that there should be (${\cal N}$+2)($\cal N -$1)/2  independent $\mathscr{
S}{}_{i \, j}^{(S)}$ tensors and $\cal N$($\cal N -$1)/2 independent $\mathscr{A}{}_{i \, j}^{[A
]}$ tensors, in the case of $\cal N$ = 2 there should exist two independent $\mathscr{S}{}_{i 
\, j}^{(S)}$ tensors and a single $\mathscr{A}{}_{i \, j}^{[A]}$ tensor.  We can thus make the 
identifications
\be
\mathscr{ S}{}_{i \, j}^{(S)} ~=~ {\{} \, ({\s}^3){}_{i \, j} , \, ({\s}^1){}_{i \, j} \, {\}} 
~~~~,~~~~ \mathscr{A}{}_{i \, j}^{[A]}  ~=~ \,  {\{} \, i  ({\s}^2){}_{i \, j}  \, {\}} ~~~~.
\ee
This implies there should be a doublet of each of the Lie algebra-valued operators ${\mathscr{
H}}{}^{(4)}$, ${{\mathscr{H}}}{}^{(5)}$, and $\mathscr{H}{}_{\m}{}^{(6)}$ where the components 
of the doublets are associated with  $({\s}^3){}_{i \, j} $, and $({\s}^1){}_{i \, j}$.  In a similar
manner since there is a single $\mathscr{A}{}_{i \, j}^{[A]}$, there must be singlets $\mathscr{H}{
}_{\m}{}^{(7)}$, and $\mathscr{H}{}_{\m \, \n}{}^{(8)}$ operators.   As we will see shortly, 
this will be born out by explicit calculations.

We now turn to conjectures about the $\cal N$ = 4 extension of our results.

Although we have no explicit calculations in this current work for the case of $\cal N$ = 4, having
discussed the results of (\ref{H1}) for the case of $\cal N$ = 2, we can make a conjecture about
the case of $\cal N$ = 4.  For $\cal N$ = 4, we note that $\mathscr{A}{}_{i \, j}^{[A]}$ tensors can 
be used to define  $\mathscr{A}{}_{i \, j}^{[A]\pm}$ via
\be
\mathscr{A}{}_{i \, j}^{[A]\pm} ~\equiv~ \fracm 12 \,{ \big [} \,  \mathscr{A}{}_{i \, j}^{[A]} 
~\pm ~ \fracm 12  \, \e{}_{i \, j \, k \, l} \, \mathscr{A}{}_{k \, l}^{[A]}   \,{ \big ]}
\label{ABs}
\ee
As we began with only six $\mathscr{A}{}_{i \, j}^{[A]\pm}$ operators, this decomposition leads
to three $\mathscr{A}{}_{i \, j}^{[A]\, +}$ and three $\mathscr{A}{}_{i \, j}^{[A]\, -}$ tensors.  In fact,
these triplets are precisely the $\a{}^{\Hat {\rm I}}$-matrices and $\b{}^{\Hat {\rm I}}$-matrices seen in appendix A.  With this recognition we are also able to make the identification
\be \eqalign{
\mathscr{ S}{}_{i \, j}^{(S)} ~&=~  {\{} \, ({\a}{}^{\Hat {\rm I}} \, {\b}{}^{\Hat {\rm K}}){}_{i \, j}
{\}} ~~,~~~ 
\mathscr{A}{}_{i \, j}^{[A]}  ~=~  \, {\{} \, i({\a}{}^{\Hat {\rm I}}){}_{i \, j}, ~ i({\b}{}^{\Hat {\rm I}}){}_{i \, j}   
\, {\}}  ~~~,
}
\ee
where the ${\Hat {\rm I}}$ type of index takes on three values.

In the case of $\cal N$ = 2, it was seen that there are three functions, one 
each associated with $\d{}_{i \, j}$ and $\mathscr{ S}{}_{i \, j}^{(S)}$ such that
\be \eqalign{
\big[ \hat{\bm h}^{(0) \m} (\Pi ) \big]_{abc}{}^{d} ~&=~ 
\big[ \hat{\bm h}^{(3) \m}  ( \Pi ) \big]_{abc}{}^{d} ~=~
\big[ \hat{\bm h}^{(1) \m}  (\Pi ) \big]_{abc}{}^{d} }   ~~~.
\ee
We conjecture that for the case of $\cal N$ = 4, there are ten functions associated
with $\d{}_{i \, j}$ and $({\a}{}^{\Hat {\rm I}} \, {\b}{}^{\Hat {\rm K}}){}_{i \, j}$ that satisfy
the same condition.  An even bolder conjecture is to assert that all six of the functions 
associated with $({\a}{}^{\Hat {\rm I}}){}_{i \, j}$ and $({\b}{}^{\Hat {\rm I}}){}_{i \, j}$
satisfy the same sort of conditions.

Let us be very clear about the result in (\ref{mtrc2}), it is ``engineered" (${\cal X}_1$
= 7${\cal X}_2$ = 7${\cal X}_3$ = 1) in (\ref{mtrcM5}) on the 4D side.  In particular, the 
appearance of the factors of ``1'' and ``3'' in (\ref{mtrc}) is controlled by a rule related to how 
the integers in $\Pi$ are multiplied by $\g$-matrices in the functions $\big[ \hat{\bm h}^{\m}  (\Pi) \big]_{abc}{}^{d}$ and $\big[ \hat{\bm h}^{(2) \m}  (\Pi ) \big]_{abc}{}^{d}$.  
If one of the integers p, q, r, or s is multiplied by a factor of [$\g{}_{\m}$ , $\g{}_{\n}$] in either of the functions $\big[ \hat{\bm h}^{\m}  (\Pi ) \big]_{abc}{}^{d}$ or $\big[ \hat{\bm h}^{(2) \m}  (\Pi ) \big]_{abc}{}^{d}$, then when the covariant Gadget is written, a corresponding factor of ``3'' should be 
engineered to appear multiplying that integer.  Otherwise, only a factor 
of ``1'' should be present.  On the adinkra side of (\ref{adnkGRR}), no engineering is necessary as there is no freedom permitted by our definitions.  The validity of  (\ref{holography}) is an example of what we have long called ``SUSY holography'' \cite{ENUF}.

Though we make this as a ``phenomenological observation'' based on the study of minimal 4D,
$\cal N$ = 1 and $\cal N$ = 2 supermultiplets, we have no deep understanding of why this
must be enforced to lead from a direct relation of the covariant 4D Gadget to the corresponding
1D Gadget.

The reader who has been following our progress will not be surprised that on the
basis of the results presented here we have a surmise to make about the lattice parameters.
If we return to the formula for the dimension of the irreducible representations of su(3)
given in (\ref{dimnum}), there it is clear the parameters $p$ and $q$ are of fundamental
significance in order to be able to express the result.  It is our assertion that the lattice
parameters ${\rm p}{}^{({{\cal R}})}$, ${\rm q}{}^{({{\cal R}})}$, ${\rm r}{}^{({{\cal R}})}$, 
and ${\rm s}{}^{({{\cal R}})}$ in their entirety provide the analogs of these su(3) parameters 
in order to write for spacetime SUSY representations the analog of Freudenthal-type formulae. 
The implication of this surmise is that a covariant spacetime Gadget $\widehat{\mathcal{G}}[({\Hat {\cal R}}),({\Hat {\cal R}})]$, for any supersymmetrical representation $({\Hat {\cal R}})$, provides
a Casimir-like value for the representation.  As our work has shown \cite{Grf5}, there are
classes of covariant spacetime Gadgets.  However, the one that satisfies the condition
\be
\widehat{\mathcal{G}}[({\Hat {\cal R}}),({\Hat {\cal R}}{}^{\prime})]  ~=~ 
{\mathcal{G}}[({ {\cal R}}),({ {\cal R}}{}^{\prime})]  ~~~,
\ee
is distinguished because the adinkra Gadget on the RHS appears parameter-free
up to an overall normalization constant.

A future ambitious goal of our work is to explore whether a concept similar to the
Gadget can be combined with the Cubic Casimir noted in (\ref{Frd4c}) to find an
expression for the spacetime SUSY anomaly written in terms of the lattice parameters.

\vspace{.05in}
\begin{center}
\parbox{4in}{{\it ``Up comes stream upon stream, hill upon hill, 
\\ When it appears there is no way ahead; 
\\ Beyond shady willows and bright flowers still ~---~ 
\\ Lies another quiet village instead.'' 
 \\ ${~}$ 
 \\ ${~}$ 
 \\ ${~}$ } \,\,-\,\, Lu You}
 \parbox{4in}{
 $~~$}  
\end{center}
{\bf Acknowledgements}\\[.1in] \indent
The research of S. J. Gates is supported by the endowment of the Ford Foundation 
Professorship of Physics at Brown University.  S.-N. Mak would like to acknowledge 
her participation in the annual Brown University Adinkra Math/Phys Hangouts" 
in 2017. This work was partially supported by the National Science 
Foundation grant PHY-1620074. We also gratefully acknowledge discussions
with X. Xiao, a participant in the 2018 Summer Student Theoretical Physics
Research Session, who was instrumental in uncovering the result shown in
(\ref{su3GdGt}).

\newpage
\appendix

\newpage
\section{Matrix Representation Results}

This appendix consists of two parts.  The first provides a presentation of conventions
and result for 3 $\times$ 3 matrices of our discussions and latter part does so for
4 $\times$ 4 matrices.

For the 3 $\times$ 3 matrix generators of su(3) we use
\begin{equation}
\begin{aligned}
 {\bm \l}_{1} =& 
 \begin{bmatrix}
   0 & 1 & 0 \\
   1 & 0 & 0 \\
   0 & 0 & 0
 \end{bmatrix} ~~, &
 {\bm \l}_{2} =& 
 \begin{bmatrix}
   0 & -i & 0 \\
   i & 0 & 0 \\
   0 & 0 & 0
 \end{bmatrix} ~~, & 
 {\bm \l}_{3} =& 
 \begin{bmatrix}
   1 & 0 & 0 \\
   0 & -1 & 0 \\
   0 & 0 & 0
 \end{bmatrix} ~~, \\
 {\bm \l}_{4} =& 
 \begin{bmatrix}
   0 & 0 & 1 \\
   0 & 0 & 0 \\
   1 & 0 & 0
 \end{bmatrix} ~~, &
 {\bm \l}_{5} =& 
 \begin{bmatrix}
   0 & 0 & -i \\
   0 & 0 & 0 \\
   i & 0 & 0
 \end{bmatrix} ~~, \\
 {\bm \l}_{6} =& 
 \begin{bmatrix}
   0 & 0 & 0 \\
   0 & 0 & 1 \\
   0 & 1 & 0
 \end{bmatrix} ~~, & 
 {\bm \l}_{7} =& 
 \begin{bmatrix}
   0 & 0 & 0 \\
   0 & 0 & -i \\
   0 & i & 0
 \end{bmatrix} ~~, &
 {\bm \l}_{8} =& 
 \frac{1}{\sqrt{3}}\begin{bmatrix}
   1 & 0 & 0 \\
   0 & 1 & 0 \\
   0 & 0 & -2
 \end{bmatrix} ~~.
\end{aligned}
\label{su3}
\end{equation}
As well with regard to explicit form of the eigenvectors we write them as $| \tfrac 12 , \, \tfrac 1{2{\sqrt 3}} \rangle$, $| - \tfrac 12 , \, \tfrac 1{2{\sqrt 3}} \rangle$, and $| 0 , \, - \tfrac 1{\sqrt 3} \rangle$ where
\be
\left|  \fracm 12 , \, \fracm 1{2 {\sqrt 3}} \right\rangle ~=~ 
{\begin{bmatrix}
1\\0\\0
\end{bmatrix}}
 ~~~,~~~
\left| - \fracm 12 , \, \fracm 1{2 {\sqrt 3}} \right\rangle ~=~  
{\begin{bmatrix}
0\\1\\0
\end{bmatrix}}
~~~,~~~ \left| 0 , \, - \fracm {1}{\sqrt 3} \right\rangle ~=~
{\begin{bmatrix}
0\\0\\1
\end{bmatrix}}
~~~.
\label{Es}
\ee
Our final discussion of explicit matrices associated with su(3) in this appendix involves
the formula presented in (\ref{quGdGt}).

It is well-known in the context of the Standard Model, the Cabibbo-Kobayashi-Maskawa
matrix  \cite{CKM1,CKM2}
plays an important role with respect to CP-violation.  The three generations of 
lowest isospin component quarks can be assembled into two triplets 
according to
\be
{\begin{bmatrix}
d{}^{\prime}\\s{}^{\prime}\\b{}^{\prime}
\end{bmatrix}}  ~ =~ 
{\begin{bmatrix}
V{}_{ud} & V{}_{us} & V{}_{ub} \\
V{}_{cd} & V{}_{cs} & V{}_{cb} \\
V{}_{td} & V{}_{ts} & V{}_{tb} \\
\end{bmatrix}}   \, 
{\begin{bmatrix}
d\\s\\b
\end{bmatrix}}  ~=~{\bm  {\cal V}} \, {\begin{bmatrix}
d\\s\\b
\end{bmatrix}}
~~~,
\label{CKM}
\ee
via use of the Cabibbo-Kobayashi-Maskawa matrix.  As the triplets $d{}^{\prime}-s{
}^{\prime}-b{}^{\prime}$ and $d-s-b$ appear, the generators of an su(3) can act
upon these.  However, this particular su(3) is related to the distinct families of 
quark, not the usual flavor symmetry.  Thus, it is possible to write currents related
to this ``family-su(3)'' symmetry and the su(3) matrices can be used to define
such currents.  Since the former triplet is related to the latter triplet via the $\bm {\cal 
V}$ matrix, it follows that
\be
{\bm \l}{}_{i}^{\prime} ~=~ {\bm {\cal V}} \, {\bm \l}{}_{i} \, {\bm {\cal V}}{}^{-1}  ~~~,
\ee
So (\ref{su3GdGt}) takes the form
\be   \eqalign{
{C}{}_{4,\, \cal G} ({\cal R}, \, {\cal R}^{\prime}) ~&=~ \fracm 1{16}   \sum_{i = 1, j= 1}^8 \,  
{\rm {Tr}} \left( \, \left\{ {\bm  \l}{}^i ~,~ {\bm  \l}{}^j  \,  \right\} \, \left\{ {\bm  \l}{}_i{}^{{\prime}} 
~,~ {\bm  \l}{}_j{}^{{\prime}}  \,  \right\} \,  \right)      \cr
~&=~  \fracm{1}{288} \big\{ ~33 ~+~ 15\,  {\big |}  \,{\rm {Tr}} \left( \,  {\bm {\cal V}}   \, \right)
{\big |}^2  \, \big\}    \cr
~&=~  \fracm{1}{288} \big\{ ~33 ~+~   15\,  {\big |} c_1 \,+\, ( c_1 - e^{i \d})  ( \, c_2
c_3 + s_2 s_3 )
{\big |}^2 \, \big\}   \cr
~&=~  \fracm{1}{288} \big\{ ~33 ~+~  15\,  {\big |} \,
( \,c_{12} ~+~ c_{23} \, ) \, c_{13} ~+~
c_{12} \, c_{23} ~-~ s_{12} \, s_{23} \, s_{13} \, e^{i \d_{13}}
\, {\big |}^2 \, \big\}  
~~~.
} \ee
On the penultimate line of this equation, the value of ${C}{}_{4,\, \cal G} ({\cal R}, \, {\cal 
R}^{\prime})$ is expressed in terms of the ``KM'' parameters of the C-K-M matrix
while the final line it is expressed in terms of the ``standard'' parameters of the C-K-M matrix.

Now we move on the relevant matrix structure for 4 $\times $ 4 matrix related to the isospin
indices in our discussions.

The ${\bm \alpha}$ and ${\bm \b}$ matrices used in this paper are:
\begin{equation}
\begin{aligned}
{\bm \a} {}^{1}=&{\begin{bmatrix}
 0 & 0 & 0 & -i \\
 0 & 0 & -i & 0 \\
 0 & i & 0 & 0 \\
 i & 0 & 0 & 0
\end{bmatrix}} 
~~, &
{\bm \a}{}^{2}=&{\begin{bmatrix}
 0 & -i & 0 & 0 \\
 i & 0 & 0 & 0 \\
 0 & 0 & 0 & -i \\
 0 & 0 & i & 0
\end{bmatrix}} 
~~, &
{\bm \a}{}^{3}=&{\begin{bmatrix}
 0 & 0 & -i & 0 \\
 0 & 0 & 0 & i \\
 i & 0 & 0 & 0 \\
 0 & -i & 0 & 0
\end{bmatrix}} 
~~, \\
{\bm \b}{}^{1}=&{\begin{bmatrix}
 0 & 0 & 0 & -i \\
 0 & 0 & i & 0 \\
 0 & -i & 0 & 0 \\
 i & 0 & 0 & 0
\end{bmatrix}} 
~~, &
{\bm \b}{}^{2}=&{\begin{bmatrix}
 0 & 0 & -i & 0 \\
 0 & 0 & 0 & -i \\
 i & 0 & 0 & 0 \\
 0 & i & 0 & 0
\end{bmatrix}} 
~~, &
{\bm \b}{}^{3}=&{\begin{bmatrix}
 0 & -i & 0 & 0 \\
 i & 0 & 0 & 0 \\
 0 & 0 & 0 & i \\
 0 & 0 & -i & 0
\end{bmatrix}} 
~~.
\end{aligned}
\end{equation}

In terms of tensor products of Pauli spin matrices ${\bm \s}^i$ and the $2\times2$ identity 
matrix ${\bm {\rm I}}{}_{2\times 2}$, this can be written as
\begin{align}
{\bm \alpha}^1 ~=&~ {\bm \s}^2 \otimes {\bm \s}^1~~~,~~~{\bm \alpha}^2 ~=~ \bm{{\rm I}} 
{}_{2 \times 2}
\otimes {\bm \s}^2~~~,~~~{\bm \alpha}^3 ~=~ {\bm \s}^2 \otimes {\bm \s}^3 ~\,~,  \\
{\bm \b}^1 ~=&~ {\bm \s}^1 \otimes {\bm \s}^2~~~,~~~{\bm \b}^2 ~=~ {\bm \s}^2 \otimes 
\bm{{\rm I}} {}_{2 \times 2}	~~~,~~~{\bm \b}^3 ~=~ {\bm \s}^3 \otimes {\bm \s}^2 ~~~.
\end{align}
These matrices form two mutually commuting su(2) algebras
\begin{align}
[ {\bm \alpha}^{\hat{a}}, {\bm \alpha}^{\hat{b}} ] ~=~ & i\, 2  \epsilon^{\hat{a}\hat{b}\hat{c
}} {\bm \alpha}^{\hat{c}}~~~,~~~[ {\bm \b}^{\hat{a}}, {\bm \b}^{\hat{b}} ] ~=~ i\,2  \epsilon^{
\hat{a}\hat{b}\hat{c}} {\bm \b}^{\hat{c}}~~~,~~~[ {\bm \alpha}^{\hat{a}}, {\bm \b}^{\hat{b}} ] 
=  0~~~.
\end{align}
Owing to the definitions above, the ${\bm \alpha}$ and ${\bm \b}$ matrices satisfy the trace 
orthogonality relationships 
\begin{align}\label{e:abtraces}
Tr( {\bm \alpha}^{\hat{a}} {\bm \b}^{\hat{b}} ) ~=&~ 0 ~~~,~~~
Tr({\bm \alpha}^{\hat{a}} {\bm \alpha}^{\hat{b}}) ~=~  Tr( {\bm 
\b}^{\hat{a}} {\bm \b}^{\hat{b}}) = 4\delta^{\hat{a}\hat{b}}~~~.
\end{align}

\newpage
\section {4D, $\mathcal{N}$~=~2 Supermultiplets}

In this paper, we consider 4D, $\mathcal{N}$=2 supermultiplets with 8 bosons and 8 fermions
as these are the minimal representation presented among the results shown in Table \ref{tab:N2susy}.  In
the work of \cite{adnkKyeoh} these results were presented.  The transformation laws and anti-commutator 
algebra are reproduced here for the convenience of the reader.

\subsection{4D, $\mathcal{N}$=2 Vector Supermultiplet}

We construct the 4D, $\mathcal{N}=2$ vector supermultiplet from the 4D, $\mathcal{N}=1$ chiral 
and vector supermultiplets. Let
\be
 \Psi_{c k} = \begin{bmatrix}
 \psi_{c} \\
 \lambda_{c}
 \end{bmatrix}   ~~~,
\ee
where $k$ is the isospin index. The Lagrangian is 
\be
\begin{split}
\mathcal{L} =& - \tfrac{1}{2} \pa_{\mu} A \pa^{\mu} A - \tfrac{1}{2} \pa_{\mu} B 
\pa^{\mu} B + \tfrac{1}{2} F^{2} + \tfrac{1}{2} G^{2} - \tfrac{1}{4} F_{\mu\nu} F^{\mu\nu} 
+ \tfrac{1}{2} d^{2}     \\
&  + i \tfrac{1}{2} \delta^{ij} (\gamma^{\mu})^{bc} \Psi_{b i} \pa_{\mu} \Psi_{c j} ~~~,
\end{split}
\ee
where
\be
 F_{\mu\nu} = \pa_{\mu} A_{\nu} - \pa_{\nu} A_{\mu}  ~~~,
\ee
 is the usual field strength.  The corresponding transformation laws are
\be
\begin{split}
 \text{D}^{i}_{a} A &= \delta^{ij} \Psi_{a j}  ~~~, \\
 \text{D}^{i}_{a} B &= i \delta^{ij} (\gamma^{5})_{a}{}^{b} \Psi_{b j}  ~~~, \\
 \text{D}^{i}_{a} F &= \delta^{ij} (\gamma^{\mu})_{a}{}^{b} \pa_{\mu} \Psi_{b j}  ~~~, \\
 \text{D}^{i}_{a} G &= i (\sigma^{3})^{ij} (\gamma^{5}\gamma^{\mu})_{a}{}^{b} \pa_{\mu} \Psi_{b j}  ~~~, \\
 \text{D}^{i}_{a} A_{\mu} &= i (\sigma^{2})^{ij} (\gamma_{\mu})_{a}{}^{b} \Psi_{b j}  ~~~, \\
 \text{D}^{i}_{a} d &= i (\sigma^{1})^{ij} (\gamma^{5}\gamma^{\mu})_{a}{}^{b} \pa_{\mu} \Psi_{b j}  ~~~, \\
 \text{D}^{i}_{a} \Psi^{j}_{b} &= \delta^{ij} \big[ i (\gamma^{\mu})_{ab} \pa_{\mu} A - 
 (\gamma^{5}\gamma^{\mu})_{ab} \pa_{\mu} B - i C_{ab} F \big] + (\sigma^{3})^{ij} 
 (\gamma^{5})_{ab} G  \\
 & + (\sigma^{1})^{ij} (\gamma^{5})_{ab} d + \tfrac{1}{4} (\sigma^{2})^{ij} ([\gamma^{\mu},\
 \gamma^{\nu}])_{ab} (\pa_{\mu} A_{\nu} - \pa_{\nu} A_{\mu})  ~~~.
\end{split}
\ee
The transformation laws satisfy the algebra
\begin{align}
 \{\text{D}^{i}_{a},\text{D}^{j}_{b}\} \chi &= i 2 \delta^{ij} (\gamma^{\mu})_{ab} \pa_{\mu} \chi  
 ~~~, \\ 
\{\text{D}^{i}_{a},\text{D}^{j}_{b}\} A_{\mu} &= i 2 \delta^{ij} (\gamma^{\nu})_{ab} F_{\nu\mu} 
+ i (\sigma^{2})^{ij} \left[ i 2 C_{ab} \pa_{\mu} A - 2(\gamma^{5})_{ab} \pa_{\mu} B  \right] ~~~,
\end{align}
where
\be
 \chi \in \{A,B,F,G,d,\Psi_{c k}\}   ~~~.
\ee


\subsection{4D, $\mathcal{N}$=2 Tensor Supermultiplet}

We construct the 4D, $\mathcal{N}=2$ tensor supermultiplet from the 4D, $\mathcal{N}=1$ 
chiral and tensor supermultiplets. Let
\be
 \Psi_{c k} = \begin{bmatrix}
 \psi_{c} \\
 \chi_{c}
 \end{bmatrix}  ~~~.
\ee
The Lagrangian is 
\be
\begin{split}
\mathcal{L} =& - \tfrac{1}{2} \pa_{\mu} A \pa^{\mu} A - \tfrac{1}{2} \pa_{\mu} B \pa^{\mu} B 
+ \tfrac{1}{2} F^{2} + \tfrac{1}{2} G^{2} - \tfrac{1}{3} H_{\mu\nu\alpha} H^{\mu\nu\alpha} - 
\tfrac{1}{2} \pa_{\mu} \varphi \pa^{\mu} \varphi    \\
 &  + i \tfrac{1}{2} \delta^{ij} (\gamma^{\mu})^{bc} \Psi_{b i} \pa_{\mu} \Psi_{c j}   ~~~.
\end{split}
\ee
where
\be
 H_{\mu\nu\alpha} = \pa_{\mu} B_{\nu\alpha} + \pa_{\nu} B_{\alpha\mu} + \pa_{\alpha} B_{
 \mu\nu} ~~~.
\ee
The corresponding transformation laws are
\be
\begin{split}
 \text{D}^{i}_{a} A &= (\sigma^{3})^{ij} \Psi_{a j}  ~~~,  \\
 \text{D}^{i}_{a} B &= i \delta^{ij} (\gamma^{5})_{a}{}^{b} \Psi_{b j}  ~~~,  \\
 \text{D}^{i}_{a} F &= \delta^{ij} (\gamma^{\mu})_{a}{}^{b} \pa_{\mu} \Psi_{b j}  ~~~,  \\
 \text{D}^{i}_{a} G &= i \delta^{ij} (\gamma^{5}\gamma^{\mu})_{a}{}^{b} \pa_{\mu} \Psi_{b j}  ~~~,  \\
 \text{D}^{i}_{a} \varphi &= (\sigma^{1})^{ij} \Psi_{a j}  ~~~,  \\
 \text{D}^{i}_{a} B_{\mu\nu} &= - i \tfrac{1}{4} (\sigma^{2})^{ij} ([\gamma_{\mu},\gamma_{\nu }])_{a}{}^{b} \Psi_{b j}  ~~~,  \\
 \text{D}^{i}_{a} \Psi^{j}_{b} &= \delta^{ij} \big[ - (\gamma^{5}\gamma^{\mu})_{ab} \pa_{\mu} 
 B - i C_{ab} F + (\gamma^{5})_{ab} G  \big] + i (\sigma^{3})^{ij} (\gamma^{\mu})_{ab} \pa_{ \mu} A  \\
 & + i (\sigma^{1})^{ij} (\gamma^{\mu})_{ab} \pa_{\mu} \varphi - i (\sigma^{2})^{ij} \epsilon_{
 \mu}{}^{\nu\alpha\beta} (\gamma^{5}\gamma^{\mu})_{ab} \pa_{\nu} B_{\alpha\beta}  ~~~.
\end{split}
\ee
The transformation laws satisfy the algebra
\begin{align}
 \{\text{D}^{i}_{a},\text{D}^{j}_{b}\} \chi =& i 2 \delta^{ij} (\gamma^{\mu})_{ab} \pa_{\mu} \chi  ~~~,  \\
\begin{split}
\{\text{D}^{i}_{a},\text{D}^{j}_{b}\} B_{\mu\nu} =& i 2 \delta^{ij} (\gamma^{\alpha})_{ab} H_{\alpha
\mu\nu}   \\ 
&  + i (\gamma_{[\mu})_{ac}\pa_{\nu]} \left[ (\sigma^{1})^{ij} \delta_{b}{}^{c} A + (\sigma^{2})^{ij} 
(\gamma^{5})_{b}{}^{c} B - (\sigma^{3})^{ij} \delta_{b}{}^{c} \varphi  \right]   ~~~,
\end{split}
\end{align}
where
\be
 \chi \in \{A,B,F,G,\varphi,\Psi_{c k}\}  ~~~.
\ee

\subsection{4D, $\mathcal{N}$=2 Axial-Vector Supermultiplet}

We construct the 4D, $\mathcal{N}=2$ axial-vector supermultiplet from the 4D, $\mathcal{N}=1$ 
chiral and axial-vector supermultiplets. Let
\be
\Psi_{c k} = \begin{bmatrix}
\psi_{c} \\
\tilde{\lambda}_{c}
\end{bmatrix}  ~~~.
\ee
The Lagrangian is 
\be
\begin{split}
\mathcal{L} =& - \tfrac{1}{2} \pa_{\mu} A \pa^{\mu} A - \tfrac{1}{2} \pa_{\mu} B \pa^{\mu} B + 
\tfrac{1}{2} F^{2} + \tfrac{1}{2} G^{2} - \tfrac{1}{4} F_{\mu\nu} F^{\mu\nu} + \tfrac{1}{2} \tilde{d}^{2} 
\\
&  + i \tfrac{1}{2} \delta^{ij} (\gamma^{\mu})^{bc} \Psi_{b i} \pa_{\mu} \Psi_{c j}  ~~~,
\end{split}
\ee
where
\be
 F_{\mu\nu} = \pa_{\mu} U_{\nu} - \pa_{\nu} U_{\mu}  ~~~.
\ee
The corresponding transformation laws are
\be
\begin{split}
 \text{D}^{i}_{a} A &= \delta^{ij} \Psi_{a j}  ~~~,  \\
 \text{D}^{i}_{a} B &= i \delta^{ij} (\gamma^{5})_{a}{}^{b} \Psi_{b j}  ~~~,  \\
 \text{D}^{i}_{a} F &= (\sigma^{3})^{ij} (\gamma^{\mu})_{a}{}^{b} \pa_{\mu} \Psi_{b j}  ~~~,  \\
 \text{D}^{i}_{a} G &= i \delta^{ij} (\gamma^{5}\gamma^{\mu})_{a}{}^{b} \pa_{\mu} \Psi_{b j}  ~~~,  \\
 \text{D}^{i}_{a} U_{\mu} &= - (\sigma^{2})^{ij} (\gamma^{5}\gamma_{\mu})_{a}{}^{b} \Psi_{b j}  ~~~,  \\
 \text{D}^{i}_{a} \tilde{d} &= - (\sigma^{1})^{ij} (\gamma^{\mu})_{a}{}^{b} \pa_{\mu} \Psi_{b j}  ~~~,  \\
 \text{D}^{i}_{a} \Psi^{j}_{b} &= \delta^{ij} \big[ i (\gamma^{\mu})_{ab} \pa_{\mu} A - (\gamma^{5}
 \gamma^{\mu})_{ab} \pa_{\mu} B + (\gamma^{5})_{ab} G \big] - i (\sigma^{3})^{ij} C_{ab} F  \\
 & + i (\sigma^{1})^{ij} C_{ab} \tilde{d} + i \tfrac{1}{4} (\sigma^{2})^{ij} (\gamma^{5}[\gamma^{\mu},
 \gamma^{\nu}])_{ab} (\pa_{\mu} U_{\nu} - \pa_{\nu} U_{\mu})   ~~~.
\end{split}
\ee
The transformation laws satisfy the algebra
\begin{align}
 \{\text{D}^{i}_{a},\text{D}^{j}_{b}\} \chi &= i 2 \delta^{ij} (\gamma^{\mu})_{ab} \pa_{\mu} \chi  ~~~,  \\ 
 \{\text{D}^{i}_{a},\text{D}^{j}_{b}\} U_{\mu} &= i 2 \delta^{ij} (\gamma^{\nu})_{ab} F_{\nu\mu} - i (
 \sigma^{2})^{ij} \left[ 2(\gamma^{5})_{ab} \pa_{\mu} A + i 2 C_{ab} \pa_{\mu} B \right]  ~~~,
\end{align}
where
\be
 \chi \in \{A,B,F,G,\tilde{d},\Psi_{c k}\}  ~~~.
\ee

\subsection{4D, $\mathcal{N}$=2 Axial-Tensor Supermultiplet}

Finally, we construct the 4D, $\mathcal{N}=2$ axial-tensor supermultiplet from the 4D, $\mathcal{N}
=1$ chiral and axial-tensor supermultiplets. Let
\be
 \Psi_{c k} = \begin{bmatrix}
 \psi_{c} \\
 \tilde{\chi}_{c}
 \end{bmatrix}    ~~~.
\ee
The Lagrangian is 
\be
\begin{split}
\mathcal{L} =& - \tfrac{1}{2} \pa_{\mu} A \pa^{\mu} A - \tfrac{1}{2} \pa_{\mu} B \pa^{\mu} B + \tfrac{1}{2} 
F^{2} + \tfrac{1}{2} G^{2} - \tfrac{1}{3} H_{\mu\nu\alpha} H^{\mu\nu\alpha} - \tfrac{1}{2} \pa_{\mu} \tilde{
\varphi} \pa^{\mu} \tilde{\varphi}   \\
 &  + i \tfrac{1}{2} \delta^{ij} (\gamma^{\mu})^{bc} \Psi_{b i} \pa_{\mu} \Psi_{c j} ~~~,
\end{split}
\ee
where
\be
 H_{\mu\nu\alpha} = \pa_{\mu} C_{\nu\alpha} + \pa_{\nu} C_{\alpha\mu} + \pa_{\alpha} C_{\mu\nu}  ~~~.
\ee
The corresponding transformation laws are
\be
\begin{split}
 \text{D}^{i}_{a} A &= \delta^{ij} \Psi_{a j}  ~~~,  \\
 \text{D}^{i}_{a} B &= i (\sigma^{3})^{ij} (\gamma^{5})_{a}{}^{b} \Psi_{b j}  ~~~,  \\
 \text{D}^{i}_{a} F &= \delta^{ij} (\gamma^{\mu})_{a}{}^{b} \pa_{\mu} \Psi_{b j}  ~~~,  \\
 \text{D}^{i}_{a} G &= i \delta^{ij} (\gamma^{5}\gamma^{\mu})_{a}{}^{b} \pa_{\mu} \Psi_{b j}  ~~~,  \\
 \text{D}^{i}_{a} \tilde{\varphi} &= i (\sigma^{1})^{ij} (\gamma^{5})_{a}{}^{b} \Psi_{b j}  ~~~,  \\
 \text{D}^{i}_{a} C_{\mu\nu} &= \tfrac{1}{4} (\sigma^{2})^{ij} (\gamma^{5}[\gamma_{\mu},\gamma_{\nu
 }])_{a}{}^{b} \Psi_{b j}  ~~~,  \\
 \text{D}^{i}_{a} \Psi^{j}_{b} &= \delta^{ij} \big[ i (\gamma^{\mu})_{ab} \pa_{\mu} A - i C_{ab} F + (
 \gamma^{5})_{ab} G  \big] - (\sigma^{3})^{ij} (\gamma^{5}\gamma^{\mu})_{ab} \pa_{\mu} B   \\
 & - (\sigma^{1})^{ij} (\gamma^{5}\gamma^{\mu})_{ab} \pa_{\mu} \tilde{\varphi} + (\sigma^{2})^{ij} 
 \epsilon_{\mu}{}^{\nu\alpha\beta} (\gamma^{\mu})_{ab} \pa_{\nu} C_{\alpha\beta}  ~~~.
\end{split}
\ee
The transformation laws satisfy the algebra
\begin{align}
\{\text{D}^{i}_{a},\text{D}^{j}_{b}\} \chi =& i 2 \delta^{ij} (\gamma^{\mu})_{ab} \pa_{\mu} \chi ~~~,  \\
\begin{split}
\{\text{D}^{i}_{a},\text{D}^{j}_{b}\} C_{\mu\nu} =& i 2 \delta^{ij} (\gamma^{\alpha})_{ab} H_{\alpha
\mu\nu}  \\ 
&  + i (\gamma_{[\mu})_{ac} \pa_{\nu]} \left[ (\sigma^{1})^{ij} \delta_{b}{}^{c} B - (\sigma^{2})^{ij} 
(\gamma^{5})_{b}{}^{c} A - (\sigma^{3})^{ij} \delta_{b}{}^{c} \tilde{\varphi}  \right] ~~~,
\end{split} 
\end{align}
where
\be
 \chi \in \{A,B,F,G,\tilde{\varphi},\Psi_{c k}\}   ~~~.
\ee

\newpage \noindent
\newpage
\section{Fermionic Holoraumy Matrices of 1D, $N$ = 8 Supermultiplets}
\label{appen:V}

In this final appendix, explicit expressions for the fermionic holoraumy matrices are
given in term of 2 $\times$ 2 matrices whose elements are themselves signed 4 $\times$ 
4 permutations as this allows for a fairly compact notation.

\subsection{Fermionic Holoraumy Matrices of 1D, $N$ = 8 Vector Supermultiplet}

\begin{equation}
\begin{aligned}
{\bm {\tilde{V}}}{}^{(2VS)}_{12} =& i \begin{bmatrix}
  \langle \bar{2}1\bar{4}3 \rangle & 0 \\ 
  0 & \langle 2\bar{1}\bar{4}3 \rangle 
 \end{bmatrix} &  ,~~ 
{\bm {\tilde{V}}}{}^{(2VS)}_{13} =& i \begin{bmatrix}
  \langle \bar{3}41\bar{2} \rangle & 0 \\ 
  0 & \langle \bar{3}\bar{4}12 \rangle 
 \end{bmatrix} & ,~~
{\bm {\tilde{V}}}{}^{(2VS)}_{14} =& i \begin{bmatrix}
  \langle \bar{4}\bar{3}21 \rangle & 0 \\ 
  0 & \langle 4\bar{3}2\bar{1} \rangle 
 \end{bmatrix} ,~~ \\ 
{\bm {\tilde{V}}}{}^{(2VS)}_{15} =& i \begin{bmatrix}
  0 & \langle \bar{1}\bar{2}3\bar{4} \rangle \\ 
  \langle 12\bar{3}4 \rangle & 0 
 \end{bmatrix} & ,~~ 
{\bm {\tilde{V}}}{}^{(2VS)}_{16} =& i \begin{bmatrix}
  0 & \langle \bar{2}143 \rangle \\ 
  \langle \bar{2}1\bar{4}\bar{3} \rangle & 0 
 \end{bmatrix} & ,~~
{\bm {\tilde{V}}}{}^{(2VS)}_{17} =& i \begin{bmatrix}
  0 & \langle \bar{3}4\bar{1}\bar{2} \rangle \\ 
  \langle 341\bar{2} \rangle & 0 
 \end{bmatrix},~~  \\ 
{\bm {\tilde{V}}}{}^{(2VS)}_{18} =& i \begin{bmatrix}
  0 & \langle \bar{4}\bar{3}\bar{2}1 \rangle \\ 
  \langle \bar{4}321 \rangle & 0 
 \end{bmatrix} & ,~~
{\bm {\tilde{V}}}{}^{(2VS)}_{23} =& i \begin{bmatrix}
  \langle \bar{4}\bar{3}21 \rangle & 0 \\ 
  0 & \langle \bar{4}3\bar{2}1 \rangle 
 \end{bmatrix} & ,~~
{\bm {\tilde{V}}}{}^{(2VS)}_{24} =& i \begin{bmatrix}
  \langle 3\bar{4}\bar{1}2 \rangle & 0 \\ 
  0 & \langle \bar{3}\bar{4}12 \rangle 
 \end{bmatrix} ,~~ \\ 
{\bm {\tilde{V}}}{}^{(2VS)}_{25} =& i \begin{bmatrix}
  0 & \langle 2\bar{1}43 \rangle \\ 
  \langle 2\bar{1}\bar{4}\bar{3} \rangle & 0 
 \end{bmatrix} & ,~~
{\bm {\tilde{V}}}{}^{(2VS)}_{26} =& i \begin{bmatrix}
  0 & \langle \bar{1}\bar{2}\bar{3}4 \rangle \\ 
  \langle 123\bar{4} \rangle & 0 
 \end{bmatrix} & ,~~
{\bm {\tilde{V}}}{}^{(2VS)}_{27} =& i \begin{bmatrix}
  0 & \langle \bar{4}\bar{3}2\bar{1} \rangle \\ 
  \langle 4\bar{3}21 \rangle & 0 
 \end{bmatrix} ,~~ \\ 
{\bm {\tilde{V}}}{}^{(2VS)}_{28} =& i \begin{bmatrix}
  0 & \langle 3\bar{4}\bar{1}\bar{2} \rangle \\ 
  \langle 34\bar{1}2 \rangle & 0 
 \end{bmatrix} & ,~~
{\bm {\tilde{V}}}{}^{(2VS)}_{34} =& i \begin{bmatrix}
  \langle \bar{2}1\bar{4}3 \rangle & 0 \\ 
  0 & \langle \bar{2}14\bar{3} \rangle 
 \end{bmatrix} & ,~~
{\bm {\tilde{V}}}{}^{(2VS)}_{35} =& i \begin{bmatrix}
  0 & \langle \bar{3}\bar{4}\bar{1}2 \rangle \\ 
  \langle 3\bar{4}12 \rangle & 0 
 \end{bmatrix} ,~~ \\ 
{\bm {\tilde{V}}}{}^{(2VS)}_{36} =& i \begin{bmatrix}
  0 & \langle \bar{4}3\bar{2}\bar{1} \rangle \\ 
  \langle 43\bar{2}1 \rangle & 0 
 \end{bmatrix} & ,~~
{\bm {\tilde{V}}}{}^{(2VS)}_{37} =& i \begin{bmatrix}
  0 & \langle 1\bar{2}\bar{3}\bar{4} \rangle \\ 
  \langle \bar{1}234 \rangle & 0 
 \end{bmatrix} & ,~~
{\bm {\tilde{V}}}{}^{(2VS)}_{38} =& i \begin{bmatrix}
  0 & \langle 21\bar{4}3 \rangle \\ 
  \langle \bar{2}\bar{1}\bar{4}3 \rangle & 0 
 \end{bmatrix} ,~~ \\ 
{\bm {\tilde{V}}}{}^{(2VS)}_{45} =& i \begin{bmatrix}
  0 & \langle 4\bar{3}\bar{2}\bar{1} \rangle \\ 
  \langle 432\bar{1} \rangle & 0 
 \end{bmatrix} & ,~~
{\bm {\tilde{V}}}{}^{(2VS)}_{46} =& i \begin{bmatrix}
  0 & \langle \bar{3}\bar{4}1\bar{2} \rangle \\ 
  \langle \bar{3}412 \rangle & 0 
 \end{bmatrix} & ,~~
{\bm {\tilde{V}}}{}^{(2VS)}_{47} =& i \begin{bmatrix}
  0 & \langle 214\bar{3} \rangle \\ 
  \langle \bar{2}\bar{1}4\bar{3} \rangle & 0 
 \end{bmatrix} ,~~ \\ 
{\bm {\tilde{V}}}{}^{(2VS)}_{48} =& i \begin{bmatrix}
  0 & \langle \bar{1}2\bar{3}\bar{4} \rangle \\ 
  \langle 1\bar{2}34 \rangle & 0 
 \end{bmatrix} & ,~~
{\bm {\tilde{V}}}{}^{(2VS)}_{56} =& i \begin{bmatrix}
  \langle 2\bar{1}\bar{4}3 \rangle & 0 \\ 
  0 & \langle \bar{2}1\bar{4}3 \rangle 
 \end{bmatrix} & ,~~
{\bm {\tilde{V}}}{}^{(2VS)}_{57} =& i \begin{bmatrix}
  \langle \bar{3}\bar{4}12 \rangle & 0 \\ 
  0 & \langle \bar{3}41\bar{2} \rangle 
 \end{bmatrix} ,~~ \\ 
{\bm {\tilde{V}}}{}^{(2VS)}_{58} =& i \begin{bmatrix}
  \langle 4\bar{3}2\bar{1} \rangle & 0 \\ 
  0 & \langle \bar{4}\bar{3}21 \rangle 
 \end{bmatrix} & ,~~
{\bm {\tilde{V}}}{}^{(2VS)}_{67} =& i \begin{bmatrix}
  \langle \bar{4}3\bar{2}1 \rangle & 0 \\ 
  0 & \langle \bar{4}\bar{3}21 \rangle 
 \end{bmatrix} & ,~~
{\bm {\tilde{V}}}{}^{(2VS)}_{68} =& i \begin{bmatrix}
  \langle \bar{3}\bar{4}12 \rangle & 0 \\ 
  0 & \langle 3\bar{4}\bar{1}2 \rangle 
 \end{bmatrix} ,~~ \\ 
{\bm {\tilde{V}}}{}^{(2VS)}_{78} =& i \begin{bmatrix}
  \langle \bar{2}14\bar{3} \rangle & 0 \\ 
  0 & \langle \bar{2}1\bar{4}3 \rangle 
 \end{bmatrix} ~~.
\end{aligned}
\end{equation}

\subsection{Fermionic Holoraumy Matrices of 1D, $N$ = 8 Tensor Supermultiplets}

\begin{equation}
\begin{aligned}
{\bm {\tilde{V}}}{}^{(2TS)}_{12} =& i \begin{bmatrix}
  \langle\bar{2}1\bar{4}3 \rangle & 0 \\ 
  0 & \langle\bar{2}14\bar{3} \rangle 
 \end{bmatrix} & ,~~ 
{\bm {\tilde{V}}}{}^{(2TS)}_{13} =& i \begin{bmatrix}
  \langle\bar{3}41\bar{2} \rangle & 0 \\ 
  0 & \langle\bar{3}\bar{4}12 \rangle 
 \end{bmatrix} & ,~~ 
{\bm {\tilde{V}}}{}^{(2TS)}_{14} =& i \begin{bmatrix}
  \langle\bar{4}\bar{3}21 \rangle & 0 \\ 
  0 & \langle\bar{4}3\bar{2}1 \rangle 
 \end{bmatrix} ,~~  \\ 
{\bm {\tilde{V}}}{}^{(2TS)}_{15} =& i \begin{bmatrix}
  0 & \langle1\bar{2}\bar{3}\bar{4} \rangle \\ 
  \langle\bar{1}234 \rangle & 0 
 \end{bmatrix} & ,~~ 
{\bm {\tilde{V}}}{}^{(2TS)}_{16} =& i \begin{bmatrix}
  0 & \langle21\bar{4}3 \rangle \\ 
  \langle\bar{2}\bar{1}\bar{4}3 \rangle & 0 
 \end{bmatrix} & ,~~ 
{\bm {\tilde{V}}}{}^{(2TS)}_{17} =& i \begin{bmatrix}
  0 & \langle341\bar{2} \rangle \\ 
  \langle\bar{3}4\bar{1}\bar{2} \rangle & 0 
 \end{bmatrix} ,~~  \\ 
{\bm {\tilde{V}}}{}^{(2TS)}_{18} =& i \begin{bmatrix}
  0 & \langle4\bar{3}21 \rangle \\ 
  \langle\bar{4}\bar{3}2\bar{1} \rangle & 0 
 \end{bmatrix} & ,~~ 
{\bm {\tilde{V}}}{}^{(2TS)}_{23} =& i \begin{bmatrix}
  \langle\bar{4}\bar{3}21 \rangle & 0 \\ 
  0 & \langle4\bar{3}2\bar{1} \rangle 
 \end{bmatrix} & ,~~ 
{\bm {\tilde{V}}}{}^{(2TS)}_{24} =& i \begin{bmatrix}
  \langle3\bar{4}\bar{1}2 \rangle & 0 \\ 
  0 & \langle\bar{3}\bar{4}12 \rangle 
 \end{bmatrix} ,~~  \\ 
{\bm {\tilde{V}}}{}^{(2TS)}_{25} =& i \begin{bmatrix}
  0 & \langle214\bar{3} \rangle \\ 
  \langle\bar{2}\bar{1}4\bar{3} \rangle & 0 
 \end{bmatrix} & ,~~ 
{\bm {\tilde{V}}}{}^{(2TS)}_{26} =& i \begin{bmatrix}
  0 & \langle\bar{1}2\bar{3}\bar{4} \rangle \\ 
  \langle1\bar{2}34 \rangle & 0 
 \end{bmatrix} & ,~~ 
{\bm {\tilde{V}}}{}^{(2TS)}_{27} =& i \begin{bmatrix}
  0 & \langle\bar{4}321 \rangle \\ 
  \langle\bar{4}\bar{3}\bar{2}1 \rangle & 0 
 \end{bmatrix} ,~~  \\ 
{\bm {\tilde{V}}}{}^{(2TS)}_{28} =& i \begin{bmatrix}
  0 & \langle34\bar{1}2 \rangle \\ 
  \langle3\bar{4}\bar{1}\bar{2} \rangle & 0 
 \end{bmatrix} & ,~~ 
{\bm {\tilde{V}}}{}^{(2TS)}_{34} =& i \begin{bmatrix}
  \langle\bar{2}1\bar{4}3 \rangle & 0 \\ 
  0 & \langle2\bar{1}\bar{4}3 \rangle 
 \end{bmatrix} & ,~~ 
{\bm {\tilde{V}}}{}^{(2TS)}_{35} =& i \begin{bmatrix}
  0 & \langle3\bar{4}12 \rangle \\ 
  \langle\bar{3}\bar{4}\bar{1}2 \rangle & 0 
 \end{bmatrix} ,~~  \\ 
{\bm {\tilde{V}}}{}^{(2TS)}_{36} =& i \begin{bmatrix}
  0 & \langle432\bar{1} \rangle \\ 
  \langle4\bar{3}\bar{2}\bar{1} \rangle & 0 
 \end{bmatrix} & ,~~ 
{\bm {\tilde{V}}}{}^{(2TS)}_{37} =& i \begin{bmatrix}
  0 & \langle\bar{1}\bar{2}3\bar{4} \rangle \\ 
  \langle12\bar{3}4 \rangle & 0 
 \end{bmatrix} & ,~~ 
{\bm {\tilde{V}}}{}^{(2TS)}_{38} =& i \begin{bmatrix}
  0 & \langle\bar{2}143 \rangle \\ 
  \langle\bar{2}1\bar{4}\bar{3} \rangle & 0 
 \end{bmatrix} ,~~  \\ 
{\bm {\tilde{V}}}{}^{(2TS)}_{45} =& i \begin{bmatrix}
  0 & \langle43\bar{2}1 \rangle \\ 
  \langle\bar{4}3\bar{2}\bar{1} \rangle & 0 
 \end{bmatrix} & ,~~ 
{\bm {\tilde{V}}}{}^{(2TS)}_{46} =& i \begin{bmatrix}
  0 & \langle\bar{3}412 \rangle \\ 
  \langle\bar{3}\bar{4}1\bar{2} \rangle & 0 
 \end{bmatrix} & ,~~ 
{\bm {\tilde{V}}}{}^{(2TS)}_{47} =& i \begin{bmatrix}
  0 & \langle2\bar{1}43 \rangle \\ 
  \langle2\bar{1}\bar{4}\bar{3} \rangle & 0 
 \end{bmatrix} ,~~  \\ 
{\bm {\tilde{V}}}{}^{(2TS)}_{48} =& i \begin{bmatrix}
  0 & \langle\bar{1}\bar{2}\bar{3}4 \rangle \\ 
  \langle123\bar{4} \rangle & 0 
 \end{bmatrix} & ,~~ 
{\bm {\tilde{V}}}{}^{(2TS)}_{56} =& i \begin{bmatrix}
  \langle\bar{2}14\bar{3} \rangle & 0 \\ 
  0 & \langle\bar{2}1\bar{4}3 \rangle 
 \end{bmatrix} & ,~~ 
{\bm {\tilde{V}}}{}^{(2TS)}_{57} =& i \begin{bmatrix}
  \langle\bar{3}\bar{4}12 \rangle & 0 \\ 
  0 & \langle\bar{3}41\bar{2} \rangle 
 \end{bmatrix} ,~~  \\ 
{\bm {\tilde{V}}}{}^{(2TS)}_{58} =& i \begin{bmatrix}
  \langle\bar{4}3\bar{2}1 \rangle & 0 \\ 
  0 & \langle\bar{4}\bar{3}21 \rangle 
 \end{bmatrix} & ,~~ 
{\bm {\tilde{V}}}{}^{(2TS)}_{67} =& i \begin{bmatrix}
  \langle4\bar{3}2\bar{1} \rangle & 0 \\ 
  0 & \langle\bar{4}\bar{3}21 \rangle 
 \end{bmatrix} & ,~~ 
{\bm {\tilde{V}}}{}^{(2TS)}_{68} =& i \begin{bmatrix}
  \langle\bar{3}\bar{4}12 \rangle & 0 \\ 
  0 & \langle3\bar{4}\bar{1}2 \rangle 
 \end{bmatrix} ,~~  \\ 
{\bm {\tilde{V}}}{}^{(2TS)}_{78} =& i \begin{bmatrix}
  \langle2\bar{1}\bar{4}3 \rangle & 0 \\ 
  0 & \langle\bar{2}1\bar{4}3 \rangle 
 \end{bmatrix} ~~.
\end{aligned}
\end{equation}

\subsection{Fermionic Holoraumy Matrices of 1D, $N$ = 8 Axial-Vector Supermultiplet}

\begin{equation}
\begin{aligned}
{\bm {\tilde{V}}}{}^{(2AVS)}_{12} =& i \begin{bmatrix}
  \langle\bar{2}1\bar{4}3 \rangle & 0 \\ 
  0 & \langle\bar{2}14\bar{3} \rangle 
 \end{bmatrix} & ,~~ 
{\bm {\tilde{V}}}{}^{(2AVS)}_{13} =& i \begin{bmatrix}
  \langle\bar{3}41\bar{2} \rangle & 0 \\ 
  0 & \langle34\bar{1}\bar{2} \rangle 
 \end{bmatrix} & ,~~ 
{\bm {\tilde{V}}}{}^{(2AVS)}_{14} =& i \begin{bmatrix}
  \langle\bar{4}\bar{3}21 \rangle & 0 \\ 
  0 & \langle4\bar{3}2\bar{1} \rangle 
 \end{bmatrix} ,~~  \\ 
{\bm {\tilde{V}}}{}^{(2AVS)}_{15} =& i \begin{bmatrix}
  0 & \langle\bar{1}2\bar{3}\bar{4} \rangle \\ 
  \langle1\bar{2}34 \rangle & 0 
 \end{bmatrix} & ,~~ 
{\bm {\tilde{V}}}{}^{(2AVS)}_{16} =& i \begin{bmatrix}
  0 & \langle\bar{2}\bar{1}\bar{4}3 \rangle \\ 
  \langle21\bar{4}3 \rangle & 0 
 \end{bmatrix} & ,~~ 
{\bm {\tilde{V}}}{}^{(2AVS)}_{17} =& i \begin{bmatrix}
  0 & \langle\bar{3}\bar{4}1\bar{2} \rangle \\ 
  \langle\bar{3}412 \rangle & 0 
 \end{bmatrix} ,~~  \\ 
{\bm {\tilde{V}}}{}^{(2AVS)}_{18} =& i \begin{bmatrix}
  0 & \langle\bar{4}321 \rangle \\ 
  \langle\bar{4}\bar{3}\bar{2}1 \rangle & 0 
 \end{bmatrix} & ,~~ 
{\bm {\tilde{V}}}{}^{(2AVS)}_{23} =& i \begin{bmatrix}
  \langle\bar{4}\bar{3}21 \rangle & 0 \\ 
  0 & \langle\bar{4}3\bar{2}1 \rangle 
 \end{bmatrix} & ,~~ 
{\bm {\tilde{V}}}{}^{(2AVS)}_{24} =& i \begin{bmatrix}
  \langle3\bar{4}\bar{1}2 \rangle & 0 \\ 
  0 & \langle34\bar{1}\bar{2} \rangle 
 \end{bmatrix} ,~~  \\ 
{\bm {\tilde{V}}}{}^{(2AVS)}_{25} =& i \begin{bmatrix}
  0 & \langle\bar{2}\bar{1}4\bar{3} \rangle \\ 
  \langle214\bar{3} \rangle & 0 
 \end{bmatrix} & ,~~ 
{\bm {\tilde{V}}}{}^{(2AVS)}_{26} =& i \begin{bmatrix}
  0 & \langle1\bar{2}\bar{3}\bar{4} \rangle \\ 
  \langle\bar{1}234 \rangle & 0 
 \end{bmatrix} & ,~~ 
{\bm {\tilde{V}}}{}^{(2AVS)}_{27} =& i \begin{bmatrix}
  0 & \langle4\bar{3}21 \rangle \\ 
  \langle\bar{4}\bar{3}2\bar{1} \rangle & 0 
 \end{bmatrix} ,~~  \\ 
{\bm {\tilde{V}}}{}^{(2AVS)}_{28} =& i \begin{bmatrix}
  0 & \langle\bar{3}\bar{4}\bar{1}2 \rangle \\ 
  \langle3\bar{4}12 \rangle & 0 
 \end{bmatrix} & ,~~ 
{\bm {\tilde{V}}}{}^{(2AVS)}_{34} =& i \begin{bmatrix}
  \langle\bar{2}1\bar{4}3 \rangle & 0 \\ 
  0 & \langle2\bar{1}\bar{4}3 \rangle 
 \end{bmatrix} & ,~~ 
{\bm {\tilde{V}}}{}^{(2AVS)}_{35} =& i \begin{bmatrix}
  0 & \langle3\bar{4}\bar{1}\bar{2} \rangle \\ 
  \langle34\bar{1}2 \rangle & 0 
 \end{bmatrix} ,~~  \\ 
{\bm {\tilde{V}}}{}^{(2AVS)}_{36} =& i \begin{bmatrix}
  0 & \langle43\bar{2}1 \rangle \\ 
  \langle\bar{4}3\bar{2}\bar{1} \rangle & 0 
 \end{bmatrix} & ,~~ 
{\bm {\tilde{V}}}{}^{(2AVS)}_{37} =& i \begin{bmatrix}
  0 & \langle\bar{1}\bar{2}\bar{3}4 \rangle \\ 
  \langle123\bar{4} \rangle & 0 
 \end{bmatrix} & ,~~ 
{\bm {\tilde{V}}}{}^{(2AVS)}_{38} =& i \begin{bmatrix}
  0 & \langle\bar{2}1\bar{4}\bar{3} \rangle \\ 
  \langle\bar{2}143 \rangle & 0 
 \end{bmatrix} ,~~  \\ 
{\bm {\tilde{V}}}{}^{(2AVS)}_{45} =& i \begin{bmatrix}
  0 & \langle432\bar{1} \rangle \\ 
  \langle4\bar{3}\bar{2}\bar{1} \rangle & 0 
 \end{bmatrix} & ,~~ 
{\bm {\tilde{V}}}{}^{(2AVS)}_{46} =& i \begin{bmatrix}
  0 & \langle\bar{3}4\bar{1}\bar{2} \rangle \\ 
  \langle341\bar{2} \rangle & 0 
 \end{bmatrix} & ,~~ 
{\bm {\tilde{V}}}{}^{(2AVS)}_{47} =& i \begin{bmatrix}
  0 & \langle2\bar{1}\bar{4}\bar{3} \rangle \\ 
  \langle2\bar{1}43 \rangle & 0 
 \end{bmatrix} ,~~  \\ 
{\bm {\tilde{V}}}{}^{(2AVS)}_{48} =& i \begin{bmatrix}
  0 & \langle\bar{1}\bar{2}3\bar{4} \rangle \\ 
  \langle12\bar{3}4 \rangle & 0 
 \end{bmatrix} & ,~~ 
{\bm {\tilde{V}}}{}^{(2AVS)}_{56} =& i \begin{bmatrix}
  \langle\bar{2}14\bar{3} \rangle & 0 \\ 
  0 & \langle\bar{2}1\bar{4}3 \rangle 
 \end{bmatrix} & ,~~ 
{\bm {\tilde{V}}}{}^{(2AVS)}_{57} =& i \begin{bmatrix}
  \langle34\bar{1}\bar{2} \rangle & 0 \\ 
  0 & \langle\bar{3}41\bar{2} \rangle 
 \end{bmatrix} ,~~  \\ 
{\bm {\tilde{V}}}{}^{(2AVS)}_{58} =& i \begin{bmatrix}
  \langle4\bar{3}2\bar{1} \rangle & 0 \\ 
  0 & \langle\bar{4}\bar{3}21 \rangle 
 \end{bmatrix} & ,~~ 
{\bm {\tilde{V}}}{}^{(2AVS)}_{67} =& i \begin{bmatrix}
  \langle\bar{4}3\bar{2}1 \rangle & 0 \\ 
  0 & \langle\bar{4}\bar{3}21 \rangle 
 \end{bmatrix} & ,~~ 
{\bm {\tilde{V}}}{}^{(2AVS)}_{68} =& i \begin{bmatrix}
  \langle34\bar{1}\bar{2} \rangle & 0 \\ 
  0 & \langle3\bar{4}\bar{1}2 \rangle 
 \end{bmatrix} ,~~  \\ 
{\bm {\tilde{V}}}{}^{(2AVS)}_{78} =& i \begin{bmatrix}
  \langle2\bar{1}\bar{4}3 \rangle & 0 \\ 
  0 & \langle\bar{2}1\bar{4}3 \rangle 
 \end{bmatrix} ~~.
\end{aligned}
\end{equation}

\subsection{Fermionic Holoraumy Matrices of 1D, $N$ = 8 Axial-Tensor Supermultiplets}

\begin{equation}
\begin{aligned}
{\bm {\tilde{V}}}{}^{(2ATS)}_{12} =& i \begin{bmatrix}
  \langle\bar{2}1\bar{4}3 \rangle & 0 \\ 
  0 & \langle2\bar{1}\bar{4}3 \rangle 
 \end{bmatrix} & ,~~ 
{\bm {\tilde{V}}}{}^{(2ATS)}_{13} =& i \begin{bmatrix}
  \langle\bar{3}41\bar{2} \rangle & 0 \\ 
  0 & \langle34\bar{1}\bar{2} \rangle 
 \end{bmatrix} & ,~~ 
{\bm {\tilde{V}}}{}^{(2ATS)}_{14} =& i \begin{bmatrix}
  \langle\bar{4}\bar{3}21 \rangle & 0 \\ 
  0 & \langle\bar{4}3\bar{2}1 \rangle 
 \end{bmatrix} ,~~  \\ 
{\bm {\tilde{V}}}{}^{(2ATS)}_{15} =& i \begin{bmatrix}
  0 & \langle\bar{1}\bar{2}\bar{3}4 \rangle \\ 
  \langle123\bar{4} \rangle & 0 
 \end{bmatrix} & ,~~ 
{\bm {\tilde{V}}}{}^{(2ATS)}_{16} =& i \begin{bmatrix}
  0 & \langle\bar{2}1\bar{4}\bar{3} \rangle \\ 
  \langle\bar{2}143 \rangle & 0 
 \end{bmatrix} & ,~~ 
{\bm {\tilde{V}}}{}^{(2ATS)}_{17} =& i \begin{bmatrix}
  0 & \langle\bar{3}412 \rangle \\ 
  \langle\bar{3}\bar{4}1\bar{2} \rangle & 0 
 \end{bmatrix} ,~~  \\ 
{\bm {\tilde{V}}}{}^{(2ATS)}_{18} =& i \begin{bmatrix}
  0 & \langle\bar{4}\bar{3}2\bar{1} \rangle \\ 
  \langle4\bar{3}21 \rangle & 0 
 \end{bmatrix} & ,~~ 
{\bm {\tilde{V}}}{}^{(2ATS)}_{23} =& i \begin{bmatrix}
  \langle\bar{4}\bar{3}21 \rangle & 0 \\ 
  0 & \langle4\bar{3}2\bar{1} \rangle 
 \end{bmatrix} & ,~~ 
{\bm {\tilde{V}}}{}^{(2ATS)}_{24} =& i \begin{bmatrix}
  \langle3\bar{4}\bar{1}2 \rangle & 0 \\ 
  0 & \langle34\bar{1}\bar{2} \rangle 
 \end{bmatrix} ,~~  \\ 
{\bm {\tilde{V}}}{}^{(2ATS)}_{25} =& i \begin{bmatrix}
  0 & \langle2\bar{1}\bar{4}\bar{3} \rangle \\ 
  \langle2\bar{1}43 \rangle & 0 
 \end{bmatrix} & ,~~ 
{\bm {\tilde{V}}}{}^{(2ATS)}_{26} =& i \begin{bmatrix}
  0 & \langle\bar{1}\bar{2}3\bar{4} \rangle \\ 
  \langle12\bar{3}4 \rangle & 0 
 \end{bmatrix} & ,~~ 
{\bm {\tilde{V}}}{}^{(2ATS)}_{27} =& i \begin{bmatrix}
  0 & \langle\bar{4}\bar{3}\bar{2}1 \rangle \\ 
  \langle\bar{4}321 \rangle & 0 
 \end{bmatrix} ,~~  \\ 
{\bm {\tilde{V}}}{}^{(2ATS)}_{28} =& i \begin{bmatrix}
  0 & \langle3\bar{4}12 \rangle \\ 
  \langle\bar{3}\bar{4}\bar{1}2 \rangle & 0 
 \end{bmatrix} & ,~~ 
{\bm {\tilde{V}}}{}^{(2ATS)}_{34} =& i \begin{bmatrix}
  \langle\bar{2}1\bar{4}3 \rangle & 0 \\ 
  0 & \langle\bar{2}14\bar{3} \rangle 
 \end{bmatrix} & ,~~ 
{\bm {\tilde{V}}}{}^{(2ATS)}_{35} =& i \begin{bmatrix}
  0 & \langle34\bar{1}2 \rangle \\ 
  \langle3\bar{4}\bar{1}\bar{2} \rangle & 0 
 \end{bmatrix} ,~~  \\ 
{\bm {\tilde{V}}}{}^{(2ATS)}_{36} =& i \begin{bmatrix}
  0 & \langle4\bar{3}\bar{2}\bar{1} \rangle \\ 
  \langle432\bar{1} \rangle & 0 
 \end{bmatrix} & ,~~ 
{\bm {\tilde{V}}}{}^{(2ATS)}_{37} =& i \begin{bmatrix}
  0 & \langle\bar{1}2\bar{3}\bar{4} \rangle \\ 
  \langle1\bar{2}34 \rangle & 0 
 \end{bmatrix} & ,~~ 
{\bm {\tilde{V}}}{}^{(2ATS)}_{38} =& i \begin{bmatrix}
  0 & \langle\bar{2}\bar{1}\bar{4}3 \rangle \\ 
  \langle21\bar{4}3 \rangle & 0 
 \end{bmatrix} ,~~  \\ 
{\bm {\tilde{V}}}{}^{(2ATS)}_{45} =& i \begin{bmatrix}
  0 & \langle\bar{4}3\bar{2}\bar{1} \rangle \\ 
  \langle43\bar{2}1 \rangle & 0 
 \end{bmatrix} & ,~~ 
{\bm {\tilde{V}}}{}^{(2ATS)}_{46} =& i \begin{bmatrix}
  0 & \langle341\bar{2} \rangle \\ 
  \langle\bar{3}4\bar{1}\bar{2} \rangle & 0 
 \end{bmatrix} & ,~~ 
{\bm {\tilde{V}}}{}^{(2ATS)}_{47} =& i \begin{bmatrix}
  0 & \langle\bar{2}\bar{1}4\bar{3} \rangle \\ 
  \langle214\bar{3} \rangle & 0 
 \end{bmatrix} ,~~  \\ 
{\bm {\tilde{V}}}{}^{(2ATS)}_{48} =& i \begin{bmatrix}
  0 & \langle1\bar{2}\bar{3}\bar{4} \rangle \\ 
  \langle\bar{1}234 \rangle & 0 
 \end{bmatrix} & ,~~ 
{\bm {\tilde{V}}}{}^{(2ATS)}_{56} =& i \begin{bmatrix}
  \langle2\bar{1}\bar{4}3 \rangle & 0 \\ 
  0 & \langle\bar{2}1\bar{4}3 \rangle 
 \end{bmatrix} & ,~~ 
{\bm {\tilde{V}}}{}^{(2ATS)}_{57} =& i \begin{bmatrix}
  \langle34\bar{1}\bar{2} \rangle & 0 \\ 
  0 & \langle\bar{3}41\bar{2} \rangle 
 \end{bmatrix} ,~~  \\ 
{\bm {\tilde{V}}}{}^{(2ATS)}_{58} =& i \begin{bmatrix}
  \langle\bar{4}3\bar{2}1 \rangle & 0 \\ 
  0 & \langle\bar{4}\bar{3}21 \rangle 
 \end{bmatrix} & ,~~ 
{\bm {\tilde{V}}}{}^{(2ATS)}_{67} =& i \begin{bmatrix}
  \langle4\bar{3}2\bar{1} \rangle & 0 \\ 
  0 & \langle\bar{4}\bar{3}21 \rangle 
 \end{bmatrix} & ,~~ 
{\bm {\tilde{V}}}{}^{(2ATS)}_{68} =& i \begin{bmatrix}
  \langle34\bar{1}\bar{2} \rangle & 0 \\ 
  0 & \langle3\bar{4}\bar{1}2 \rangle 
 \end{bmatrix} ,~~  \\ 
{\bm {\tilde{V}}}{}^{(2ATS)}_{78} =& i \begin{bmatrix}
  \langle\bar{2}14\bar{3} \rangle & 0 \\ 
  0 & \langle\bar{2}1\bar{4}3 \rangle 
 \end{bmatrix} ~~.
\end{aligned}
\end{equation}

\newpage
$$~~$$

\end{document}